\documentclass{iopart}
\usepackage{iopams} 
\usepackage{setstack}
\usepackage{geometry}
\usepackage{color}
\usepackage{amssymb}
\usepackage{mathrsfs}

\pdfoutput=1

\usepackage{graphicx,epsfig}
\usepackage{bm}% bold math

\usepackage{braket}

\newcommand{\be}{\begin{equation}}
\newcommand{\ee}{\end{equation}}
\newcommand{\bea}{\begin{eqnarray}}
\newcommand{\eea}{\end{eqnarray}}
\newcommand{\nn}{\nonumber\\}
\newcommand\eps{\epsilon}
\newcommand\veps{\varepsilon}

\def\fr#1{(\ref{#1})}
\def\sumprime{\sideset{}{'}\sum}

\newcommand{\s}{\sigma}
\newcommand{\R}{{\mathrm{R}}}
\newcommand{\NS}{{\mathrm{NS}}}
\renewcommand{\Eref}[1]{Eq.~(\ref{#1})}

\newcommand{\lrsub}[3]{{\vphantom{#2}}_{#1}{\!\!\; #2}_{#3}}
\newcommand{\binom}[2]{\Bigl(\begin{array}{c}#1\\#2
\end{array}\Bigr)
}

\begin{document}
\title[Quantum Quench in the Transverse Field Ising chain I]{Quantum
Quench in the Transverse Field Ising chain I: Time evolution of
order parameter correlators} 
\author{Pasquale Calabrese$^1$, Fabian H.L. Essler$^2$, and Maurizio
  Fagotti$^{2}$} 
\address{$^1$ Dipartimento di Fisica dell'Universit\`a di Pisa and INFN - Pisa 56127, Italy}
\address{$^2$ The Rudolf Peierls Centre for Theoretical Physics, University of Oxford -  Oxford OX1 3NP, United Kingdom}
\begin{abstract}
We consider the time evolution of order parameter correlation
functions after a sudden quantum quench of the magnetic field in the
transverse field Ising chain. Using two novel methods based on
determinants and form factor sums respectively, we derive
analytic expressions for the asymptotic behaviour of one and two point
correlators. We discuss quenches within the ordered and disordered
phases as well as quenches between the phases and to the quantum
critical point. We give detailed account of both methods.
\end{abstract}

\maketitle

%%%%%%%%%%%%
\section{Introduction}%
%%%%%%%%%%%%

The non-equilibrium dynamics of isolated quantum systems after a
sudden ``quench'' of a parameter characterizing their respective
Hamiltonians is a subject currently under intensive theoretical and
experimental investigation.
Recent experiments on trapped ultra-cold atomic gases
\cite{uc,tc-07,tetal-11,cetal-12,getal-11,kww-06} have established that
these systems are sufficiently weakly coupled to their environments as 
to allow the observation of essentially unitary non-equilibrium time
evolution on very long time scales. This in turn provides the
opportunity of investigating fundamental questions of many-body
quantum mechanics, which in standard condensed matter systems are
obscured by decoherence and dissipation.
Two of the main questions raised in these works are
(i) how fast correlations spread in quantum many-body systems and
(ii) whether observables such as (multi-point) correlation functions 
generically relax to time independent values, and if they do, what
principles determine their stationary properties.
The first issue was addressed in a seminal work by Lieb and Robinson
\cite{lr-72}, which established that in lattice many-body systems 
information has a finite speed of propagation and provided a bound on
the maximal group velocity. In recent years an effective light-cone
scenario has been proposed theoretically
\cite{cc-05,cc-06,cc-07l,sc-10,ds-11}, was tested in many numerical
computations \cite{mwnm-09,lk-08,bpck-12,gree-12,ham-12,es-12} and
was finally observed in cold-atom experiments \cite{cetal-12}.  

The relaxation towards stationary behaviour at first appears very
surprising, because unitary time evolution maintains the system in a
pure state at all times. The resolution of this apparent paradox is that
in the thermodynamic limit, (finite) \emph{subsystems} can and do display
correlations characterized by a mixed state, namely the one obtained
by tracing out the degrees of freedom outside the subsystem itself. In
other words, the system acts as its own bath.

In groundbreaking (``quantum Newton's cradle'') experiments on the
relaxation towards stationary states in systems of ultra-cold atoms
Kinoshita, Wenger and Weiss \cite{kww-06} demonstrated the importance
of dimensionality and conservation laws for many-body quantum dynamics
out of equilibrium. In essence, these experiments show that
three dimensional condensates relax quickly to a stationary state that
is characterized by an effective temperature, a process known as
``thermalization'', whereas the relaxation of quasi one-dimensional
systems is slow and towards an unusual non-thermal distribution.
This difference has been attributed to the existence of additional 
(approximate) local conservation laws in the quasi-1D case, which are
argued to constrain the dynamics in analogy with classical integrable
systems. The findings of Ref.~\cite{kww-06} sparked a tremendous
theoretical effort aimed at clarifying the effects of quantum
integrability on the non-equilibrium evolution in many-particle
quantum systems \cite{rev,gg,rdo-08,cc-06,cc-07,caz-06,mw-07,cdeo-08,bs-08,fcm-08,ke-08,scc-09,roux-09,sfm-11,fm-10,bdkm-11,kla-06,bkl-10,bhc-10,gce-10,rf-11,ccrss-11,cic-11,rs-11,zch-11,spr-11,mc-12,gm-12,ck-12,mc-12b,can}. 
Many of these studies are compatible with the widely held belief (see
e.g. \cite{rev} for a comprehensive summary) that the reduced density
matrix of any finite subsystem (which determines correlation functions
of any local observables within the subsystem) of an infinite system
can be described in terms of either an effective thermal (Gibbs)
distribution or a so-called generalized Gibbs ensemble
\cite{gg}. It has been conjectured that the latter arises for integrable
models, while the former is obtained for generic systems and evidence
supporting this view has been obtained in a number of examples
\cite{gg,rdo-08,cc-07,caz-06,mw-07,cdeo-08,bs-08,scc-09,sfm-11,fm-10,bdkm-11}.
On the other hand, several numerical studies
\cite{kla-06,bhc-10,gce-10,mc-12,gm-12} suggest that the full picture
may well be more complex.

Moreover, open questions remain even with regards
to the very existence of stationary states. For example, the order
parameter of certain mean-field models have been shown to
display persistent oscillations \cite{MF1,MF2,MF3,MF4,MF5,MF6}.  
Non-decaying oscillations have also been observed numerically \cite{bhc-10}
in some non-integrable one-dimensional systems. This has given rise to
the concept of ``weak thermalization'', which refers to a situation where
time-averaged quantities are thermal, while oscillations persist
on short time-scales.  
We note that questions related to thermalization and spreading of
correlations have also been studied using holographic techniques
\cite{ads1,ads2,ads3}. 

Given that stationary behaviour is strictly speaking a property at
infinite times in the thermodynamic limit (and these limits do not
commute) it is important to have available analytic results that become
exact in certain limits. To that end we consider here the
non-equilibrium dynamics of the transverse field Ising chain  
%\footnote{In Ref. \cite{CEF} we fixed $J=1/2$, thus keep this in mind when comparing the formulas here with \cite{CEF}.}
\be
H(h)=-J\sum_{j=1}^L\Bigl[\sigma_j^x\sigma_{j+1}^x+h\sigma_j^z\Bigr]\, ,
\label{Hamiltonian}
\ee
where $\sigma_j^\alpha$ are the Pauli matrices at site $j$, $J>0$ and we
impose periodic boundary conditions 
$\sigma_{L+1}^\alpha=\sigma^\alpha_1$. 
%As the sign of $J$ can be changed by a unitary transformation we
%assume without loss of generality $J>0$ from here on. 
The Hamiltonian \fr{Hamiltonian} exhibits a
$\mathbb{Z}_2$ symmetry of rotations around the z-axis in spin space
by 180 degrees 
\be
\sigma_j^{\alpha}\rightarrow-\sigma_j^\alpha\ ,\quad
\alpha=x,y\ ,\qquad
\sigma_j^z\rightarrow\sigma_j^z.
\label{Z2}
\ee

The model \fr{Hamiltonian} is a crucial paradigm of quantum critical
behaviour and quantum phase transitions \cite{sach-book}. 
At zero temperature and in the thermodynamic limit it exhibits
ferromagnetic ($h < 1$) and paramagnetic ($h > 1$) phases, separated
by a quantum critical point at $h_c=1$. 
For $h<1$ and $L\rightarrow\infty$ there are two degenerate ground
states related by the $\mathbb{Z}_2$ symmetry \fr{Z2}. Spontaneous
symmetry breaking selects a unique ground state, in which spins align
along the $x$-direction. On the other hand, for magnetic fields $h>1$
the ground state is non-degenerate and as the magnetic field $h$ is
increased spins align more and more along the $z$-direction. 
The order parameter for the quantum phase transition is the ground
state expectation value $\braket{\sigma^x_j}$.
In the following we will refer to the ferromagnetic phase as the
ordered phase, and to the paramagnetic one as the disordered phase.
We note that the model \fr{Hamiltonian} is (approximately) realized
both in solids \cite{ising-solids} and in systems of cold Rubidium
atoms confined in an optical lattice \cite{ising-ca}. In the latter
realization it is possible to investigate its non-equilibrium dynamics
experimentally.

As shown in \ref{app:diag}, the model can be diagonalized by a
Jordan-Wigner transformation which maps the model to spinless fermions
with local annihilation operators $c_j$, followed by a Fourier
transform and a Bogolioubov transformation. In terms of the
momentum space Bogoliubov fermions $\alpha_k$ the Hamiltonian reads
\be
H(h)=\sum_{k}
\veps_h(k) {\alpha}^\dagger_{k}{\alpha}_{k}\,, \qquad
%\ee
%where we dropped an unimportant constant and we defined
%\bea
%\tan \theta_k&=&\left[\frac{\sin(k)}{\cos(k)-h}\right] ,\nn
\veps_h(k)=2J\sqrt{1+h^2-2h\cos(k)}.
\label{Hbog}
\ee
For details and more precise definitions see \ref{app:diag}.

%%%%%%%%%%%%%%%%%%%%%
\subsection{Quench protocol and observables}
\label{s:quench}%
%%%%%%%%%%%%%%%%%%%%%

In the following we focus on a {\it global quantum quench} of the
magnetic field. We assume that the many-body system is prepared in the
ground state $|\Psi_0\rangle$ of Hamiltonian $H(h_0)$. At time $t=0$
the field $h_0$ is changed instantaneously to a different value $h$ and
one then considers the unitary time evolution of the system
characterized by the new Hamiltonian $H(h)$, i.e. the initial state
$|\Psi_0\rangle$ evolves as
\be
|\Psi_0(t)\rangle=e^{-itH(h)}|\Psi_0\rangle.
\ee
The above protocol corresponds to an experimental situation, in which
a system parameter has been changed on a time scale that is small
compared to any other time scale in the system. We note that this can
be achieved in cold-atom experiments \cite{uc,tetal-11,cetal-12}.

In this paper we study
%Some of the most natural observables to be studied are 
the one and two-point functions of the order parameter
\bea
\rho^x(t)&=&\frac{\langle\Psi_0(t)|\sigma^x_\ell|\Psi_0(t)\rangle}
{\langle\Psi_0(t)|\Psi_0(t)\rangle} ,\\
\rho^{xx}(\ell,t)&=&\frac{\langle\Psi_0(t)|\sigma^x_{j+\ell}\sigma^x_j
|\Psi_0(t)\rangle}{\langle\Psi_0(t)|\Psi_0(t)\rangle}, \qquad  \rho^{xx}_c(\ell,t)=\rho^{xx}(\ell,t)-(\rho^x(t))^2.
\eea
%and of the transverse spins 
%\bea
%\rho^z(t)=\frac{\langle\Psi_0(t)|\sigma^z_\ell|\Psi_0(t)\rangle}
%{\langle\Psi_0(t)|\Psi_0(t)\rangle},
%\qquad  \rho^{zz}_c(\ell,t)=\frac{\langle\Psi_0(t)|\sigma^z_{j+\ell}\sigma^z_j
%|\Psi_0(t)\rangle}{\langle\Psi_0(t)|\Psi_0(t)\rangle}
%-(\rho^z(t))^2.
%\eea
%A global quantum quench of the transverse field in the Ising model
%is an ideal testing ground for many thermalization ideas. An example
%is to what extent thermalization could depend on the observable itself.
%As was pointed out in \cite{rsms-08,bhc-10,can}, one may expect the
%\emph{locality} of the observable with respect to the elementary
%excitations of the system to affect the late time behaviour of an
%observable after a quantum quench. In the Ising chain, the transverse
%spin operator $\sigma^z_j$ is local with respect to the fermionic
%degrees of freedom $c_j$, while the order parameter $\sigma^x_j$ is
%non-local. Hence the Ising chain is a perfect testing ground
%for analyzing the role played by locality \cite{rsms-08}. 
%

Although quantum quenches in the 1D Ising model have been the subject
of many works \cite{rsms-08,mc,ir-00,ir-10,sps-04,fc-08,s-08,cz-10,fcg-11,ri-11},
results on the time evolution of order parameter correlation functions were
reported only very recently by us in a short communication \cite{CEF}. 
In the following we give detailed derivations on the {\it full
  asymptotic time and distance dependence} of one- and two-point order
parameter correlation functions in the thermodynamic limit for
quenches within the two phases. These have been obtained by two novel,
complementary methods, and we discuss both of them in detail. 
The first method is based on the determinant representation of
correlation functions characteristic of free-fermionic theories. The
second is based on the form-factor approach
\cite{Smirnov92book,FF,Delfino04,mussardobook,review} and is
applicable more generally to integrable quenches in interacting
quantum field theories \cite{bpgda-10,fm-10}. We stress that this
approach is qualitatively different from numerical approaches based on
quantum integrability \cite{INT,grd-09,ns-11,mc-12,ck-12}. In particular it
allows us to obtain analytical results in the thermodynamic limit.
New results not previously reported in \cite{CEF} include expressions
for the two-point function of the order parameter for quenches within
the disordered phase and for quenches across the critical point.

%Unlike order parameter correlators the time evolution of one and two
%point functions of the transverse spins is straightforward to analyze
%(as the latter are fermion bilinears) and has been known since 1970
%\cite{mc}. Despite of their simplicity, the connected two-point
%function $\rho^{zz}_c(\ell,t)$ (corresponding to density
%correlations in gases) exhibits an interesting and quite general
%phenomenon, namely a cross-over in the relaxational behaviour at an
%exponentially large time scale $ t_{\rm cross}\sim e^\ell$. As a consequence
%the truly stationary regime of $\rho^{zz}_c(\ell,t)$ is observed
%only at very late times $t>t_{\rm cross}$, which in general is a
%serious limitation.

%%%%%%%%%%%%%%%%%%%%%%%%%%%%%%%%%%%%%%%%%%%%%%%%%%%%%
\subsection{The quench variables}\label{s:The model}%
%%%%%%%%%%%%%%%%%%%%%%%%%%%%%%%%%%%%%%%%%%%%%%%%%%%%%

As shown in \ref{app:diag} both the initial and final
Hamiltonians can be diagonalized by combined Jordan-Wigner
and Bogoliubov transformations with Bogoliubov angles
$\theta_k^0$ and $\theta_k$ respectively. The corresponding Bogoliubov
fermions are related by a linear transformation \fr{relation}
characterized by the difference $\Delta_k=\theta_k-\theta^0_k$. 
In order to parametrize the quench it is useful to introduce
the quantity
\be
\cos \Delta_k=\frac{h h_0- (h+h_0) \cos k+1}{\sqrt{1+h^2-2h\cos(k)} \sqrt{1+h_0^2-2h_0\cos(k)}}.
\ee
We note that $\cos \Delta_k$ is invariant under the two transformations
\bea
(h_0,h)&\rightarrow (h,h_0)\quad  {\rm and}\quad
(h_0,h)\rightarrow \Big(\frac{1}{h_0},\frac{1}{h}\Big)\ .
\label{maps}
\eea
However, we stress that the quantum quench itself is not invariant
under the maps \fr{maps}. In the form factor approach a more natural
quantity to consider is 
\be
K(p)=\tan[\Delta_p/2].
\label{Kofp}
\ee

%%%%%%%%%%%%%%%%%%%%%%%%%%%%%%%%%%%%%%%%%%%
\subsection{Time scales for two-point correlators.}
%%%%%%%%%%%%%%%%%%%%%%%%%%%%%%%%%%%%%%%%%%%

In this manuscript we are mainly concerned with equal time correlation
functions of spins separated by a distance $\ell$, which we take
to be much larger than the lattice spacing, i.e. $\ell\gg 1$.
For fixed $\ell$ the time evolution is naturally divided into three
regimes, which are determined by the propagation velocity of
elementary excitations of the post-quench Hamiltonian 
$v(k)=\frac{d\veps_h(k)}{dk}$. For a given final magnetic field $h$, the
maximal propagation velocity is
\be
v_{\rm max}= \max_{k\in [-\pi,\pi]} |\veps'_h(k)|= 2J\min[h,1]\,.
\ee
The three different regimes are:
\begin{itemize}
\item Short-times $v_{\rm max} t\ll \ell$. 
\item Intermediate times $v_{\rm max} t\sim \ell$. This regime is of
particular importance for both experiments and numerical computations.
A convenient way of describing this regime is to consider evolution
along a particular ``ray'' $\kappa\ell=v_{\rm max}t$ in space-time,
see Fig.~\ref{fig:STSL}.
In order to obtain a very accurate description of the dynamics at a
particular point along this ray, one may then construct an asymptotic
expansion in the single variable $\ell$ around the 
{\it space-time scaling limit} $v_{\rm max}t,\ell\to\infty$, $\kappa$
fixed.
\begin{figure}[t]
\includegraphics[width=0.75\textwidth]{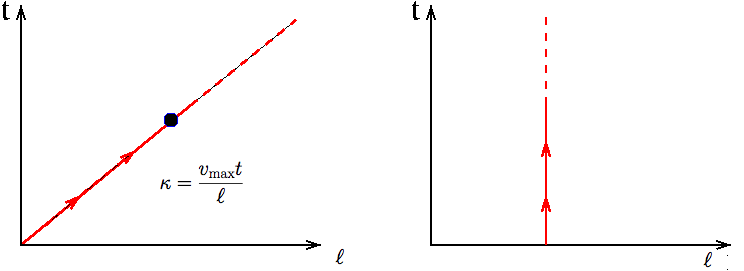}
\caption{Left panel: for intermediate times $v_{\rm max}t\sim\ell$ the
behaviour of $\rho^{\alpha\alpha}(\ell,t)$ is most conveniently
determined by considering its asyptotic expansion around infinity
(``space-time scaling limit'') along the ray $v_{\rm
  max}t=\kappa\ell$. This viewpoint is appropriate for any large,
finite $t$ and $\ell$. Right panel: the asymptotic late-time regime
is reached by considering time evolution at fixed $\ell$. To describe
this regime one should consider an asymptotic expansion of
$\rho^{\alpha\alpha}(\ell,t)$ around $t=\infty$ at fixed $\ell$.}
\label{fig:STSL}
 \end{figure}

\item Late times $v_{\rm max} t\gg \ell$. This includes the limit
  $t\to\infty$ at fixed but large $\ell$. In this regime it is no
longer convenient to consider evolution along a particular ray in
space-time. In order to obtain accurate results for the late time
dynamics, one should construct an asymptotic expansion in $t$
around infinity, see Fig.~\ref{fig:STSL}.
\end{itemize}
Because of the horizon effect \cite{cc-06,cc-07} the short-time regime
does not display interesting features. % for connected correlators. 
We therefore focus on intermediate and late times. 
The most convenient way of analyzing the intermediate time regime is
via the space-time scaling limit. It is important to note that taking the
limit $\kappa\to 0$ of results obtained in the space-time scaling
limit is \emph{not} expected to reproduce the short-time regime.
Similarly, in general taking $\kappa\to\infty$ in the space-time
scaling limit does not necessarily reproduce the late time behaviour.
%i.e. in general we have
%\be
%\lim_{\kappa\to\infty}\lim_{v_{\rm max}t,\ell\to\infty\atop \kappa\ {\rm
%  fixed}}\rho^{\alpha\alpha}_x(\ell,t)\neq
%\lim_{\ell\to\infty}\lim_{v_{\rm max}t\to\infty}\rho^{\alpha\alpha}_x(\ell,t).
%\ee
%We will show that the limits \emph{do commute} for $\rho^{xx}(\ell,t)$
%for most but not all quenches, but they never commute for $\rho^{zz}(\ell,t)$.

\subsection{Organization of the manuscript.}
This is the first in a series of two papers of the dynamics in the
transverse field Ising chain after a sudden quench of the magnetic
field. The second paper, henceforth referred to as ``paper II'',
gives a detailed account of properties in the stationary state,
i.e. at $t=\infty$. The present manuscript deals with the time
evolution of observables and is organized in the following way.
In section \ref{sec:summaryANDdiscussion} we present a detailed
summary of our main results. In section \ref{s:Determinant
  approach} we discuss the determinant approach for calculating
two-point functions after quantum quenches in models with free fermionic
spectra. Section \ref{sec:FF} introduces the form factor approach to
correlation functions in integrable models after quantum quenches and
gives a detailed account of its application to one and two point
functions in the Ising chain. The scaling limit for quenches close to
the critical point is constructed in section \ref{sec:scaling}, and
section \ref{sec:Conclusions} contains our conclusions. Our
conventions for diagonalizing the Ising Hamiltonian with periodic
boundary conditions are summarized in
\ref{app:diag} and the initial state is expressed in terms of
eigenstates of the post-quench Hamiltonian in \ref{app:initial}.
Finally, \ref{App:prod} and \ref{a:useful} deal with
certain technical issues.

\section{Summary and discussions of the results}
\label{sec:summaryANDdiscussion}

In this section we present a comprehensive summary of our main
results. Our analytic results are valid in the thermodynamic limit
$L\to\infty$ and are obtained by two different
methods.
\begin{enumerate}
\item{} The first is based on representing correlation functions as
determinants and then determining the asymptotic behaviour in the
space-time scaling limit
\be
t,\ell\to\infty\ , \frac{v_{\rm max}t}{\ell}\ {\rm fixed}.
\ee
Results obtained by this methods are exact.
\item{} 
The second method employs a Lehmann representation for
correlation functions, which provides an expansion of the correlator
in powers of the functions $K(k)$ \fr{Kofp}. The Lehmann
representation is recast as a \emph{low-density expansion} and then
the dominant terms at late times and large distances are summed to all
orders in $K(k)$. The method exploits the existence of a small
parameter, namely the average densities $n(k)$ of elementary
excitations of the post-quench Hamiltonian $H(h)$ with momentum $k$ in
the initial state $|\Psi_0(0)\rangle$ 
\be
\frac{\langle\Psi_0(0)| n(k)|\Psi_0(0)\rangle}
{\langle\Psi_0(0)| \Psi_0(0)\rangle}
=\frac{K^2(k)}{1+K^2(k)}\ .
\ee
A \emph{small quench} is defined as being such that these densities
are small for all $k$, i.e. ${\rm max}_kK^2(k)\ll 1$ (this does not
necessarily imply that $h$ and $h_0$ have to be very close to each
other). The form factor approach provides accurate results for small
quenches at late times and large distances. In practice the form
factor method provides very good approximations to the exact result,
except for quenches to or from the close vicinity of the critical
point. 
\end{enumerate}
Analytic results obtained by the two methods are compared to a direct
numerical evaluation of the determinant representations of correlation
functions in the thermodynamic limit. Finite-size effects are
concomitantly absent and for the purposes of the comparisons shown in
the following figures the numerical results can therefore be
considered to be exact. 

\subsection{One-point correlation function}

For quenches starting in the disordered phase, i.e. $h_0>1$, the order
parameter expectation value is zero for all times because the
$\mathbb{Z}_2$ symmetry remains unbroken.
For quenches that start and end in the ordered phase, i.e. $h_0, h<1$,
exact calculations based on the determinant approach show that
the order parameter relaxes to zero exponentially at late times ($t\gg
1$) (note that since $\ln \left |\cos\Delta_k\right|<0$ the exponential is
always decreasing) 
\begin{equation}
\label{eq:OP}
\langle \sigma^x_i(t)\rangle \simeq 
\left({\cal C}^x_{\rm FF}\right)^\frac{1}{2}
\exp\left[t \int_{0}^\pi
  \frac{\mathrm d k}{\pi} \veps'_h(k) 
\ln\left |\cos\Delta_k\right| \right],
\end{equation}
where
\be
{\cal C}^x_{\rm FF}=\frac{1-hh_0+\sqrt{(1-h^2)(1-h_0^2)}}
{2\sqrt{1-hh_0}(1-h_0^2)^\frac{1}{4}}\ .
\label{CFF}
\ee
This result is obtained by applying the cluster decomposition
principle to the two-point function \fr{eq:predictionintr}
(see \Eref{eq:prediction} in Section \ref{s:Determinant approach} for
the proof of \Eref{eq:predictionintr}). The form factor approach gives
(see Section \ref{ss:1point})
\be
\label{eq:OPFF}
\langle \sigma^x_i(t)\rangle \simeq
(1-h^2)^\frac{1}{8}\exp\left[-t \int_{0}^\pi \frac{\mathrm d k}{\pi}
  \veps'_h(k)\ {2K^2(k)}\right].
\ee
We note that
\bea
\ln\left|\cos\Delta_k\right|
=\ln\left[\frac{1-K^2(k)}{1+K^2(k)}\right]=-2K^2(k)-\frac{2}{3}K^6(k)+\ldots\ ,\nn
\left({\cal C}^x_{\rm FF}\right)^\frac{1}{2}\simeq (1-h^2)^\frac{1}{8}+
\frac{(h-h_0)^4}{64(1-h^2)^\frac{31}{8}}+\ldots,
\eea
so that \fr{eq:OPFF} is indeed the ``low-density'' approximation to
\fr{eq:OP}.
\begin{figure}[t]
\includegraphics[width=0.48\textwidth]{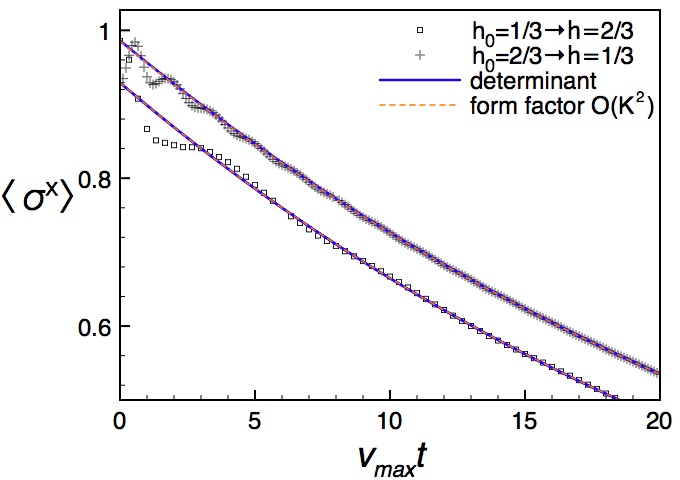}
\includegraphics[width=0.48\textwidth]{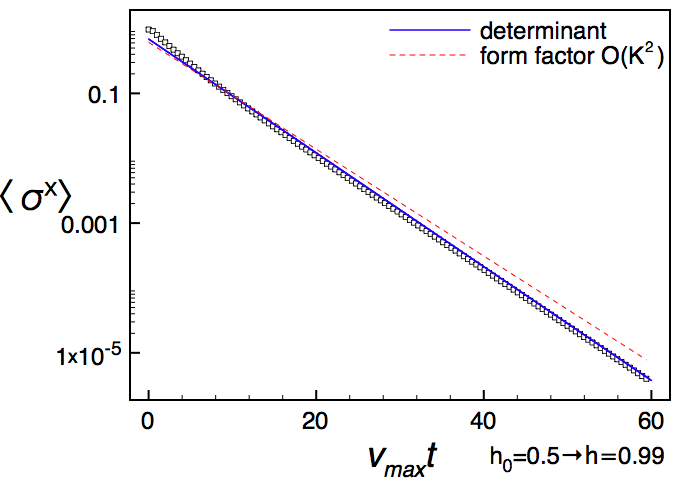}
\caption{Expectation value of the order parameter after a quench
within the ferromagnetic phase. Numerical data are compared with the
asymptotic predictions from both determinants and form factors at
order $O(K^2)$. The right panel shows the accuracy of the determinant
result even very close to the critical point, while the form factor
result ceases to be an accurate approximation because the density of
excitations is no longer small. Numerical results are obtained by
considering the cluster decomposition of the two-point function at
distance $\ell=180$. The left hand panel shows that the short-time
behaviour after the quench is sensitive to whether the transverse
field is increased or decreased. If $1>h>h_0$ then $\braket{\sigma^x(t)}$
decreases for short times, while it initially increases if $1>h_0>h$.
}
\label{Fig:1P}
 \end{figure}

Fig.~\ref{Fig:1P} shows the comparison of the asymptotic result
against exact numerical computation (obtained from cluster
decomposition of the two-point function). It is evident that the
asymptotic results become accurate already for small values of $t$.
Eqn (\ref{eq:OP}) is asymptotically valid also for quenches to
the critical point as is shown in the left panel of Fig.~\ref{Fig:1P}.

\begin{figure}[t]
\begin{center}
\includegraphics[width=0.5\textwidth]{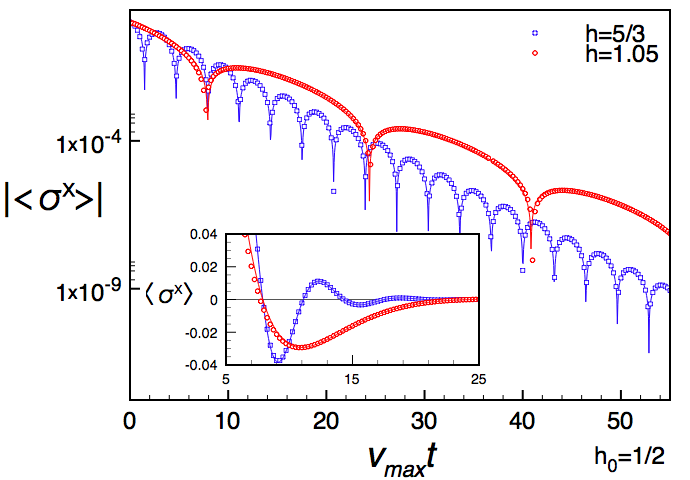}
\caption{Expectation value of the order parameter after a quench from the ferromagnetic to the paramagnetic phase (inset) 
and its absolute value (main plot) in logarithmic scale to show the exponential decay of the correlation. 
Numerical data from $h_0=1/2$ to $h=5/3$ and $h=1.05$ are compared with 
the asymptotic conjecture in Eq. (\ref{eq:OP2}) which is reported as a 
continuous line.  The asymptotic formula is valid all the way up to the critical point.
Data are obtained by considering the cluster decomposition of the two-point function at distance $\ell=180$.}
\label{Fig:1PFP}
\end{center}
\end{figure}

For a  quench from the ferromagnetic $h_0<1$ to the paramagnetic $h>1$
phase, we \emph{conjecture} that the expectation value of the order
parameter at late times $t$ is given by 
\begin{equation}
\label{eq:OP2}
\fl\qquad
\langle \sigma^x_i(t)\rangle=\left({\cal C}^x_{\rm FP}\right)^\frac{1}{2}
\left[1+\cos(2\veps_h(k_0)t+\alpha)+\ldots\right]^\frac{1}{2}
\exp\left[t \int_{0}^\pi \frac{\mathrm d k}{\pi} \veps'_h(k)
\ln\left|\cos\Delta_k\right| \right],
\end{equation}
where $k_0$ is a solution of the equation $\cos\Delta_{k_0}=0$,
$\alpha(h,h_0)$ is an unknown constant, and
\be
{\cal C}^x_{\rm FP}=\left[\frac{h\sqrt{1-h_0^2}}{h+h_0}\right]^\frac{1}{2}.
\label{CxFP}
\ee
The dots in eqn \fr{eq:OP2} indicate subleading contributions.
The conjecture
\fr{eq:OP2} is compared to the numerically calculated one-point
function (the 1-point function is obtained by applying the cluster
decomposition principle to the two-point function
\Eref{eq:semipredictionintr}, see Section \ref{ss:FtoP})
in
Fig.~\ref{Fig:1PFP}. The agreement is clearly excellent.
From a mathematical point of view the oscillating factor
is a correction to the asymptotic behaviour, as is most clearly seen
by considering $\log|\langle \sigma^x_i(t)|\rangle$.
However, by virtue of its oscillatory nature, its presence obscures
the leading behaviour and needs to be included in order to have a
good description of the quench dynamics. 
In the limit $h\to1$, $k_0$ goes to $0$ and $\veps_1(k_0)=0$,
signaling that the crossover between (\ref{eq:OP}) and  (\ref{eq:OP2})
is smooth. In particular, when approaching the critical point, the
oscillation frequency decreases as shown in Fig.~\ref{Fig:1PFP}.

%%%%%%%%%%%%%%%%%%%%%%%%%%%%%%%%%%%%%%%%%%%%%%%%%%%%%%%%
\subsection{Equal time two-point correlation function}
%%%%%%%%%%%%%%%%%%%%%%%%%%%%%%%%%%%%%%%%%%%%%%%%%%%%%%%%%

%%%%%%%%%%%%%%%%%%%%%%%%%%%%%%%%%%%%%%%%%%%%%%%%%%%%%%%
\subsubsection{Quench within the ferromagnetic phase.}
%%%%%%%%%%%%%%%%%%%%%%%%%%%%%%%%%%%%%%%%%%%%%%%%%%%%%%%

\begin{figure}[t]
\begin{center}
\includegraphics[width=0.48\textwidth]{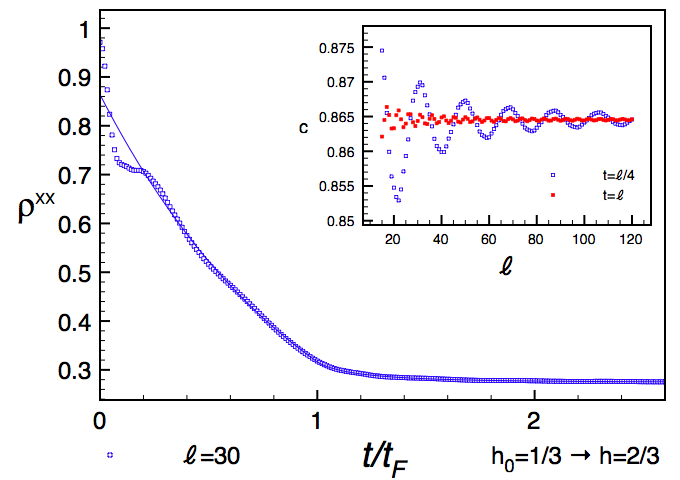}
\includegraphics[width=0.48\textwidth]{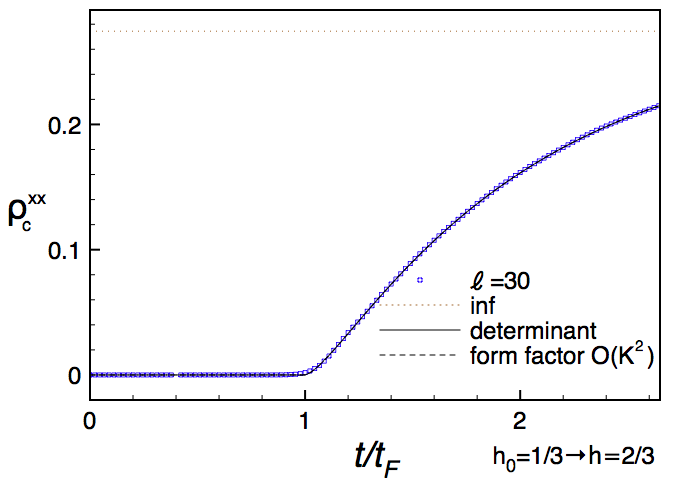}
\caption{
Numerical data for a quench within the ferromagnetic phase from
$h_0=1/3$ to $h=2/3$. Left: The two-point function against the
asymptotic prediction \Eref{eq:predictionintr} for $\ell=30$ (up to a
multiplicative factor) showing excellent agreement in the scaling
regime. Inset: Ratio between the numerical data and asymptotic
prediction (\ref{eq:prediction}). The leading correction is time
independent, but subleading contributions oscillate. Right: The
connected correlation function for the same parameters as on the left.
For $t<t_F$, $\rho^{xx}_c(\ell,t)$ vanishes identically in the scaling
regime.}
\label{fig:OO}
\end{center}
\end{figure}

For a quench within the ordered phase, the determinant approach
(detailed in Section \ref{s:Determinant approach})
leads to the following result for the two-point function 
in the space-time scaling limit ($t,\ell\to\infty$ with $v_{\rm
  max}t/\ell$ fixed)  
\bea
\rho^{xx}_{FF}(\ell,t)&\simeq& {\cal C}^x_{\rm FF}
\exp\Big[\ell \int_0^\pi \frac{\mathrm d k}{\pi}\ln\left|\cos\Delta_k\right|
\theta_H\big(2\veps'_h(k)t-\ell\big)\Big] 
\nn \fl &&\times
\exp\Big[
2t \int_0^\pi\frac{\mathrm d k}{\pi} \veps'_h(k)
\ln\left |\cos\Delta_k\right|\theta_H\big(\ell-2\veps'_h(k)t\big)\Big].
\label{eq:predictionintr}
\eea
Here $\theta_H(x)$ is the Heaviside step function
\be
\theta_H(x)=
\cases{1 & if $x>0$,\cr 0 & else .\cr}
\ee
The constant ${\cal C}^x_{\rm FF}$ \fr{CFF} is fixed by matching
\fr{eq:predictionintr} to the corresponding result at infinite time
$t=\infty$, which is derived in paper II \cite{CEF_paper2}.

In the limit $\ell\to\infty$ \fr{eq:predictionintr} gives the square of
the result \fr{eq:OP} for the one-point function. For times smaller
than the Fermi time 
\be
t_F= \frac{\ell}{2v_{\rm max}},
\ee
the first exponential factor in \fr{eq:predictionintr} equals 1.
Thus, in the space-time scaling limit,  connected correlations {\it
  vanish identically} for times $t<t_F$ and begin to form only after
the Fermi time.
This is a general feature of quantum quenches \cite{cc-06,cc-07} and
has been recently observed in experiments on one dimensional cold-atomic 
gases \cite{cetal-12}. We stress that this by no means implies that
the connected correlations are exactly zero for $t<t_F$: in any model,
both on the lattice or in the continuum there are exponentially
suppressed terms (in $\ell$) which vanish in the scaling limit.  
The form factor approach gives the following result for large $t$ and
$\ell$ (see Section \ref{sec:2spoint})
\bea
\rho^{xx}_{FF}(\ell,t)&\simeq&(1-h^2)^\frac{1}{4}
\exp\Big[-2\ell \int_0^\pi \frac{\mathrm d k}{\pi}{K^2(k)}
\theta_H\big(2\veps'_h(k)t-\ell\big)\Big] 
\nn \fl &&\times
\exp\Big[
-4t \int_0^\pi\frac{\mathrm d k}{\pi} \veps'_h(k)
{K^2(k)}\theta_H\big(\ell-2\veps'_h(k)t\big)\Big].
\label{eq:predictionintrFF}
\eea
As expected, it gives the low density approximation to the full result
\fr{eq:predictionintr}.

A comparison (for a typical quench from $h_0=1/3$ to $h=2/3$) 
between the asymptotic results (\ref{eq:predictionintr}) 
(\ref{eq:predictionintrFF}) and numerical results for the correlation
function at a finite but large distance ($\ell=30$) is shown in
Fig.~\ref{fig:OO}. The numerical results are obtained by expressing
the two-point correlator in the thermodynamic limit as the determinant
of an $\ell\times \ell$ matrix (see section \ref{s:Determinant approach}) 
and then evaluating the determinant for
different times. 
As we are concerned with equal time correlators only 
we do not need to extract the two-point function from a cluster
decomposition of the 4-point function \cite{rsms-08}. 
The agreement is clearly
excellent. The ratio between the exact numerics and the analytic
result (\ref{eq:predictionintr}) in the space-time scaling limit is
shown in the inset of Fig.~\ref{fig:OO} for two values of
$\kappa=v_{\rm max}t/\ell$. We see the ratio approaches a constant for
large $\ell$. The corrections to this constant are seen to be oscillating.
The right panel of Fig.~\ref{fig:OO} shows results for the connected
two-point function for the same quench ($h_0=1/3\to h=2/3$). We see
that the numerical data are fit very well by both 
(\ref{eq:predictionintr}) and the form factor result
(\ref{eq:predictionintrFF}). The connected correlator is exponentially
small for $t<t_F$ and correlations start forming at $t_F$.   

In order to elucidate the relaxational behaviour of the two point
function it is useful to follow Refs \cite{cc-06,cc-07} and rewrite
(\ref{eq:predictionintr}) in the form 
\be
\frac{\rho^{xx}_{FF}(\ell,t)}
{\big(\rho^x_{FF}(t)\big)^2}\sim \exp\left[\int_0^\pi\frac{dk}{\pi}
{\Big(\frac{2t}{\tau(k)}-\frac{\ell}{\xi(k)}\Big)}
\theta_H\Big(\frac{2t}{\tau(k)}-\frac{\ell}{\xi(k)}\Big) %(2\veps'_h(k)t-\ell)
\right].
\label{2point-sc}
\ee
Here we have defined mode-dependent correlation lengths $\xi(k)$ and
decay times $\tau(k)$ by
\be
\xi^{-1}(k)=-\ln |\cos \Delta_k|\ ,\quad
\tau^{-1}(k)=-\veps'_h(k)\ln |\cos \Delta_k|\ .
\ee
We observe that these quantities are related by the velocity
$v_k=\veps'_h(k)$ of the momentum $k$ mode
\be
v_k\ \tau(k)=\xi(k),
\label{vk}
\ee
which allows us to rewrite the theta-function in \fr{2point-sc} as
$\theta_H(2v_kt-\ell)$. The physical interpretation of \fr{2point-sc}
is then clear: a given mode contributes to the relaxational behaviour
only if the distance $\ell$ lies within its forward ``light cone''
(the factor of two multiplying the velocity is explained in Refs
\cite{cc-06,cc-07}). The form of the remaining factor then follows
from the stationary behaviour: the time dependent piece compensates
the factor $\big(\rho^x_{FF}\big)^{-2}$, while the time-independent
part is fixed by the $t\to\infty$ value of the correlator. As we
already pointed out in our letter \cite{CEF}, this implies that the
generalized Gibbs ensemble that characterizes the stationary state
in fact determines the relaxational behaviour at late times as well.

\paragraph{\it Approach to infinite times within the space-time scaling regime.}
For the quench within the ferromagnetic phase, the infinite time limit
at fixed $\ell$ gives the same result as the infinite time limit
within the space-time scaling regime, i.e.
\be
\lim_{t\to\infty}\frac{1}{\ell}\ln\left|\rho_{FF}^{xx}(\ell, t)\right|=
\lim_{\kappa\to 0}\lim_{\ell,t\to\infty\atop{t/\ell}\ {\rm fixed}}
\frac{1}{\ell}\ln\left|\rho_{FF}^{xx}(\ell, t)\right|.
\ee
It is then useful to consider the approach to the stationary value
within the space-time scaling regime result (\ref{eq:predictionintr}).  
In the limit $t\to\infty$, we have 
 \be
\rho^{xx}_{FF}(\ell,t=\infty)\propto
\exp\Big[\ell \int_0^\pi \frac{\mathrm d k}{\pi}\ln\left|\cos\Delta_k\right| \Big]\,.
\label{tinfty}
\ee
The corrections to \fr{tinfty} for large, finite times arise
from the modes with $\veps'_h(k)\sim 0$. For any $h\neq 1$, both modes
with $k=0$ and $k=\pi$ contribute to this correction and at the  same
order since both $\ln\left|\cos\Delta_k\right|$ and the dispersion
relation itself are quadratic at both points. 
A straightforward calculation gives for any $h,h_0<1$
\be
\frac{\rho^{xx}_{FF}(\ell,t\gg\ell)}{\exp\Big[\ell \int_0^\pi \frac{\mathrm d k}{\pi}\ln\left|\cos\Delta_k\right| \Big]}\propto
1+ \frac{(h - h_0)^2 (1 - 2 h h_0 + h_0^2)}{96\pi (1 - h_0^2)^2} \frac{\ell^4}{(v_{\rm max}t)^3}+\dots\,.
\ee

%%%%%%%%%%%%%%%%%%%%%%%%%%%%%%%%%%%%%%%%%%%%%%%%%%%%%%%%%%%%%%%%%%%%
\subsubsection{Quenches from the ferromagnetic phase to the quantum
  critical point.}
\begin{figure}[t]
\begin{center}
\includegraphics[width=0.48\textwidth]{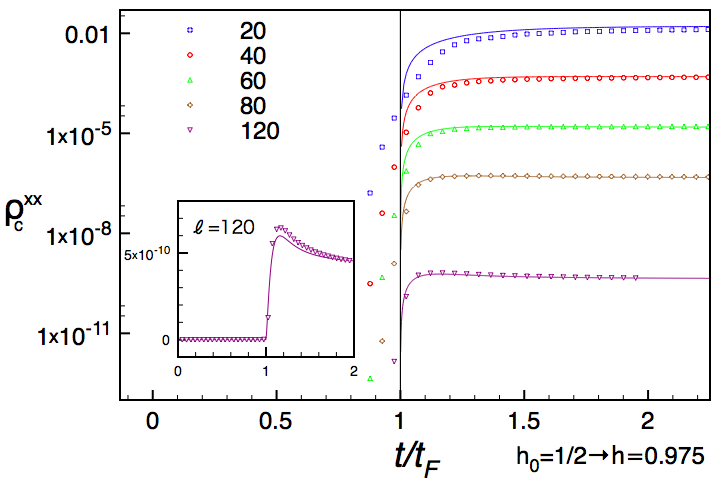}
\includegraphics[width=0.48\textwidth]{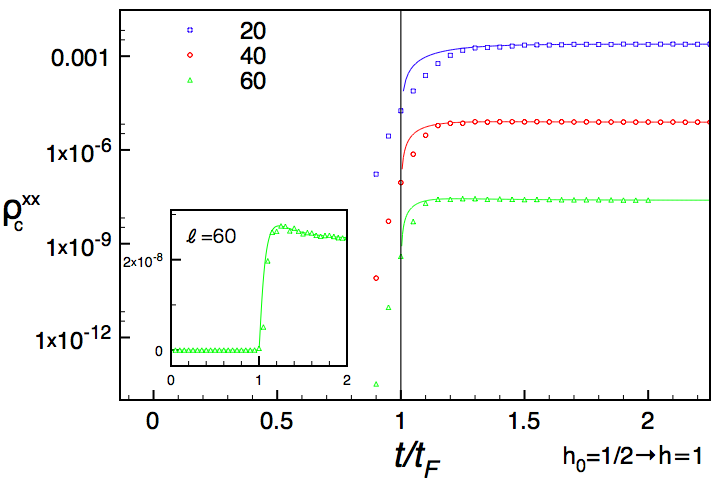}
\caption{
Numerical results for the connected two-point function for a quench
from the ferromagnetic phase to a magnetic field very close to (left) and
exactly at the critical point (right). The asymptotic behaviour
agrees with the prediction (\ref{eq:predictionintr}) shown as
continuous lines, but corrections are visible for smaller values of
$\ell$.
 }
\label{fig:OC}
\end{center}
\end{figure}

%%%%%%%%%%%%%%%%%%%%%%%%%%%%%%%%%%%%%%%%%%%%%%%%%%%%%%%%%%%%%%%%%%%%
If we adjust the constant ${\cal C}^x_{\rm FF}$ appropriately, equation
(\ref{eq:predictionintr}) holds even for quenches to the quantum
critical point. In Fig.~\ref{fig:OC} numerical 
results for the connected correlation function are compared with the
asymptotic prediction for quenches very close to the critical point
(left) and exactly to the critical point (right). 
In both cases, the asymptotic prediction is seen to become more
accurate when $\ell$ is increased. However, it is clear from the
figure that the asymptotic prediction works better for quenches  
exactly to the critical point than for quenches very close to it. 
This somewhat counterintuitive behaviour is readily understood
as a property of subleading contributions to
(\ref{eq:predictionintr}). We have already seen in
the inset of Fig.~\ref{fig:OO} that there are subleading oscillating
corrections. A more detailed analysis shows that they are power 
laws with an exponent that tends to zero upon approaching the critical
point. This explains why the agreement of (\ref{eq:predictionintr})
with the numerical results is worse in Fig.~\ref{fig:OC} (left) than
in Fig.~\ref{fig:OO}. At the same time as the exponent of the
power-law correction tends to zero upon quenching ever closer to the
critical point, the oscillation frequency approaches zero as well.
For quenches exactly to the critical point the leading oscillating
corrections are therefore absent: they have morphed into a
renormalization of the constant amplitude multiplying
(\ref{eq:predictionintr}). This is the reason why
(\ref{eq:predictionintr}) works better for quenches exactly to the
critical point than for quenches close to it.

For a quench to the critical point at $h_c=1$, the result 
(\ref{eq:predictionintr}) should be compared with conformal field
theory (CFT) predictions of Refs \cite{cc-06,cc-07}, which give the
following result valid in the scaling limit of the Ising model
\be
\lim_{h\to 1}\ \frac{\rho^{xx}(r,t)}{(1-h^2)^{\frac{1}{4}}}
\propto
\left\{\begin{array}{ll}
e^{- \pi t/8\tau_0}  \quad &{\rm for}\;v t<r/2\\
e^{- \pi r/16v\tau_0} \quad &{\rm for}\; vt>r/2\, .\\
\end{array}\right.
\label{freeboson}
\ee 
Here $r$ is the physical distance, $v$ the velocity characterizing
the (strictly) linear dispersion relation $\veps_q=vq$ at the critical
point, and $\tau_0$ the so-called extrapolation time. Eqn
\fr{freeboson} is valid for times and distances such that
$(x/v),t,(x/v)-t\gg\tau_0$. The scaling
limit of the result (\ref{eq:predictionintr}) is constructed in
section \ref{sec:scaling}, and considering a quench to the critical point we
obtain from \fr{rhoscaling} 
\bea
\fl\qquad
\lim_{h\to 1}\ \frac{\rho^{xx}(r,t)}{(1-h^2)^{\frac{1}{4}}}
\propto&
%\exp\left(\int_0^\infty\frac{dq}{\pi}\ \ln\left[
%\frac{vq}{\sqrt{v^2q^2+\Delta_0^2}}\right]
%\left[r\theta(2vt-r)+2vt\theta(r-2vt)\right]\right)\nn
&\exp\left(-\frac{\Delta_0}{2v}
\Big[r\theta(2vt-r)+2vt\theta(r-2vt)\Big]\right)
,
\label{rhoscaling_QCP}
\eea
where $\Delta_0=J(1-h_0)$. Comparing \fr{rhoscaling_QCP} to
\fr{freeboson} we see that the two expressions agree if we take the
extrapolation time to be
\be
\tau_0^{-1}=\frac{8\Delta_0}{\pi}.
%\frac{16}{\pi}\int_0^\infty\frac{dq}{\pi}\ \ln\left[
%\frac{vq}{\sqrt{v^2q^2+\Delta_0^2}}\right]\ .
\ee
Outside the scaling limit the CFT expression \fr{freeboson} is not
expected to provide a good approximation to the full result
(\ref{eq:predictionintr}) because the nonlinearity of the dispersion
relation becomes important.

%%%%%%%%%%%%%%%%%%%%%%%%%%%%%%%%%%%%%%%%%%%%%%%%%%%%%%%%%%%%%%%%%%%%%%%%%%%
\subsubsection{Quenches from the ferromagnetic to the paramagnetic phase.}
%%%%%%%%%%%%%%%%%%%%%%%%%%%%%%%%%%%%%%%%%%%%%%%%%%%%%%%%%%%%%%%%%%%%%%%%%%%
\begin{figure}[t]
\begin{center}
\includegraphics[width=0.48\textwidth]{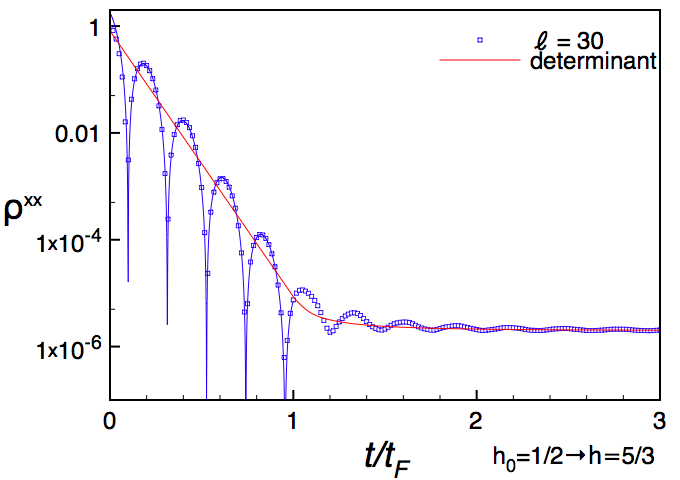}
\includegraphics[width=0.48\textwidth]{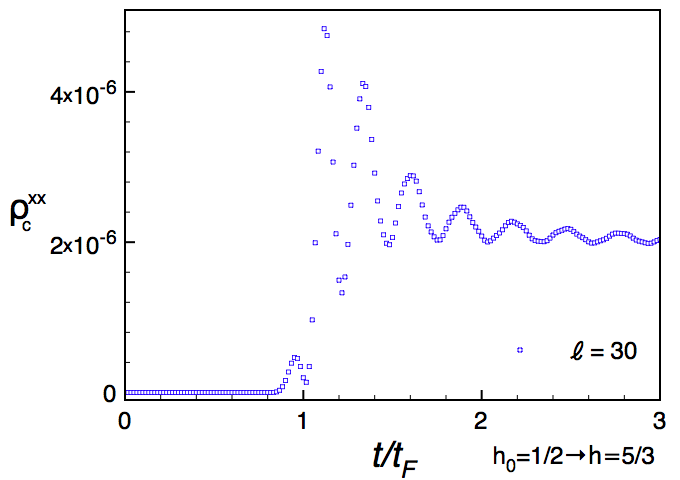}
\caption{
Numerical data for a quench from the ferromagnetic to the paramagnetic phase. 
Left: The two-point function against the asymptotic prediction \Eref{eq:semipredictionintr} (valid for $t<t_F$)
showing excellent agreement. 
Right: The connected correlation function for the same quench as on the left. 
For $t<t_F$, $\rho^{xx}_c$ vanishes identically in the scaling regime. 
For $t>t_F$, $\rho^{xx}$ (and so $\rho^{xx}_c$) approaches the asymptotic value  \Eref{eq:predictionintr}, but subleading oscillating 
corrections are visible.
 }
\label{fig:OP}
\end{center}
\end{figure}

We have not been able to carry out a full analytical calculation
of the time evolution of the two point function for a quench across
the critical point. However, we conjecture\footnote{Our conjecture is
  based on properties of the spectrum of the matrix $\Gamma$
\fr{eq:Gamma0}.} that (see Section \ref{ss:FtoP})
\bea
\label{eq:semipredictionintr}
\fl\qquad
\rho^{xx}_{FP}(\ell,t)\simeq {\cal C}^x_{\rm FP}
\exp\Bigl[\int_{0}^\pi \frac{\mathrm d  k}{\pi}2 \veps'_h(k)
  t\ln|\cos\Delta_k|\Bigr] \nn
\fl\qquad\qquad\qquad\times\ 
\cases{\exp\Big[\int_0^\pi \frac{\mathrm d k}{\pi}
\big(\ell-2t\veps'_h(k)\big)
\ln\left|\cos\Delta_k\right|
\theta_H\big(2\veps'_h(k)t-\ell\big)\Big] & $t>t_F$\ ,\\
1+\cos(2\veps_h({k_0}) t+\alpha)+\ldots & $t<t_F$\ ,\cr}
\eea
where $\alpha$ and $k_0$ are the same as in (\ref{eq:OP2}) (see also
Ref. \cite{CEF_paper2}) and the constant factor 
${\cal C}^x_{\rm FP}$ is given in \fr{CxFP}.
%\be
%{\cal C}^x_{\rm FP}=\left[\frac{h\sqrt{1-h_0^2}}{h+h_0}\right]^\frac{1}{2}.
%\ee
This prediction is compared with the numerically calculated correlation
function in Fig.~\ref{fig:OP} (left) and the agreement is clearly very
good. For $t<t_F$ \fr{eq:semipredictionintr} is simply the square of
the corresponding one-point function, which ensures that connected
correlations vanish for $t<t_F$ in the space-time scaling regime.
This is in agreement with numerical results for the connected
two-point function as shown in the right hand panel of
Fig.~\ref{fig:OP}. As can be seen in Fig.~\ref{fig:OP}
oscillations are present for $t>t_F$ as well, but they
display a rather fast decay in time towards the determinant result
(\ref{eq:predictionintr}).
Like for the one-point function, the oscillating
factor is a correction to the leading asymptotic behaviour in the
space-time scaling limit, but it needs to be included to give a good
description of the numerical data.  
Finally, we note that for $h=1$ we have $\veps_1(k_0)=0$ and
(\ref{eq:semipredictionintr}) reduces to (\ref{eq:predictionintr}).

%%%%%%%%%%%%%%%%%%%%%%%%%%%%%%%%%%%%%%%%%%%%%%%%%%%%%%%
\subsubsection{Quench within the paramagnetic phase.}
%%%%%%%%%%%%%%%%%%%%%%%%%%%%%%%%%%%%%%%%%%%%%%%%%%%%%%%
For quenches within the paramagnetic phase the form factor approach
gives the following result (see Section \ref{ss:PtoP})
\bea
\fl\quad
\label{eq:preddisdisintr}
\rho^{xx}_{PP}(\ell,t)
&\simeq\rho^{xx}_{PP}(\ell,\infty)+
(h^2-1)^\frac{1}{4}\sqrt{4J^2h}
\int_{-\pi}^\pi\frac{dk}{\pi}\frac{K(k)}{\veps_k}
\sin(2t\veps_k-k\ell)\nn
\fl
&\qquad\times\ \exp\bigg[-2\int_0^\pi\frac{dp}{\pi}
K^2(p)\left(
\ell+\theta_H(\ell-2t\veps'_{p})[2t\veps'_{p}-\ell]\right)
\bigg]+\ldots
\eea
The regime of validity of \fr{eq:preddisdisintr} is sufficiently large
values of $\ell$ and $t$ and ``small'' quenches in the sense discussed
in the beginning of section \ref{sec:summaryANDdiscussion}. We have
not attempted to calculate the infinite time limit
$\rho^{xx}_{PP}(\ell,\infty)$ within the framework of the form factor
approach, because its exact large-$\ell$ asymptotics is known from the
determinant approach to be \cite{CEF,CEF_paper2}
\bea
\rho^{xx}_{PP}(\ell,\infty)\simeq {\cal C}^x_{\rm PP}(\ell) e^{-\ell/\xi}\ ,
\label{rhoxxinfty}
\eea
where ${\cal C}^x_{\rm PP}(\ell)$ is determined in paper II \cite{CEF_paper2} and
\bea
\fl\qquad
\xi^{-1}=\ln\left( {\rm min}[h_0,h_1]\right)
-\ln\left[h_1\frac{h+h_0}{2hh_0}\right],\quad
h_1=\frac{1+h h_0+\sqrt{(h^2-1)(h_0^2-1)}}{h+h_0}.
\eea
As discussed in our previous letter \cite{CEF}, \fr{rhoxxinfty} is
described by a general Gibbs ensemble. Based on the form factor 
result \fr{eq:preddisdisintr} one may speculate that the full answer
may have the structure
\bea
\fl\qquad
\rho^{xx}(\ell,t)&\simeq\left[{\cal C}^x_{\rm PP}(\ell)+
(h^2-1)^\frac{1}{4}\sqrt{4J^2h}\int_{-\pi}^\pi\frac{dk}{\pi}\frac{K(k)}{\veps_k}
\sin(2t\veps_k-k\ell)+\ldots \right]\nn\fl
&\times
\exp\bigg[-\int_0^\pi\frac{dp}{\pi}
\ln\left[\frac{1+K^2(p)}{1-K^2(p)}\right]\left(
\ell+\theta_H(\ell-2t\veps'_{p})[2t\veps'_{p}-\ell]\right)
\bigg]+\ldots\ .
\eea

\begin{figure}[t]
\begin{center}
\includegraphics[width=0.46\textwidth]{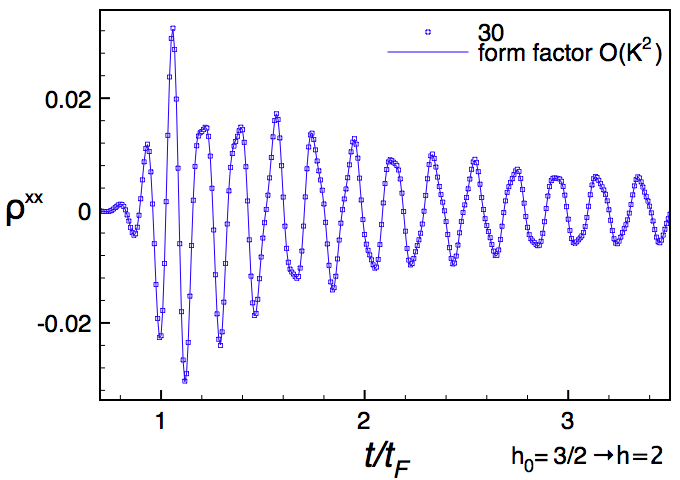}
\includegraphics[width=0.46\textwidth]{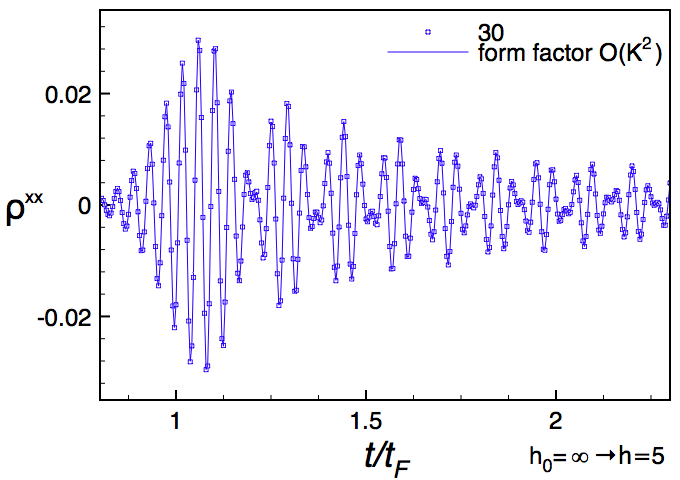}
\caption{Time evolution of order parameter correlators for quenches
  within the paramagnetic phase. 
Left: $\rho^{xx}_{PP}(\ell=30,t)$ for a quench from
  $h_0=\frac{3}{2}$ to $h=2$.
Right: $\rho^{xx}_{PP}(\ell=30,t)$ for a quench from
  $h_0=\infty$ to $h=5$.
The two-point function is seen to exhibit oscillatory power-law decay
at late times. The form factor result (solid lines) is seen to give a
very good description of the numerical data (points). The short time
regime is not shown as the correlators are exponentially small by
virtue of the horizion effect.
 }
\label{fig:PP}
\end{center}
\end{figure}

In Fig.~\ref{fig:PP} we compare the analytic result
\fr{eq:preddisdisintr} to numerical results obtained for two different
quenches within the paramagnetic phase. The agreement is seen to be
excellent. 
\begin{figure}[t]
\begin{center}
\includegraphics[width=0.46\textwidth]{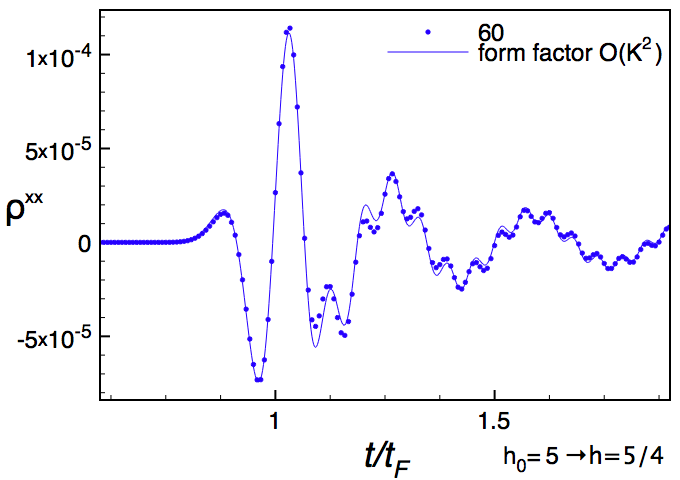}
\caption{Time evolution of $\rho^{xx}_{PP}(\ell=60,t)$ for a quench
from $h_0=5$ to $h=5/4$. As the quench is no longer small, deviations
from the form factor result become more pronounced.}
\label{fig:PP2}
\end{center}
\end{figure}

For strong quenches the form factor result is not
expected to be quantitatively accurate. This can be seen in
Fig.~\ref{fig:PP2}. In all cases, the two-point function is seen to
display slowly decaying oscillatory behaviour on the time scales
shown. At sufficiently large $t$ the decay is proportional to $t^{-3/2}$.
This is in marked contrast to quenches within the ordered
phase. The origin of this difference lies in the nature of the
relaxational processes that drive the time evolution. The oscillatory
behaviour seen in the paramagnetic phase arises from processes
involving the annihilation of spin-flip excitations, while the
smooth exponential behaviour seen in the ferromagnetic phase is
related to the ballistic motion of domain wall excitations.

The structure of \fr{eq:preddisdisintr} implies the existence of a
late time crossover scale $t^*$, at which the second contribution
becomes smaller than $\rho_{PP}^{xx}(\ell,\infty)$. 
Using that $\rho_{PP}^{xx}(\ell,\infty)\propto e^{-\ell/\xi}$ and that
the second contribution decays like $(Jt)^{-3/2}$ at late times we may
estimate $t^*$ as
\be
Jt^*\sim e^{2\ell/(3\xi)}\ .
\ee
For the cases shown in Fig.~\ref{fig:PP} this gives $Jt^*\sim 4114$ and 
$Jt^*\sim 10^{20}$ respectively ($Jt_F=7.5$ in both cases). This means
that in both cases the stationary behaviour characterized by the
generalized Gibbs ensemble is revealed only at \emph{very late times}.

Finally, we note that so far we have not been able to analyze the time
evolution of order parameter correlators for quenches within the
paramagnetic phase by means of the determinant approach. 

%%%%%%%%%%%%%%%%%%%%%%%%%%%%%%%%%%%%%%%%%%%%%%%%%%%%%%%%%%%%%%%%%%%%
\subsubsection{Quenches from the paramagnetic to the ferromagnetic phase.}
\begin{figure}[t]
\begin{center}
\includegraphics[width=0.48\textwidth]{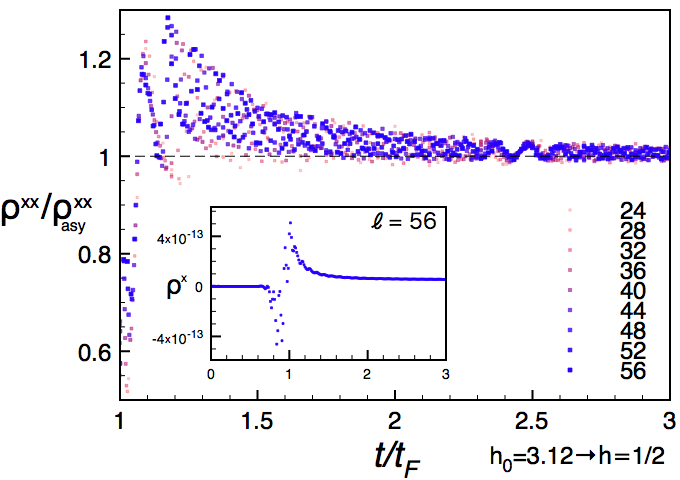}
\caption{
Order parameter two point function for a quench from the paramagnetic
to the ferromagnetic phase (inset) and its ratio with the asymptotic
prediction $\rho_{\rm PF}^{xx}$ given in eqn (\ref{eq:predictionintr_PF})
(main plot).  
We consider a quench from  $h_0=(2+3\sqrt{2})/2$ to $h=1/2$.
For any $t>t_F$, the ratio approaches $1$ with increasing $\ell$. }
\label{fig:PO}
\end{center}
\end{figure}

%%%%%%%%%%%%%%%%%%%%%%%%%%%%%%%%%%%%%%%%%%%%%%%%%%%%%%%%%%%%%%%%%%%%
Here we have not been able to obtain analytic results by either
the determinant or the form factor approach. However, observations
within the framework of the determinant approach (see
  Section \ref{ss:applicability}) suggest that for late 
times $t>t_F$ the leading asymptotic behaviour of the two-point
function should be given by 
\bea
\rho^{xx}_{PF}(\ell,t)&\simeq& {\cal C}^x_{\rm PF}(\ell)
\exp\Big[\ell \int_0^\pi \frac{\mathrm d k}{\pi}\ln\left|\cos\Delta_k\right|
\theta_H\big(2\veps'_h(k)t-\ell\big)\Big] 
\nn \fl &&\times
\exp\Big[
2t \int_0^\pi\frac{\mathrm d k}{\pi} \veps'_h(k)
\ln\left |\cos\Delta_k\right|\theta_H\big(\ell-2\veps'_h(k)t\big)\Big]
\ ,\quad t>t_F.
\label{eq:predictionintr_PF}
\eea
Here the factor ${\cal C}^x_{\rm PF}(\ell)$ is given by the $t=\infty$
results of \cite{CEF_paper2} 
\be
{\cal C}^x_{\rm PF}(\ell)=
\left[\frac{h_0-h}{\sqrt{h_0^2-1}}\right]^\frac{1}{2}
\cos\Big(\ell\ {\rm arctan}\Big[\frac{\sqrt{(1-h^2)(h_0^2-1)}}{1+h_0h}
\Big]\Big).
\ee
We have
tested this conjecture by comparing it to the numerical results and
found it to hold. An example is shown in Fig.~\ref{fig:PO}, where we plot
the ratio of the numerically calculated correlation function 
and the analytic expression (\ref{eq:predictionintr}) for several
values of the distance $\ell$. The ratio clearly approaches $1$ at
late times. We note that the values of $\ell$ have been chosen in a way such
that for the particular quench considered (i.e. $h_0=3.12$ to $h=0.5$)
oscillations in $\ell$ are suppressed in the $t\to\infty$ limit.
Our analytic methods do not currently provide an understanding of the
quench dynamics for times shorter than the Fermi time $t<t_F$.

%%%%%%%%%%%%%%%%%%%%%%%%%%%%%%%%%%%%%%%%%%%%%%%%%%%%%%%%
\subsection{Quenches to $h=0$ or $h=\infty$.}
%%%%%%%%%%%%%%%%%%%%%%%%%%%%%%%%%%%%%%%%%%%%%%%%%%%%%%%%%

These quenches are special, because the post-quench Hamiltonian is
classical in both cases. As a result the dynamics is anomalous.
For example, for quenches to $h=0$ correlation functions involving
only $\sigma^x_j$ operators are time independent, because the final
Hamiltonian $H(h)$ commutes with all $\sigma^x_j$. Correlation
functions involving other operators do depend on time, but generally
exhibit persistent oscillations. For instance, we have that
\be
\fl\quad
e^{iH(h=0)t}\sigma^z_je^{-iH(h=0)t}=\cos^2(2Jt)
\sigma^z_j+\frac{1}{2}\sin(4Jt)
\big(\sigma^x_{j-1}+\sigma^x_{j+1}\big)\sigma^y_j
-\sin^2(2Jt)\sigma^x_{j-1}\sigma^z_j\sigma^x_{j+1},
\ee
and as a result $\langle\Psi(t)|\sigma^z_j|\Psi(t)\rangle$ does not
approach a stationary value at late times.

%%%%%%%%%%%%%%%%%%%%%%%%%%%%%%%%
\section{Determinant approach: Analytic derivation of the asymptotic two-point function 
for a quench within the ordered phase.}\label{s:Determinant approach}%
%%%%%%%%%%%%%%%%%%%%%%%%%%%%%%%%

In this section we present the analytic derivation of
Eq. (\ref{eq:predictionintr}) for the asymptotic behaviour in the
space-time scaling regime ($t,\ell\gg1$, but $t/\ell$ arbitrary) of
the two-point correlation function $\rho^{xx}(\ell,t)$. Within this
approach, it is convenient to replace the fermions $c_j$ in
\ref{app:diag} with the Majorana fermions 
\be
a_j^x=c_j^\dag+c_j\qquad a_j^y=i(c^\dag_j-c_j)\, ,
\label{majo}
\ee
which satisfy the algebra $\{a_l^x,a_n^x\}=2\delta_{l n}$, $\{a_l^y,a_n^y\}=2\delta_{l n}$, $\{a_l^x,a_n^y\}=0$.
In terms of these Majorana fermions, the operator $\sigma^x_j$ has the nonlocal representation 
\be
\sigma^x_\ell=\prod_{j=1}^{\ell-1}(i a_j^y a_j^x)a_\ell^x\, .
\ee
The two-point function of $\sigma^x$ is then the expectation value of a string of Majorana fermions
\be
\rho^{xx}(\ell,t)=\braket{\prod_{j=1}^\ell (-i a_j^y(t) a_{j+1}^x(t))}\, .
\label{rhoxxF}
\ee
The Ising Hamiltonian \fr{Hamiltonian} can be written as (see \ref{app:diag})
\be
H=\frac{1}{2}\left[1-\prod_{l=1}^L\sigma_l^z\right]
H_{o}+
\frac{1}{2}\left[1+\prod_{l=1}^L\sigma_l^z\right]
H_{e}\, ,
\ee
where $H_{o}$ and $H_e$ are quadratic Hamiltonians of the
Jordan-Wigner fermions in the sectors with odd and even fermion number
respectively ($H_{o/e}$ both commute with $\prod_{l=1}^L\sigma_l^z$).  
For finite chains, the ground state of the Ising Hamiltonian is an
eigenstate of $H_e$. However, in the thermodynamic limit, the
$\mathbb{Z}_2$ symmetry of $H$ is spontaneously broken in the ordered
phase $h_0<1$, where the ground state is a linear superposition of
the ground states of $H_{e}$ and $H_o$
\be
|\Psi(0)\rangle=\frac{|B(0)\rangle_\NS+|B(0)\rangle_\R}{2}.
\ee
Thus, for a quench starting in the ordered phase free fermion
techniques cannot be straightforwardly applied to the calculation of
correlation functions involving generic operators.
However, here we are interested only in the expectation value of
even operators $\hat O_e$ characterized by
\be
\left[\prod_{j=1}^L\sigma^z_j\right]\hat O_e
\left[\prod_{l=1}^L\sigma^z_l\right]=\hat O_e.
\ee
Hence we have 
\bea
\label{eq:SD}
\fl\qquad
\lim_{L\to\infty}
\braket{\Psi(t)|\hat O_e|\Psi(t)}&=
\lim_{L\to\infty}
\frac{{}_\NS\braket{B(t)|\hat O_e|B(t)}_\NS
+{}_{\rm R}\langle B(t)|\hat O_e|B(t)\rangle_\R}{2}\nn
&=\lim_{L\to\infty}
{}_\NS\langle B(t)|\hat O_e|B(t)\rangle_\NS\ ,
\eea
Crucially, the expectation value ${}_\NS\langle B(t)|\hat
O_e|B(t)\rangle_\NS$ can be evaluated using Wick's theorem.
In contrast to even operators, expectation values of odd operators
\be
\left[\prod_{j=1}^L\sigma^z_j\right]\hat O_o
\left[\prod_{l=1}^L\sigma^z_l\right]=-\hat O_o
\ee
are significantly more difficult to determine \cite{mc}.
The particular case of interest, $\rho^{xx}(\ell,t)$, involves an even
operator (cf. Eq. (\ref{rhoxxF})) and by straightforward application of
Wick's theorem one obtains a representation as the Pfaffian of a
$2\ell\times 2\ell$ antisymmetric matrix  \cite{mc} 
\be\label{eq:rhoxxPf}
\rho^{xx}(\ell,t)=\mathrm{pf}(\bar{\Gamma})\, ,
\ee
where $\mathrm{pf}$ denotes the Pfaffian, $\bar{\Gamma}$ is given by
\be\label{eq:Gamma0}
\bar{\Gamma} = \left[
 \begin{array}{ccccc}
\mathtt\Gamma_{0}  & \mathtt\Gamma_{-1}   &   \cdots & \mathtt\Gamma_{1-\ell}  \\
\mathtt\Gamma_{1} & \mathtt\Gamma_{0}   & &\vdots\\
\vdots&  & \ddots&\vdots  \\
\mathtt\Gamma_{\ell-1}& \cdots  & \cdots  &\mathtt\Gamma_{0}
\end{array}
\right], ~~~ 
\mathtt\Gamma_{l}=\left(
\begin{array}{cc}
-f_{l}&g_{l}\\
-g_{-l}&f_{l}
\end{array}
\right)\,.
\ee
The elements of this matrix are the fermionic
correlations\footnote{The identity in the second line of
Eq. (\ref{corrferm}) is a consequence of reflection symmetry, which
indeed implies that correlations should be invariant under the
transformation 
\be
a_l^x\rightarrow i \bigl(\prod_l\sigma_l^z\bigr) a_{L+1-l}^y\qquad a_l^y\rightarrow -i \bigl(\prod_l\sigma_l^z\bigr) a_{L+1-l}^x\, .
\ee
For example, the Dzyaloshinskii-Moriya interaction breaks this symmetry.} 
\begin{eqnarray}
g_n&\equiv& i \braket{a_l^x a_{l+n-1}^y},\nonumber\\
f_n+i\delta_{n 0}&\equiv& i \braket{a_l^x a_{l+n}^x}=i \braket{a_{l+n}^y a_{l}^y}\qquad \forall l\, .
\label{corrferm}
\end{eqnarray}
The matrix $\bar\Gamma$ is a block Toeplitz matrix because its constituent
$2\times 2$ blocks depend only on the difference between row and column
indices. It is customary to introduce the (block) symbol of the matrix
$\bar\Gamma$ via Fourier transform as follows 
\be
\label{eq:Gamma}
\fl \qquad
\mathtt\Gamma_{l}=\left(
\begin{array}{cc}
-f_{l}&g_{l}\\
-g_{-l}&f_{l}
\end{array}
\right)=
\int_{-\pi}^\pi\frac{\mathrm d k}{2\pi} e^{i l k} \hat\mathtt\Gamma(k)
\,,\quad {\rm with}\quad  \hat\mathtt\Gamma(k)=
%=\int_{-\pi}^\pi\frac{\mathrm d k}{2\pi} e^{i l k}
\left(
\begin{array}{cc}
-f(k)&g(k)\\
-g(-k)&f(k)
\end{array}
\right),
\ee
where the functions $f(k)$ and $g(k)$ are 
\begin{eqnarray}\label{eq:fg}
f(k)&=i\sin\Delta_k \sin(2\veps_h(k) t),\nn
g(k)&=-e^{i \theta_k-ik}\Bigl[\cos\Delta_k-i\sin\Delta_k \cos(2\veps_h(k) t)\Bigr]\, .
\end{eqnarray}

The spectrum of  block Toeplitz matrix $\bar \Gamma$ is the same of an $\ell\times \ell$ Toeplitz+Hankel matrix (Hankel matrices 
have elements which depend only on the sum of row and column indices, instead of the difference as in Toeplitz matrices). 
Indeed, being $\bar \Gamma$ a real antisymmetric matrix, its eigenvalues are complex conjugate pairs $\pm i\lambda_j$.
Given an eigenvector $\vec V$ of $\bar \Gamma$ with eigenvalue $i\lambda$,
we can define the vector $\vec w$ with $\ell$ components   $w_i=V_{2i-1}$ (since 
 $\mathtt\Gamma_{-l}=\sigma_y\mathtt\Gamma_l\sigma_y$, taking the even components 
 would change the sign of $\lambda$).
% so that the $\Gamma$'s eigenvectors  split in 
%two subspaces specified by the relation $\vec v_{L+1-l}=\pm \sigma_y \vec v_l$, where $\vec v_l^T=(V_{2l-1},V_{2l})$. 
Thus, the $\ell$ vectors $\vec w_i$ are solutions of the eigenvalue problem
\be
(i T\pm H)\vec w_j=\mp i \lambda_j \vec w_j,\qquad 
\left\{\begin{array}{l}
T_{l n}=f_{l-n}\\
H_{l n}=g_{l+n-L-1}
\end{array}\right.\, ,
\ee
%where $\pm \lambda$ are the corresponding eigenvalues of $\bar\Gamma$. 
In particular, this means that
\be
|\mathrm{pf}(\bar{\Gamma})|=| \det[i T\pm H]|.
\ee
%The sign of the Pfaffian can be obtained by imposing that the time evolution is smooth, 
%and hence we only need to fix the sign at the initial time (when the matrix $T$ is zero), obtaining
The sign of the Pfaffian can be fixed by observing that, for any
given $\ell$, both the determinant of the Toeplitz+Hankel matrix and
the Pfaffian of the block Toeplitz matrix are polynomials  in $f_l$
and $g_l$ of the same degree. Thus, the sign is independent of the
actual values of the matrix elements and can be determined by
considering the simplest case $T=0$ (which occurs at the initial time
and at asymptotically late times after the quench), obtaining
\be\label{eq:Pfafsign}
\rho^{xx}(\ell,t)=\mathrm{pf}(\bar{\Gamma})=(-1)^{\frac{\ell(\ell-1)}{2}}\det(H+i
T)=(-1)^{\frac{\ell(\ell-1)}{2}} \det W\, , 
\ee
where $W=H+iT$.
The matrix $\bar\Gamma$ is real and antisymmetric, so its eigenvalues
are complex conjugate pairs of the form $\pm i \lambda_j$ with
$j=1\dots \ell$. By construction $\lambda_j \in [-1,1]$ (see
e.g. \cite{vidal}) and they are also the eigenvalues of the matrix
$W=H+iT$. Thus we have 
\be
(\rho^{xx})^2=\det \bar{\Gamma} = \det W^2= \prod_{j=1}^\ell \lambda_j^2\, .
\label{WGamma}
\ee

The evaluation of the correlation function
$\rho^{xx}(\ell,t)$ for large $\ell$ is then equivalent to the
asymptotic evaluation of the determinant of a block Toeplitz matrix or
the sum of a Toeplitz and a Hankel matrices. Such matrices have been
the subject of intense study by mathematicians and physicists for more
than a century and standard, rigorous techniques for calculating their
determinants such as  Sz\"ego's lemma and the Fisher-Hartwig
conjecture are available, see e.g. \cite{FH-gen,sviaj}.  
However, these methods have been specifically designed for the
evaluation of determinants of matrices whose elements do not depend
explicitly on the matrix size. This is in contract to our case, where 
we are interested in the scaling limit $\ell,t\to\infty$ with finite
ratio $\ell/t$. Under these circumstances, each element of the matrix
$\Gamma$ in the scaling limit depends on a parameter (namely $t$)  
which is proportional to the matrix dimension $2\ell$. This precludes
the application of the aforementioned techniques for the asymptotic
evaluation of these determinants. The two exceptions are the $t=0$
case, where we recover the known equilibrium results, and the limit
$t=\infty$. In order to deal with large
values of $t$ we developed a completely novel approach, which 
follows the one we proposed in Ref. \cite{fc-08} for the entanglement
entropy  and it is based on a multi-dimensional stationary phase
approximation.

In the following we derive a rather general result valid for any block
$2\times 2$ block Toeplitz matrix $\Gamma$ with a symbol $\hat{t}(k)$
that can be cast in the form 
\be\label{eq:symbol}
 \hat{t}(k)=n_x(k)\sigma_x^{(k)}+\vec n_\perp(k) \cdot \vec\sigma^{(k)} e^{2 i \veps(k) t \sigma_x^{(k)}},
\qquad \vec n_\perp(k)\cdot \hat x=0\, .
\ee
Here the time $t$ is the only parameter proportional to the matrix
size $2\ell$, $n_x,n_\perp$ are fixed but otherwise arbitrary and
$\sigma^{( k)}$ denotes a local rotation of the Pauli matrices 
\be\label{eq:rotationsigma}
\sigma_\alpha^{( k)}\sim e^{i \vec w( k)\cdot \sigma}\sigma_\alpha e^{-i \vec w( k)\cdot \sigma}\, .
\ee
Our particular case of interest (\ref{eq:Gamma}) corresponds to having
$n_x^2+|\vec n_\perp|^2=1$.

We first consider $\Tr \Gamma^{2n}$ for positive integer $n$.
We note that the traces of odd powers of $\Gamma$ vanish, because
$\Gamma$ is a real antisymmetric matrix ($\Gamma$ is diagonalizable and
for each non-zero eigenvalue there is another one with opposite sign).
Our main result is that 
\begin{eqnarray}
\label{eq:powers}
\fl\qquad
\lim_{t,\ell\to\infty\atop t/\ell\ {\rm fixed}}
\frac{\mathrm{Tr}[\Gamma^{2n}]}{2\ell}=
\int_{-\pi}^{\pi}\frac{\mathrm d  k_0}{2\pi}\max\Bigl(1-2|\veps'(k_0)| \frac{t}\ell,0\Bigr) \Bigl(n_x( k_0)^2+|\vec n_\perp( k_0)|^2\Bigr)^{n}
\nonumber\\
\qquad+\int_{-\pi}^\pi \frac{\mathrm d k_0}{2\pi}\min\Bigl(2|\veps'(k_0)|\frac{t}\ell,1\Bigr)n_x( k_0)^{2n}\, .
\end{eqnarray}
Given the result \fr{eq:powers} it is possible to infer the asymptotic
behaviour of more complicated quantities such as
\be
\Tr {\cal F}(\Gamma^2)=\sum_n {\cal F}_n \Tr [\Gamma^{2n}]
\,,
\label{generalfunction}
\ee
where ${\cal F}(z)$ is an analytic function with power series expansion
${\cal F}(z)=\sum_n {\cal F}_n z^n$ around $z=0$.
Using Eq. (\ref{eq:powers}) and interchanging the order of sum and integration, we have
\begin{eqnarray}\label{eq:analytic}
\fl\qquad
\lim_{t,\ell\to\infty\atop t/\ell\ {\rm fixed}}
\frac{\mathrm{Tr}[{\cal F}(\Gamma^2)]}{2\ell}=
\int_{-\pi}^{\pi}\frac{\mathrm d
  k_0}{2\pi}\max\Bigl(1-2|\veps'(k_0)| \frac{t}\ell,0\Bigr)  
{\cal F}\bigl(n_x( k_0)^2+|\vec n_\perp( k_0)|^2\bigr)\nonumber\\
\qquad+\int_{-\pi}^\pi \frac{\mathrm d  k_0}{2\pi}\min\Bigl(2|\veps'(k_0)|\frac{t}\ell,1\Bigr){\cal F}(n_x( k_0)^{2})\, .
\end{eqnarray}
Clearly this approach is fully justified only as long as all
eigenvalues of $\Gamma$ fall within the radius of convergence of the
expansion of the function ${\cal F}(z)$ around $z=0$.
For example, in the case of the entanglement entropy, we have ${\cal F}(z)= -\frac{1+i z}2 \ln \frac{1+iz}2$ 
\cite{fc-08} and since  the eigenvalues of $\Gamma$ are of the form
$\pm i\lambda_j$ with $\lambda_j\in[-1,1]$
(\ref{eq:analytic}) holds (further generalizations to the entanglement
of two blocks have also been considered \cite{fc-10}). 

In the case at hand we have
\be
(\rho^{xx})^2=\det \Gamma= \exp [\Tr \ln \Gamma]
\,,\quad {\rm i.e.}\quad \ln (\rho^{xx})^2= \frac 12 \Tr \ln \Gamma^2\,.
\ee
In order to use \fr{generalfunction}, \fr{eq:analytic} we are
therefore led to consider the function ${\cal F}(z)=\frac12 \ln z$.
The latter has a branch point at $z=0$ and the previous approach appears
not to be applicable. In order to circumvent this problem we employ
a power series expansion of the logarithm around $z=1$
\be
\qquad
\fl \ln (\rho^{xx})^2= \frac 12 \Tr \ln \Gamma^2= \frac12 \Tr \ln [1+(\Gamma^2-1)]= 
\frac12 \sum_{m=1}^\infty \frac{(-1)^{m+1}}{m} \Tr [(\Gamma^2-1)^m]\,.
\label{sum2}
\ee
Application of (\ref{eq:analytic}) to the function ${\cal
  F}(z)=(z-1)^m$ then results in
\begin{eqnarray}
\fl\qquad
\lim_{t,\ell\to\infty\atop t/\ell\ {\rm fixed}}
\frac{\mathrm{Tr}[(\Gamma^2-1)^m]}{2\ell}= \int_{-\pi}^{\pi}\frac{\mathrm d  k_0}{2\pi}\max\Bigl(1-2|\veps'(k_0)| \frac{t}\ell,0\Bigr) 
\bigl(n_x( k_0)^2+|\vec n_\perp( k_0)|^2-1\bigr)^m\nonumber\\
\qquad+\int_{-\pi}^\pi \frac{\mathrm d  k_0}{2\pi}\min\Bigl(2|\veps'(k_0)|\frac{t}\ell,1\Bigr)(n_x( k_0)^{2}-1)^m\, .
\end{eqnarray}
Finally we can carry out the sum over $m$ 
\begin{eqnarray}
\fl\qquad
\lim_{t,\ell\to\infty\atop t/\ell\ {\rm fixed}}
\frac{\mathrm{Tr}[\ln\Gamma^2]}{2\ell}= \int_{-\pi}^{\pi}\frac{\mathrm d  k_0}{2\pi}\max\Bigl(1-2|\veps'(k_0)| \frac{t}\ell,0\Bigr) 
\ln \bigl(n_x( k_0)^2+|\vec n_\perp( k_0)|^2\bigr)\nonumber\\
\qquad+\int_{-\pi}^\pi \frac{\mathrm d  k_0}{2\pi}\min\Bigl(2|\veps'(k_0)|\frac{t}\ell,1\Bigr) \ln (n_x( k_0)^{2})\, ,
\label{luck}
\end{eqnarray}
which is exactly the same result we would have obtained applying
(\ref{eq:analytic}) directly to the function ${\cal F}(z)=\frac12 \ln
z$. The reason for this lies in the simple algebraic dependence 
of (\ref{eq:powers}) on $n$. The specialization of \fr{luck} to the
longitudinal correlator $\rho^{xx}$ is now straightforward.
The symbol $\ \hat\mathtt\Gamma(k)$ of the block Toeplitz matrix
$\bar{\Gamma}$ can be cast in the form (\ref{eq:symbol}) with 
\be
\vec w(k)=[\theta(k)-k]\hat x\ ,\quad  n_x^2(k)=\cos^2\Delta_k\ ,\quad
|\vec n_\perp(k)|^2=\sin^2 \Delta_k\ .
\ee
Eq. (\ref{luck}) then gives the asymptotic behaviour 
\be\label{eq:prediction}
%\fl\qquad 
\lim_{t,\ell\to\infty\atop t/\ell\ {\rm fixed}}\frac1\ell \ln |\rho^{xx}(\ell,t)|=
\int_{-\pi}^\pi \frac{\mathrm d  k_0}{2\pi}\min\Bigl(2|\veps'(k_0)|\frac{t}\ell,1\Bigr)\ln|\cos\Delta_{k_0}| \, .
\ee
This is a rewriting of (\ref{eq:predictionintr}) and represents one of
our main results.

In the reasoning leading up to \fr{luck}
we have assumed that the eigenvalues of $\Gamma^2-1$ lie within the
unit circle so that we can expand the logarithm in a power series. This
is equivalent to the requirement that the eigenvalues $\lambda_j$ of
$\Gamma$ are such that $0<\lambda_j^2\leq 1$. More precisely, if there
exists a $x_0>0$ such that for all $\ell$ the eigenvalues $\lambda_j$
of $\Gamma$ fulfil
\be
x_0<\lambda_j^2\leq 1\ ,
\ee
then the results \fr{luck} and (\ref{eq:prediction}) hold. In some
cases of interest we find that even though $\lambda_j$ are different
from zero for any finite $\ell$, one or several eigenvalues approach
zero in the limit $\ell\to\infty$. In these circumstances the steps
leading up to \fr{luck} and (\ref{eq:prediction}) are no longer
justified. However, even then (\ref{eq:prediction}) can provide useful
information about the asymptotics of $\rho^{xx}(\ell,t)$. 
On general grounds \footnote{This form has been observed is various
examples, see e.g. \cite{cc-07,kla-06,fc-08,rsms-08,bpgda-10,bhc-10,ads2}, even with oscillating factors. }
we expect the latter to be of the form
\be
\label{eq:structure}
\lim_{t,\ell\to\infty\atop t/\ell\ {\rm fixed}}\ln |\rho^{xx}(\ell,t)|=
-\ell \mu ({t}/{\ell})-\alpha\ln(\ell)+c(t/\ell)+\ldots\ ,
\ee
and the question is under what circumstances $\mu(t/\ell)$ is given by
(\ref{eq:prediction}). Under the assumption that
the general structure \fr{eq:structure} holds, this depends on the
number $N_0$ of eigenvalues $\lambda_j$ that approach zero for
$\ell\to\infty$, as well as on how quickly they tend to zero.
If $N_0$ is finite and the corresponding
eigenvalues approach zero \emph{sufficiently slowly} with $\ell$,
(\ref{eq:prediction}) will still be applicable. As an example let us
consider there case where only a single pair $\pm i\lambda_1$ of
eigenvalues approaches zero for $\ell\to\infty$ in such a way that for
large $\ell$ we have $|\lambda_{1}| \sim \ell^{-\beta}$. Since
$\rho^{xx}(\ell,t)=\prod _j \lambda_j^2$, the eigenvalue pair will
contribute to subleading logarithmic term in (\ref{eq:structure}), but
will leave the function $\mu(t/\ell)$ unchanged. On the other hand, if
our pair of eigenvalues were to approach zero exponentially fast
$|\lambda_{1}| \sim e^{-\ell \bar \mu(t/\ell)}$, then it would
contribute additively to the function $\mu(t/\ell)$ and
(\ref{eq:prediction}) would cease to hold.

As far as our quench problem is concerned we do not have a criterion
that would establish a priori whether eigenvalues exponentially close
to zero will be present. We find that for quenches within the
ferromagnetic phase they are always absent. In the other cases,
the picture emerging from numerical studies of the spectrum of $\bar
\Gamma$ suggests that only for quenches starting in the disordered phase
eigenvalues approach zero exponentially fast, and then only a single
pair $\pm i\lambda_0$ does so. Interestingly, we find that the
contribution of all other eigenvalues is again captured by
\fr{luck}. At present we are not able to determine $\lambda_0$
analytically. 

%%%%%%%%%%%%%%%%%%%%%%%%%%
\subsection{Proof of the formula for the trace of integer powers of $\Gamma$.}%
%%%%%%%%%%%%%%%%%%%%%%%%%%

The first step of our calculation is to derive an appropriate integral
representation for ${\rm Tr}\, \Gamma^{2n}$ by trading matrix
multiplications for additional integrations. The basic idea is easily
explained for a product of two $\Gamma$ matrices. Each block element
is given by 
\be
\Gamma_{ln}=\int_{-\pi}^\pi\frac{\mathrm d k}{2\pi} e^{i (l-n) k} \hat t(k)\,,
\ee
where the $2\times 2$ block $\hat t(k)$ is of the form (\ref{eq:symbol}).
Hence the block matrix elements of $\Gamma^2$ can be written as
\be\fl\qquad
(\Gamma^2)_{lm}=\sum_{m=1}^\ell \Gamma_{lm}\Gamma_{mn}=
\int_{[-\pi,\pi]^2}\frac{\mathrm d k_1}{2\pi}\frac{\mathrm d k_2}{2\pi} e^{i (l k_1-n k_2) } \hat{t}(k_1)  \cdot \hat t(k_2)  
\sum_{m=1}^\ell e^{im (k_1-k_2)}\,.
\ee
The sum over $m$  can now be replaced by an integral using
\be
e^{-i (\ell+1)  k/2} \sum_{m=1}^{\ell}e^{imk}=\frac{\ell}{2}\int_{-1}^1\mathrm d \xi \frac{ k}{2\sin (k/2)}e^{i\ell\xi {k}/{2}}\, .
\ee
The generalization to $\Gamma^{2n}$ is straightforward, and replacing
the trace in an analogous way we obtain the following integral
representation 
\begin{eqnarray}
\label{eq:1}
\fl\qquad
  \mathrm{Tr}[\Gamma^{2n}]=
\Bigl(\frac{\ell}{2}\Bigr)^{2n}\!\!\!\!\int\limits_{[-\pi,\pi]^n}\frac{\mathrm d^{2n}  k}{(2\pi)^{2n}}
\int\limits_{[-1,1]^{2n}}\mathrm d^{2n}\xi \, C(\{k\})
e^{i\ell\sum_{j=0}^{2n-1}\xi_j  (k_{j+1}- k_{j})/2} F(\{k\})\ .
\end{eqnarray}
Here the functions appearing under the integrals are
\bea\label{eq:Ck}
C(\{k\})&\equiv \prod_{j=0}^{2n-1}\frac{ k_j- k_{j-1}}{2\sin[( k_j-
    k_{j-1})/2]}\,, \\
F(\{k\})&=\mathrm{Tr}\Bigl[{\prod_{i=0}^{2n-1}n_x( k_i)\sigma_x^{( k)}+\vec n_\perp( k_i) \cdot \vec\sigma^{( k)} 
e^{2 i \veps_i t \sigma_x^{( k)}}}\Bigr]\,,
\eea
the trace is over the remaining $2\times 2$ blocks and $\veps_i=\veps( k_i)$.
Performing the change of variables
\be
\begin{array}{lr}
\zeta_0=\xi_{1},&\\
\zeta_i=\xi_{i+1}-\xi_{i}&i\in[1,n-1],
\end{array}
\ee
Eq. \fr{eq:1} can be rewritten in the form
\begin{eqnarray}\label{eq:2}
\fl\qquad 
\mathrm{Tr}[\Gamma^{2n}]=
\Bigl(\frac{\ell}{2}\Bigr)^{2n}\!\!\!\!\!\!\int\limits_{[-\pi,\pi]^{2n}}\frac{\mathrm
  d^{2n}  k}{(2\pi)^{2n}} 
\int\limits_{R_\zeta}\mathrm d^{2n}\zeta\  C(\{k\})\ 
e^{-i\ell\sum_{j=1}^{2n-1}\zeta_j ( k_{j}- k_{0})/2}\ F(\{k\}). 
%\nonumber\\
%\times \mathrm{Tr}\Bigl[{\prod_{i=0}^{2n-1}n_x( k_i)\sigma_x^{( k)}+\vec n_\perp( k_i) \cdot \vec\sigma^{( k)} 
%e^{2 i \veps_i t \sigma_x^{( k)}}}\Bigr]\,,
\end{eqnarray}
Here the domain of integration \(R_\xi\) is determined by the conditions
\be\label{eq:conditionzeta}
R_\xi:\quad-1\leq \sum_{j=0}^{k-1}\zeta_j\leq 1\qquad \forall k\in[1,2n]\, .
\ee
As the integrand in \Eref{eq:2} is independent of $\zeta_0$, we can
carry out the $\zeta_0$ integration, which gives
\begin{eqnarray}
\fl\qquad
\mathrm{Tr}[\Gamma^{2n}]=
\Bigl(\frac{\ell}{2}\Bigr)^{2n}\!\!\!\!\!\!\!\!\int\limits_{[-\pi,\pi]^{2n}}\frac{\mathrm d^{2n}  k}{(2\pi)^{2n}}\!\!
\int \mathrm d^{2n-1}\zeta\ \mu(\{\zeta\})\ C(\{k\})\
 e^{-i\ell\sum_{j=1}^{2n-1}\zeta_j ( k_{j}- k_{0})/2}\ F(\{k\}).
%\nonumber\\ \times
%\mathrm{Tr}\Bigl[{\prod_{i=0}^{2n-1}n_x( k_i)\sigma_x^{( k)}+\vec n_\perp( k_i) \cdot \vec\sigma^{( k)} 
%e^{2 i \veps_i t \sigma_x^{( k)}}}\Bigr]\, ,
\end{eqnarray}
Here the function $\mu(\{\zeta\})$ is the measure 
of the domain of $\zeta_0$ under the constraints
(\ref{eq:conditionzeta})
\be\label{eq:mu}
\fl\qquad
\mu(\{\zeta\})=\max\Bigl[0,\min_{j\in\{0,2n-1\}}\Bigl[1-\sum_{k=1}^{j}\zeta_k\Bigr]+\min_{j\in\{0,2n-1\}}\Bigl[1+\sum_{k=1}^{j}\zeta_k\Bigr]\Bigr].
\ee
It can be shown that $\mu(\{\zeta\})$ is symmetric with respect to an
arbitrary permutation of the variables $\zeta_i$.  
Since we are interested in the behaviour for \(\ell\gg 1\) and the
phase in the integral is proportional to the large parameter $\ell$,
the asymptotic behaviour can be obtained using a multi-dimensional
stationary phase approximation. The main idea behind this method is
that the leading contribution to the integral arises from the
neighborhoods of the points in which the phase is stationary.
As the symbol is independent of the integration variables $\zeta_i$,
the stationarity of these variables implies
\be
k_j\approx k_0\ ,\quad  j=1,\ldots, 2n-1.
\ee
We may replace any $k_j$ with $k_0$ everywhere except
in rapidly oscillating terms such as the $e^{2i\veps_i  t}$ factors in the
symbol. We call this the \emph{localization rule}. This rule allows us
in particular to drop the factor $C(\{k\})$, cf. (\ref{eq:Ck}), as it
is equal to one at the stationary points $C(\{k\})\simeq1$
\be
\fl\qquad
\mathrm{Tr}[\Gamma^{2n}]\simeq \Bigl(\frac{\ell}{2}\Bigr)^{2n}\!\!\!\!\!\!\int\limits_{[-\pi,\pi]^{2n}}\frac{\mathrm d^{2n}  k}{(2\pi)^{2n}}\int \mathrm d^{2n-1}\zeta\ \mu(\{\zeta\})e^{-i\ell\sum_{j=1}^{2n-1}\zeta_j (k_j- k_0)/2}F(\{k \})\, .
\label{intermediate1}
\ee 
The localization rule furthermore allows us to substitute the
integration variables in $n_x$ and $\vec n_\perp$ with $k_0$, 
and to remove the local rotation in $f(\{k\})$, i.e. under the
integral we may replace 
\be\label{fdef}
F(\{k\})\to G(\{k\})\equiv
\mathrm{Tr}{\prod_{i=0}^{2n-1}\Bigl[n_x( k_0)\sigma_x+\vec n_\perp( k_0) \cdot \vec\sigma e^{2 i \veps_i t \sigma_x}\Bigr]}\, .
\ee
As is shown in \ref{App:prod}, the product under the trace can be
rewritten as 
\be\fl\qquad
\sum_{p=0}^{2n} n_{x}^{2n-p} |n_\perp|^p (i\s_z)^p
\!\!\sum_{1\leq j_1<j_2<\dots j_p\leq 2n} \!\!
(-1)^{\sum_{m=1}^p j_m} e^{2i t  \s_x \sum_{m=1}^p (-1)^{p-m}   \veps_{j_{m-1}}}.
\label{prod1}
\ee
Each term in the sums over $j_1\dots j_p$ gives the same contribution
to the integral \fr{intermediate1} as can be shown by changing
integration variables
\bea
k'_1&=k_{j_1},\
k'_2=k_{j_2},\cdots, k'_{j_p}=k_{p},\ 
k'_l=k_l\ {\rm for\ all}\ l\notin\{j_1,\ldots,j_p\}\ ,\nn
\zeta'_1&=\zeta_{j_1},\
\zeta'_2=\zeta_{j_2},\cdots, \zeta'_{j_p}=\zeta_{p},\ 
\zeta'_l=\zeta_l\ {\rm for\ all}\ l\notin\{j_1,\ldots,j_p\},
\eea
and then invoking the invariance of $\mu(\{\zeta\})$ in
\Eref{eq:mu} under any permutation of its variables. When evaluating
the trace in (\ref{prod1}) we may therefore replace $j_m\to m$.  
We call this the \emph{contraction rule}. 
Noting that only the terms with even $p$ have a non-vanishing trace
(because $\Tr \s_z e^{i a \s_x}=0$) and then applying the contraction
rule allows us to replace
\be
G(\{k\})\rightarrow
%g( k_0,\cdots, k_{2n-1})=
\sum_{p=0}^{n} \binom{n}{p} n_{x}^{2n-2p} n_\perp^{2p} 
2\cos\Big(2t  \sum_{m=1}^{2p} (-1)^{m}  \veps_{m-1} \Big).
\ee
Here the binomial takes into account the number of terms giving
identical contributions to the integral and we have used that $\Tr
e^{i a \s_x}=2 \cos a$.  
Inserting this expression into \fr{intermediate1} we arrive at
\begin{eqnarray}\label{eq:e1}
\fl\qquad\mathrm{Tr}[\Gamma^{2n}]=
\ell \Bigl(\frac{\ell}{2}\Bigr)^{2n-1}
\sum_{p=0}^{ n}\binom{n}{p} \int\limits_{[-\pi,\pi]^{2n}}\frac{\mathrm d^{2n}  k}{(2\pi)^{2n}}\int \mathrm d^{2n-1}\zeta\ \mu(\{\zeta\})\nn
\qquad \times\ n_x^{2n-2k}n_\perp^{2 k}e^{-i\ell \sum_{j=1}^{2p-1}\zeta_j \frac{ k_{j}- k_{0}}{2}+2 i t\sum_{j=0}^{2p-1}(-1)^j \veps_j }\, ,
\end{eqnarray}
where we replaced the cosine with a complex exponential using the
symmetry of the integral under the change of variables
$\vec k\to -\vec k$ and  $\vec\zeta\to -\vec\zeta$.

We are now in a position to employ a stationary phase approximation in order
to extract an exact asymptotic result in the limit of large $\ell$. As
the phase of the integral is stationary along a one-dimensional smooth
curve, application of the stationary phase method is not a simple matter.
For a two-dimensional integral whose phase is stationary on a
one-dimensional variety, the solution can be found in Ref. \cite{Wong}.
Higher-dimensional integrals with a curve of stationary points are
only partially treated in Ref. \cite{serv-88}, where it is
demonstrated that it is possible to isolate the integration in $k_0$
and perform a standard multi-dimensional stationary phase
approximation for the remaining integrals. To apply this idea to our
case,  we rewrite \Eref{eq:e1} as  
\be\label{GamvsLam}
\mathrm{Tr}[\Gamma^{2n}]=\ell
\Bigl(\frac{\ell}{2}\Bigr)^{2n-1}\sum_{l=0}^{ n}\binom{n}{l}  
\int_{-\pi}^\pi\frac{\mathrm d  k_0}{2\pi}n_x( k_0)^{2n-2l}n_\perp(
k_0)^{2 l}\Lambda_{n;l}(k_0)\, , 
\ee
where the functions $\Lambda_{n;l}$ are given by
\be\label{Lambda}
\fl\quad\Lambda_{n;l}(k_0)=\!\!\!\!\!\!\int\limits_{[-\pi,\pi]^{2n-1}}\!\!\!\!\!\!\frac{\mathrm
  d^{2n-1}  k}{(2\pi)^{2n-1}} 
\int \mathrm d^{2n-1}\zeta \ \mu(\{\zeta\})\
e^{-i\ell \sum_{j=1}^{2n-1}\zeta_j \frac{ k_{j}- k_{0}}{2}+2 i t\sum_{j=0}^{2l-1}(-1)^j \veps_j }\, .
\ee
We now use the standard multi-dimensional phase approximation to evaluate
$\Lambda_{n;l}(k_0)$.  

The general result for the large-$\ell$ asymptotic behaviour
of a multi-dimensional integral over rapidly oscillating functions,
whose phase is stationary at an {\it isolated point} $\vec x_0$
detached from the boundary is \cite{Wong} 
\be\label{eq:SPA}
\fl\qquad\int_D \mathrm d ^N x\ p(\vec x)\ e^{i\ell q(\vec x)}=
\Bigl(\frac{2\pi}{\ell}\Bigr)^{{N}/{2}} 
p(\vec x_0)|\det A|^{-{1}/{2}}\exp\Bigl[i \ell q(\vec x_0)+\frac{i \pi \sigma_A}{4}\Bigr]\, ,
\ee
where $A_{i j}=\partial_{x_i}\partial_{x_j}q|_{\vec x_0}$ is the
Hessian matrix of $f(\vec x)$ evaluated in $\vec x_0$, and $\sigma_A$
is the signature of the matrix $A$, \emph{i.e.} the difference between
the number of the positive and negative eigenvalues.  
The stationarity conditions for the phases of the integrands in
$\Lambda_{n;l}(k_0)$ are
\be
\begin{array}{ll}
\bar{k}_j= k_0,&j=1,\dots ,2n-1,\\
\bar{\zeta}_j=4\frac{t}{\ell} (-1)^j\veps'_j,&j=1,\dots,2l-1,\\
\bar{\zeta}_j=0,&j=2l,\dots,2n-1\, .\\
\end{array}
\ee
Using an ordering of integration variables where
the $\zeta$'s are placed before the $k$'s, the Hessian is of the form
\be
A=\frac{1}{2}\left(\begin{array}{cc}
\mathrm 0&\mathrm I\\
\mathrm I&M
\end{array}\right)\, .
\ee
Hence the eigenvalues $a^{(i)}_\pm$ of $A$ are related to the eigenvalues
$\mu_i$ of the matrix $M$ by $a_\pm^{(i)}=(\mu_i\pm\sqrt{\mu_i^2+4})/4$.
The signature of $A$ is $\sigma_A=0$ and its determinant is $\det
A=-4^{1-2n}$. At the stationary point the phase in the integral
vanishes, because there is an even number of $\veps$'s with alternating
signs. Finally, the value of $\mu(\{\zeta\})$  at the stationary point
is found to be
\be
\mu(\{\bar \zeta \})=\left\{\begin{array}{ll}
\frac{2}{\ell} \max[0,\ell-2|\veps'( k_0)|t]&l\neq 0\,,\\
2&l=0\, .
\end{array}\right.
\label{intermediate2}
\ee
Putting everything together, the stationary phase approximation gives
the following result for the functions
$\Lambda_{n;l}(k_0)$
\be
\Lambda_{n;l}(k_0)=\left(\frac2\ell\right)^{2n-1} \mu(\{\bar \zeta
\})\, .
\label{intermediate3}
\ee
Inserting \fr{intermediate3} into (\ref{GamvsLam}) then gives
\be
\mathrm{Tr}[\Gamma^{2n}]= 
\ell \sum_{l=0}^{ n}\binom{n}{l} 
\int_{-\pi}^\pi\frac{\mathrm d  k_0}{2\pi}n_x( k_0)^{2n-2l}n_\perp( k_0)^{2 l}\mu(\{\bar \zeta \})\,,
\ee
and substituting the value \fr{intermediate2} for $\mu(\{\bar \zeta
\})$ we arrive at
\begin{eqnarray}
\fl\qquad\mathrm{Tr}[\Gamma^{2n}]=
2 \int_{-\pi}^{\pi}\frac{\mathrm d  k_0}{2\pi}\max\bigl(\ell-2|\veps'(k_0)| t,0\bigr) \Bigl(n_x( k_0)^2+|\vec n_\perp( k_0)|^2\Bigr)^{n}\nn
 \qquad+2\int_{-\pi}^\pi \frac{\mathrm d  k_0}{2\pi} (\ell- \max(0,\ell-2|\veps'( k_0)|t))n_x( k_0)^{2n}\, .
\label{eq:powers2}
\end{eqnarray}
This is equivalent to (\ref{eq:powers}). 
A less general version of this result was previously presented in
Ref. \cite{fc-08}.

\subsection{On the applicability of stationary phase method.}\label{ss:applicability}

\begin{figure}[t]
\includegraphics[width=0.48\textwidth]{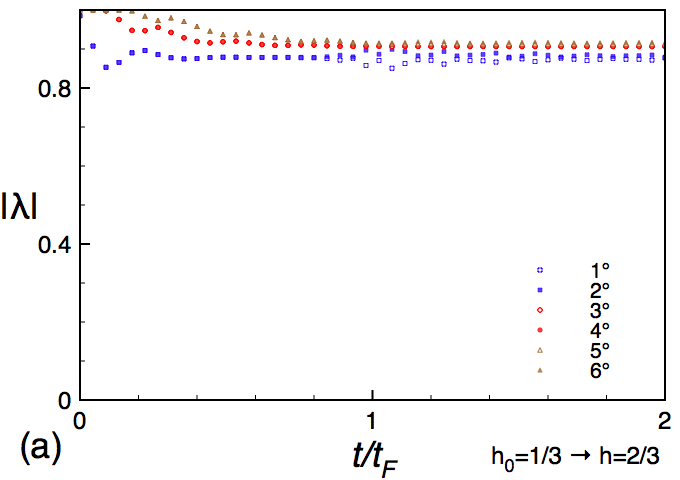}
\includegraphics[width=0.48\textwidth]{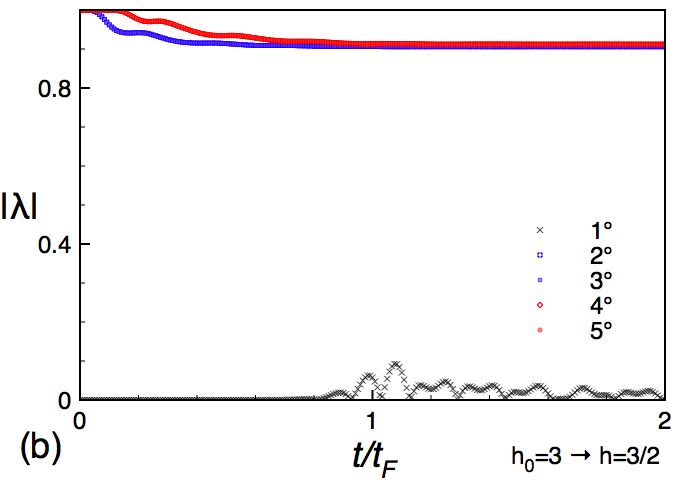}\\
\includegraphics[width=0.48\textwidth]{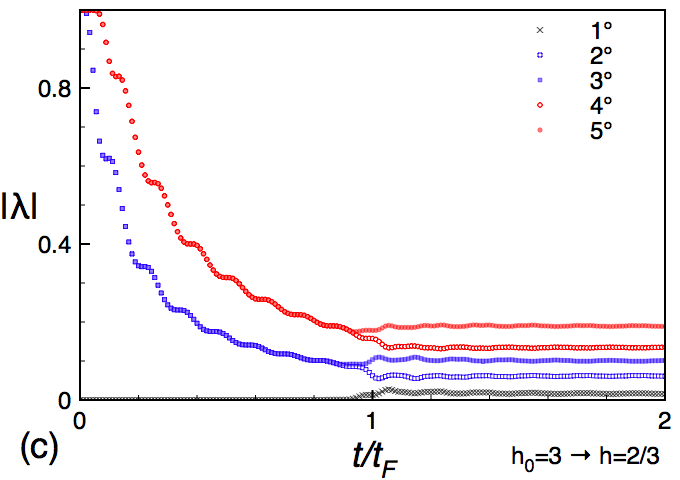}
\includegraphics[width=0.48\textwidth]{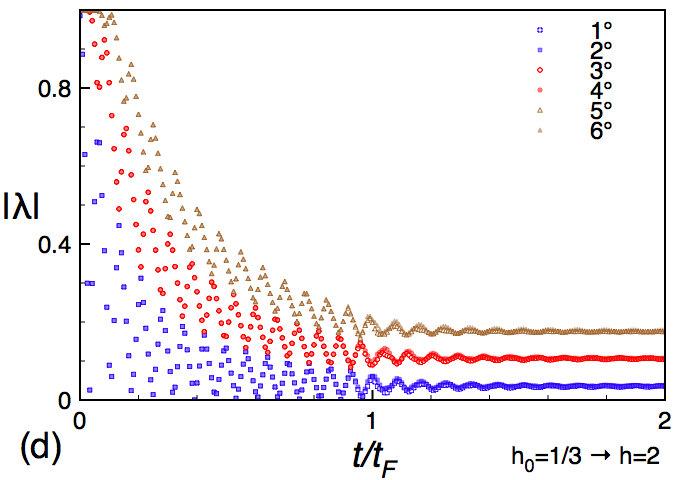}
\caption{Typical absolute values of the low lying eigenvalues of the matrix $W$ in the four kind of quenches.
In (a) and (b) $\ell=30$, while in (c) and (d) $\ell=60$.
(a) For a quench within the ordered phase, all eigenvalues are well separated from zero.
Before the Fermi time (up to finite distance corrections to $t_F$) the eigenvalues are almost degenerate in pairs.
(b) For a quench within the disordered phase, a {\it single} eigenvalue is very close to the real axis (exponentially close 
for $t<t_F$ and as a power law for $t>t_F$).
(c) Quench from the disordered to the ordered phase. For $t<t_F$
a {\it single} eigenvalue is exponentially close to the real axis. For
$t>t_F$ the distance of the smaller eigenvalue from the real axis scales
like a power law.
(d) Quench from the ordered to the disordered phase. All eigenvalues are doubly degenerate.
All the smallest eigenvalues approach zero as a power in $\ell$ for any time. 
For $t<t_F$ the smallest eigenvalue crosses periodically zero.
 }
\label{Fig:spectrum}
\end{figure}
\label{sec32}

Eq. (\ref{eq:prediction}) is our main result obtained with the
determinant approach. It is based on a multi-dimensional stationary
phase approximation for $\mathrm{Tr}[\bar{\Gamma}^{2n}]$.
The result (\ref{eq:powers2}) is then used in combination
with a series expansion in order to obtain the asymptotic behaviour
of $\det \bar{\Gamma}$, which in turn gives the square of the longitudinal
correlation function $\rho^{xx}$. As we have already discussed, this
series expansion is possible only if there is no eigenvalue 
of the matrix $\bar{\Gamma}$ that approaches zero in the large $\ell$
limit. While this restriction appears to be quite simple, we do not
have an analytic method that allows us to predict for what kind of
quenches zero eigenvalues exist in the $\ell\to\infty$ limit. To
address this question we therefore have carried out numerical studies
of the spectrum of $W$ (which is related to the spectrum
of $\bar{\Gamma}$ by \fr{WGamma}) for different quenches. We have to 
distinguish between four cases: 
(1) FM$\longrightarrow$ FM;
(2) PM$\longrightarrow$ PM;
(3) PM$\longrightarrow$ FM and
(4) FM$\longrightarrow$ PM, where FM and PM denote the ferromagnetic
and paramagnetic phases respectively. In Fig.~\ref{Fig:spectrum} we
report the (absolute values of) smallest eigenvalues of the matrix
$W$ for particular examples of the four cases. Extensive
numerical studies suggest that the spectra are similar for all
quenches of a given type, i.e. (1)-(4). The qualitative features
emerging from Fig.~\ref{Fig:spectrum} can be summarized as  
\begin{enumerate}
\item[(1)] FM$\longrightarrow$ FM:
%If $|h|,|h_0< 1$, 
all eigenvalues are well separated from zero;
\item[(2)] PM$\longrightarrow$ PM:
%If $|h_0|,|h|>1$,  
a {\it single} eigenvalue of $W$ is close to
zero. For $t<t_F$ it approaches zero exponentially fast in $\ell$,
while for $t>t_F$ it tends to zero only as a power-law; 
\item[(3)] PM$\longrightarrow$ FM:
for both $t<t_F$ and $t>t_F$ several eigenvalues scale to zero like
power laws in $\ell$. For $t<t_F$ the smallest eigenvalue approaches
zero exponentially fast in $\ell$.
\item[(4)] FM$\longrightarrow$ PM: several eigenvalues tend to
zero in a power-law fashion in $\ell$ for both $t>t_F$ and $t<t_F$.
For $t<t_F$ the smallest eigenvalue crosses zero periodically.
\end{enumerate}
The numerical analysis suggests that (\ref{eq:prediction}) can be
applied for all quenches of type (1) for large $t$ and all corrections
to (\ref{eq:prediction}) are small. In case (4) our result
(\ref{eq:prediction}) still gives the dominant behaviour, but there
are additional important oscillating power-law corrections to $\rho^{xx}(\ell,t)$
that we have not been able to determine analytically. For quenches
originating in the paramagnetic phase (\ref{eq:prediction}) gives the
dominant contribution only if $t>t_F$, but there are 
important power-law corrections to $\rho^{xx}(\ell,t)$ beyond the accuracy
of our analysis.

\begin{figure}[t]
\begin{center}
\includegraphics[width=0.48\textwidth]{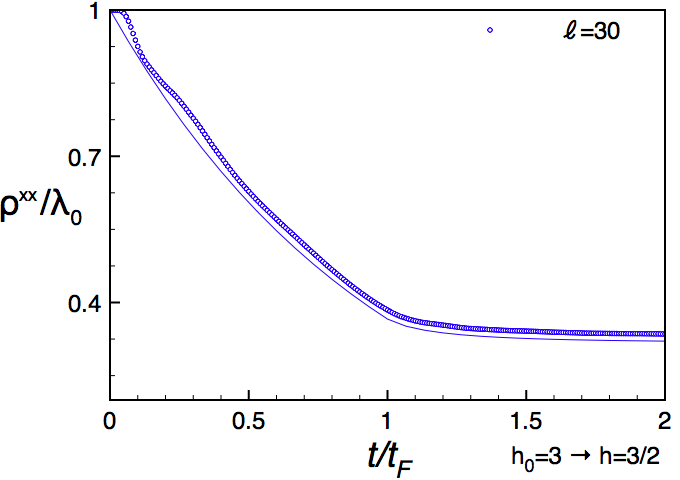}
\includegraphics[width=0.48\textwidth]{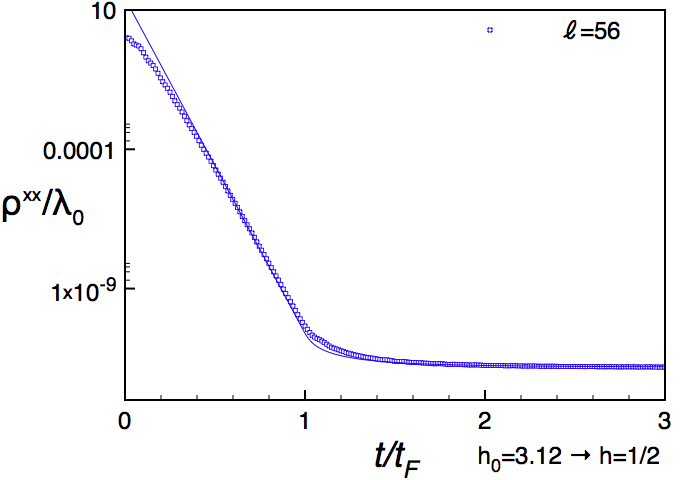}
\caption{Ratio between the numerically evaluated two point function $\rho^{xx}(\ell,t)$ and the 
smallest eigenvalue of the matrix $W$.
Left: Numerical data for a quench within the paramagnetic phase from $h_0=3$ to $h=3/2$.
The data are in perfect agreement with the prediction (\ref{eq:prediction}) 
and the leading correction is a stationary multiplicative factor.
Right: The same plot for a quench from the paramagnetic $h_0=(2+3\sqrt{2})/2$ to the ferromagnetic $h=1/2$ phase. 
}\label{fig:DDwo}
\end{center}
\end{figure}

These properties of the spectrum suggest that, for quenches starting
from the disordered phase, the product of all eigenvalues except the
only one which is close to zero (i.e. $\rho^{xx}/\lambda_0$) should be
given by Eq. (\ref{eq:prediction}). Fig.~\ref{fig:DDwo} shows that
indeed this is the case for quenches to both disordered and ordered
phases for any $t$. Thus, if we were able to obtain a prediction for
the smallest eigenvalue, we would completely characterize these
quenches as well.

Finally, we note that the gross features of the spectrum of
$W$ (and hence $\bar\Gamma$) appear to be related to the winding
number (around zero) of the (block) symbol (\ref{eq:Gamma}). We
observe that the winding number of $g(k)$ (which is only a part of
$\hat{\Gamma}(k)$) depends on the quench parameters in the following
way:  
\begin{enumerate}
\item[(1)] If $|h|,|h_0< 1$, the winding number is zero;
\item[(2)] if $|h_0|,|h|>1$, the winding number is equal to $1$;
\item[(3)] If $|h_0|<1$ and $|h|>1$, the winding number oscillates
  with $t$ between $0$ and $1$. In particular it remains $0$ for any $t>t_F$.
\item[(4)] If $|h_0|>1$ and $|h|<1$,  the winding number oscillates with $t$ between $-1$ and $1$.
\end{enumerate}
We see that the winding number vanishes when there is no eigenvalue
exponentially close to zero, but the reverse does not hold.
For (block) Toeplitz matrices with $\ell$ independent elements
the generalization of the Sz\"ego lemma to symbols with non vanishing
winding number is complicated and not known in general (see
e.g. \cite{FH-gen,sviaj}).

%%%%%%%%%%%%%%%%%%%%%%%%%%%%%%%%%
\subsubsection{A closer look to quenches from the ordered to the disordered phases.}\label{ss:FtoP}
%%%%%%%%%%%%%%%%%%%%%%%%%%%%%%%%%
For quenches from the ordered to the disordered phase, the last panel
of Fig.~\ref{Fig:spectrum} shows that all eigenvalues of $W$ move
coherently in time. In particular, the eigenvalues closest to zero
oscillate with the same frequency about zero until the Fermi time
$t_F$. After $t_F$ they do not cross zero anymore and the smallest one
remains separated from zero as a power law in $\ell$. 
These qualitative features of the spectrum would suggest that for $t<t_F$ $\rho^{xx}(t,\ell)$ should oscillate in time with the 
same frequency of the smallest eigenvalues. 
While it is not easy to put this on a solid ground, we have been able
to identify %exactly this frequency from the numerical data.  
it as the frequency with which the winding number of $g(k)$ changes.
We indeed find that the frequency is determined by the mode $k_0$ such that $\cos\Delta_{k_0}=0$.
Thus for $t<t_F$ we heuristically predict
\be\label{eq:semiprediction}
\fl\quad \braket{\sigma_1^x(t)\sigma_{\ell+1}^x(t)}\sim  
{\cal C}^x_{\rm FP}\left[1+\cos(2\veps_{k_0}t+\alpha)\right]\exp\Bigl[\int_{-\pi}^\pi
  \frac{\mathrm d  k_0}{2\pi}2|\veps'_k|t\ln|\cos\Delta_k|\Bigr]\qquad
t<t_F\, , 
\ee
where $\alpha$ is an undetermined phase shift. 
The comparison of this prediction with the direct computation of the determinant has already been shown in Fig.~\ref{fig:OP}. 
For times larger than $t_F$, \Eref{eq:prediction} matches well the evolution of the two-point function, 
up to oscillatory decaying corrections (see Fig.~\ref{fig:OP}) which is consistent with the presence of one 
eigenvalue exponentially close to zero. 

%%%%%%%%%%%%%%%%%
\section{Form-factor approach}  %
\label{sec:FF}
%%%%%%%%%%%%%%%%%
While the procedure set out in \Sref{s:Determinant approach} can be
generalized to any model with a free-fermion
representation~\cite{fagotti_unpublished}, it cannot be applied to
quenches in \emph{interacting} integrable models. In order to overcome
these limitations we have generalized the form-factor approach to
correlation functions in integrable quantum field theories
\cite{Smirnov92book,FF,Delfino04,mussardobook,review}
to quantum quenches. A characteristic feature of integrable models is
the existence of a basis of scattering states of elementary excitations,
which are simultaneous exact eigenstates of the Hamiltonian and the
momentum operator. These states can be characterized by $n$ momenta
$k_1,\ldots, k_n$ and in general also other quantum numbers
$a_1,\ldots, a_n$ 
\bea
H|k_1,\ldots,k_n\rangle_{a_1,\ldots,a_n}&=&
\Big[\sum_{j=1}^n\varepsilon_{a_j}(k_j)\Big]
|k_1,\ldots,k_n\rangle_{a_1,\ldots,a_n}\ ,\nn
P|k_1,\ldots,k_n\rangle_{a_1,\ldots,a_n}&=&
\Big[\sum_{j=1}^nk_j\Big]
|k_1,\ldots,k_n\rangle_{a_1,\ldots,a_n}\ .
\eea
Ground state correlators of local operators ${\cal O}$
can be expressed in a Lehmann representation
\bea
\langle 0|{\cal O}(t,x){\cal O}^\dagger(0,0)|0\rangle&=&
%
% typo corrected: factorial
%
\sum_{n=0}^\infty\frac{1}{n!}\sum_{k_1,\ldots,k_n\atop
a_1,\ldots,a_n}
|\langle 0|{\cal O}(0,0)|k_1,\ldots,k_n\rangle_{a_1,\ldots,a_n}|^2\nn
&&\qquad\qquad\times\ e^{-i\sum_{j=1}^n\veps_{a_j}(k_j)t-k_jx},
\eea
where $|0\rangle$ denotes the ground state and in our notations
corresponds to particle number zero.
Using the known expressions for the form factors $\langle 0|{\cal
  O}(0,0)|k_1,\ldots,k_n\rangle_{a_1,\ldots,a_n}$ the spectral sum can
be calculated to very high accuracy (for large $t$ and $x$) by 
taking into account only terms involving a small number $n$ of
particles. Recently the form factor approach has been generalized to
low temperature correlation functions \cite{ek09,finiteT} (see also
\cite{doyon,finiteT2} for related work) 
\bea
\frac{{\rm Tr}\left[e^{-\beta H}{\cal O}(t,x){\cal
    O}^\dagger(0,0)\right]}{{\rm Tr}\left[e^{-\beta H}\right]}.
\label{finiteT_eq1}
\eea
The numerator in \fr{finiteT_eq1} can now be expanded in a Lehmann
representation as follows 
\bea
&&\sum_{n,m=0}^\infty\sum_{k_1,\ldots,k_n\atop
a_1,\ldots,a_n}
\sum_{p_1,\ldots,p_m\atop
b_1,\ldots,b_m}
|{}_{b_m,\ldots,b_1}\langle p_m,\ldots,p_1|{\cal O}(0,0)|k_1,\ldots,k_n\rangle_{a_1,\ldots,a_n}|^2\nn
&&\times\ e^{-\beta\sum_{l=1}^m\veps_{b_l}(p_l)}\
e^{-i\sum_{j=1}^n\veps_{a_j}(k_j)t-k_jx}
\ e^{i\sum_{l=1}^m\veps_{b_l}(p_l)t-p_lx}.
\eea
Using exact results for the form factors
${}_{b_m,\ldots,b_1}\langle p_m,\ldots,p_1|{\cal
  O}(0,0)|k_1,\ldots,k_n\rangle_{a_1,\ldots,a_n}$
\cite{Smirnov92book,takacs}, one needs to sum an infinite number of
terms in the Lehmann expansion in order to obtain the correlation
function at late times and large distances. Such a resummation is
possible at low temperatures $T\ll \Delta$, where $\Delta$ is the
spectral gap, because the density of excitations in the state of
thermal equilibrium constitutes a natural small parameter in this case
\cite{ek09}. The situation after a quantum quench bears many
similarities to the finite temperature case. When dealing with a
``small'' quench in a gapped theory there exists a regime in which the
density of excitations (of the post-quench Hamiltonian) in the initial
state is small. It is then natural to use this small parameter in
order to carry out a low-density expansion for the observables of
interest. Let us consider a quantum quench $H_0\rightarrow H$, where
$H$ describes an integrable scattering theory: we prepare the system
in the ground state $|\Psi_0\rangle$ of $H_0$ and then consider time
evolution by $H$. We are interested in observables such as
\bea
\frac{\langle\Psi_0(t)|{\cal O}(x){\cal
    O}^\dagger(0)|\Psi_0(t)\rangle}{\langle\Psi_0|\Psi_0\rangle}.
\eea
Any translationally invariant initial state can be expressed
in terms of the eigenstates of $H$ as
\bea
\fl\quad
|\Psi_0(t)\rangle=\sum_{n=0}^\infty\sum_{k_1,\ldots,k_n\atop
a_1,\ldots,a_n} 
e^{-i\sum_{j=1}^n\veps_{a_j}(k_j)t}
F_n(\{k_j,a_j\})
|k_1,\ldots,k_n\rangle_{a_1,\ldots,a_n},\nn
\fl\quad
F_n(\{k_j,a_j\})=
{}_{a_n,\ldots,a_1}\langle k_n,\ldots,k_1|\Psi(0)\rangle.
\eea
A particular class of initial states is given by
\bea
\fl\quad
|\Psi_0(0)\rangle=\sum_{n=0}^\infty\frac{i^n}{n!}\sum_{p_1,\ldots,p_n}
\left[\prod_{j=1}^nK^{a_jb_j}(p_j)\right]|-p_1,p_1,\ldots,
-p_n,p_n\rangle_{a_1,b_1,\ldots,a_n,b_n},
\label{generalboundarystate}
\eea
where summation over the indices $a_j$, $b_k$ is implied.
As is shown in \ref{app:initial} quantum quenches of the transverse
field in the TFIM automatically lead to initial states of the form
\fr{generalboundarystate}. For general integrable quantum field theories
\fr{generalboundarystate} are known to describe boundary-states
\cite{boundarystate} and correspond to situations where the initial
state can be viewed as a boundary condition (in Euclidean space) 
that is compatible with quantum integrability. Given an initial state
of the form \fr{generalboundarystate}, the basic idea is to employ
a Lehmann representation in terms of the exact eigenstates of the
post-quench Hamiltonian $H$. For a two-point function we have
\bea\fl
\langle\Psi_0(t)|{\cal O}(x){\cal O}^\dagger(0)|\Psi_0(t)\rangle=
\sum_{m=0}^\infty\sum_{n=0}^\infty\frac{i^{n-m}}{n!m!}\sum_{p_1,\ldots,p_n\atop
k_1,\ldots,k_m}
% typo corrected
%
\left[\prod_{j=1}^nK^{a_jb_j}(p_j)\right]
\left[\prod_{l=1}^mK^{c_ld_l}(k_l)\right]^*\nn
\fl\qquad
\times\sum_{s=0}^\infty\sum_{q_1,\ldots,q_s}\frac{1}{s!}\
{}_{c_m,d_m,\ldots,c_1,d_1}\langle k_m,-k_m,\ldots,k_1-k_1|{\cal O}(t,x)
|q_1,\ldots,q_s\rangle_{f_1,\ldots,f_s}\nn
\times\ {}_{f_s,\ldots,f_1}\langle q_s,\ldots,q_1|{\cal O}^\dagger(t,0)
|-p_1,p_1,\ldots, -p_n,p_n\rangle_{a_1,b_1,\ldots,a_n,b_n}.
\label{lehmannquench}
\eea
The form factors entering this expression are known, both for the TFIM
and a variety of integrable quantum field theories
\cite{Smirnov92book,takacs}. The normalization itself has the
following Lehmann representation
\bea
\fl\qquad
\langle \Psi_0(t)|\Psi_0(t)\rangle=
\sum_{n=0}^\infty\frac{(-1)^{n}}{(n!)^2}\sum_{p_1,\ldots,p_n\atop
k_1,\ldots,k_n}
\prod_{j=1}^nK^{a_jb_j}(p_j)
\left[K^{c_jd_j}(k_j)\right]^*\nn
\fl\qquad\qquad\times\
{}_{c_n,d_n,\ldots,c_1,d_1}\langle k_n,-k_n,\ldots,k_1-k_1|
-p_1,p_1,\ldots, -p_n,p_n\rangle_{a_1,b_1,\ldots,a_n,b_n}.
\label{normalization_lehmann}
\eea

In order to extract the large time and
distance asymptotics of \fr{lehmannquench} it is necessary to sum an
infinite number of terms. In analogy to the finite temperature case
we proceed as follows.
\begin{itemize}
\item{} Consider all contributions at a given order in the functions
$K^{ab}(q)$.
\item{} At each order we encounter two types of divergences: (i)
infinite volume divergences that arise because in the thermodynamic
limit the scattering states are normalized to delta functions. These
singular contributions are ultimately compensated by corresponding
divergences in the normalization \fr{normalization_lehmann} and are
dealt with by an appropriate subtraction procedure, cf. \cite{ek09}
for the finite temperature case; 
(ii) ``infrared'' divergences, i.e. contributions that diverge as
$t\to\infty$ or $x\to\infty$. 
\item{} We isolate the terms with the strongest infrared divergences
  at each order in   $K^{ab}(q)$ and then sum up all these
  contributions, {\sl c.f.} \cite{AKT} for a similar calculation in
  the finite-temperature case. 
\end{itemize}

%%%%%%%%%%%%%%%%%%%%%%%%%%%%%%%%%%%%%%%%%%%%%%%%%%%%%%%%%%%%%%%%%%%%%%
\subsection{Finite Volume Form Factors for the TFIM}
%%%%%%%%%%%%%%%%%%%%%%%%%%%%%%%%%%%%%%%%%%%%%%%%%%%%%%%%%%%%%%%%%%%%%%
As is clear from the above discussion the basic building blocks are
the form factors. For the case of the transverse field Ising chain
these naturally depend on the precise choice of basis of free fermion
scattering states $|k_1,\ldots,k_n\rangle_{\R,\NS}$. For a particular
such choice, the finite-volume form factors of the spin operators
$\sigma^x_\ell$ have been determined in
Refs~\cite{bugrij,Gehlen:2008,iorgov}. The non-vanishing matrix
elements are
\begin{eqnarray}\label{eq:formfactors}
\fl\quad
\lrsub{\NS}{\braket{q_1,\dots,q_{2n}|\sigma_\ell^x|p_1,\dots,p_m}}{\R}=
e^{-i\ell\left[\sum_{j=1}^{2n} q_j-\sum_{l=1}^mp_m\right]}\nn
\fl\qquad\qquad\times\
i^{\lfloor n+\frac{m}{2}\rfloor}
(4J^2 h)^\frac{(m-2n)^2}{4}\sqrt{\xi\xi_T}
\prod_{j=1}^{2n}\left[\frac{e^{\eta_{q_j}}}
{L\varepsilon_{q_j}}\right]^\frac{1}{2}
\prod_{l=1}^m\left[\frac{e^{-\eta_{p_l}}}{L\varepsilon_{p_l}}\right]^\frac{1}{2}
\nonumber\\
\fl\qquad\qquad\times
\prod_{j<j'=1}^{2n}\left[\frac{\sin\bigl(\frac{q_j-q_{j'}}{2}\bigr)}
{\varepsilon_{q_j q_{j^\prime}}}\right]
\prod_{l<l'=1}^m\left[\frac{\sin\bigl(\frac{p_l-p_{l'}}{2}\bigr)}
{\varepsilon_{p_l p_{l^\prime}}}\right]
\prod_{j=1}^{2n}\prod_{l=1}^m\left[\frac{\varepsilon_{q_j p_l}}
{\sin\bigl(\frac{q_j-p_l}{2}\bigr)}\right]
%\left\{
%\begin{array}{ll}
%m\ \mathrm{even}&h<1\\
%m\ \mathrm{odd}&h>1\, ,
%\end{array}
%\right.
\end{eqnarray}
where $m$ is even (odd) for $h<1$ ($h>1$) and
\bea
\fl\qquad
\varepsilon_k=2J\sqrt{1+h^2-2 h \cos k}\ ,\quad
\epsilon_{k,k^\prime}\equiv
\frac{\varepsilon_k+\varepsilon_{k^\prime}}{2}\ ,\quad
\xi=|1-h^2|^{1/4}\, ,\nn
\fl\qquad
\xi_T=\prod\limits_{q\in\NS\atop p\in\R}\left(\epsilon_{q,p}\right)^\frac{1}{2}
\prod\limits_{q,q^\prime\in\NS}\big(\epsilon_{q,q^\prime}\big)^{-\frac{1}{4}}
\prod\limits_{p,p^\prime\in \R}\big(\epsilon_{p,p^\prime}\big)^{-\frac{1}{4}}\ ,
\quad
e^{\eta_k}=\frac{\prod\limits_{q\in\NS}\epsilon_{k,q}}{\prod\limits_{p\in
    R}\epsilon_{k,p}}.  
\eea
For large $L$ we have with exponential accuracy (in $L$)
\be
\xi_T\approx 1\qquad e^{\eta_k}\approx 1 \, .
\ee
As we are interested in the thermodynamic limit we will set these
terms equal to 1 in what follows (their deviations from 1 give rise to
subleading contributions). We stress that as a consequence of the 
different quantization conditions of $\rm R$ and $\rm NS$ momenta
there are no singularities in \Eref{eq:formfactors} as long as $L$ is
(large but) finite. The free fermionic basis used in Ref.~\cite{iorgov}
differs from the one discussed in \ref{app:diag}. However, the initial
state can still be represented in terms of eigenstates of the post
quench Hamiltonian $H(h)$ using \fr{BS_NS} and \fr{BS_R}, provided
that $K(k)$ is chosen appropriately.

%%%%%%%%%%%%%%%%%%%%%%%%%%%%%%%%%%%%%%%%%%%%%%%%%%%%%%%%%%%%%%%%%%%%%%
\subsection{Quench within the ferromagnetic phase:  time evolution of
  the order parameter}\label{ss:1point} %
%%%%%%%%%%%%%%%%%%%%%%%%%%%%%%%%%%%%%%%%%%%%%%%%%%%%%%%%%%%%%%%%%%%%%%

We now consider a quench within the ordered phase: we prepare the
system in the ground state of the Hamiltonian $H(h_0)$ and at time $t=0$
suddenly change the magnetic field from $h_0$ to $h$, where
$h_0,h<1$. We are interested in the case where the $\mathbb{Z}_2$
symmetry of $H(h_0)$ is broken spontaneously in the ground state
$|\Psi_0\rangle$. This is possible only in the thermodynamic limit. 
On the other hand, we would like to keep the length $L$ of the system
very large but finite in our calculations for technical reasons. In
order to be able to work in a large, finite volume, we therefore
take our initial state to be of the form
\be
|\Psi_0\rangle=\frac{1}{\sqrt{2}}\Big\{|0;h_0\rangle_{\rm
    NS}+|0;h_0\rangle_{\rm R}\Big\},
\label{Psi0}
\ee
where $|0;h_0\rangle_{\rm NS}$ and $|0;h_0\rangle_{\rm R}$ are the ground
states of $H(h_0)$ in the sectors with even/odd numbers of fermions
respectively, see  \ref{ferro}. For finite system size $L$, the state
\fr{Psi0} is a particular linear combination of the ground state and
first excited state of the Hamiltonian $H(h_0)$. On the other hand, in
the thermodynamic limit $L\to\infty$ \fr{Psi0} becomes the
symmetry broken ground state of $H_0$. As is shown in
\ref{app:initial} the time-evolved initial state 
\be
|\Psi_0(t)\rangle=e^{-iH(h)t}|\Psi_0\rangle
\ee
can be expressed in terms of eigenstates of $H(h)$ as
\be
|\Psi_0(t)\rangle=\frac{1}{\sqrt{2}}\left[
\frac{|B(t)\rangle_{\rm NS}}{\sqrt{{}_{\rm NS}\langle B|B\rangle_{\rm NS}}}
+\frac{|B(t)\rangle_{\rm R}}{\sqrt{{}_{\rm R}\langle B|B\rangle_{\rm
    R}}}\right],
\ee
where
\begin{eqnarray}\label{eq:boundary}
\fl\qquad
\ket{B(t)}_{\tt a}&= e^{-i E_{0}^{\tt a} t}\exp\Bigl[i \sum_{0<p\in {\tt
    a}} K(p) e^{-2i\varepsilon_p t}b_{-p}^\dag
  b_p^\dag\Bigr]\ket{0;h}_{\tt a}\ ,\quad {\tt a}={\rm R,NS}.
\end{eqnarray}
For the choice of basis underlying \fr{eq:formfactors}
the function $K(k)$ is given by 
\be
\label{eq:k_ordered}
\fl\qquad
K(k)=\frac{\sin(k)\ (h-h_0)}{\veps_{h_0}(k)
\veps_{h}(k)\big(2J\big)^{-2}+1+hh_0-(h+h_0)\cos(k)}\, . 
\ee
We note that this agrees with what would be obtained using the choice
of fermions presented in \fr{app:diag}.
The expectation value of $\sigma^x_\ell$ is then given by
\be\label{eq:order}
\langle\Psi_0(t)|\sigma^x_\ell|\Psi_0(t)\rangle=\frac{
{}_{\rm R}\langle B(t)|\sigma^x_\ell|B(t)\rangle_{\rm NS}
+{}_{\rm NS}\langle B(t)|\sigma^x_\ell|B(t)\rangle_{\rm R}
}{2\sqrt{{}_{\rm R}\langle B|B\rangle_{\rm R}\
{}_{\rm NS}\langle B|B\rangle_{\rm NS}}}.
\ee
We note that the diagonal contributions vanish
\be
{}_{\tt a}\langle B(t)|\sigma^x_\ell|B(t)\rangle_{\tt a}=0,\quad {\tt
  a}={\rm R,NS}\ ,
\ee
because
\be
e^{-i\pi{\hat N}}\sigma^x_\ell e^{i\pi{\hat N}}=-\sigma^x_\ell\ ,
\ee
where $\hat{N}$ is the fermion number operator \fr{fermionnumber}.
%
%The form-factor approach entails the expansion of
%$\lrsub{\NS}{\braket{B(t)|\sigma^x_\ell|B(t)}}{\R}$ in a Lehmann
%representation. We view $K$ as a formal ``expansion parameter'' and we
%series expand the states~(\ref{eq:boundary}) in \Eref{eq:order};
%finally, we substitute  the form factors of $\sigma^x_\ell$, \emph{i.e.}
%the matrix elements in the basis that diagonalizes the final
%Hamiltonian. To summarize, in order to compute the order parameter (or
%the two-point correlation function $\rho_\ell^{x}\equiv
%\braket{\sigma_l^x\sigma_{l+\ell}^x}$) we need two ingredients: the
%normalizations ${}_{\tt a}\langle B|B\rangle_{\tt a}$, ${\tt
%  a}=\NS,\R$ (see \Eref{eq:order}) and the form factors. 
%From a physical perspective, the small parameter in our expansion is
%the density of excitations of the post-quench Hamiltonian in the
%initial state, as is shown in \fr{densityofexc}.

The normalization is readily evaluated using the explicit
representation~(\ref{eq:boundary}) 
\begin{eqnarray}\label{eq:norm}
\lrsub{\NS}{\braket{B|B}}{\NS}=\exp\Bigl[\sum_{0<q\in\NS}\log\bigl(1+K^2(q)\bigr)\Bigr]\ ,\nn
\lrsub{\R}{\braket{B|B}}{\R}=\exp\Bigl[\sum_{0<p\in\R}\log\bigl(1+K^2(p)\bigr)\Bigr]\ ,
\end{eqnarray}
so that, for large $L$, the norms
$\lrsub{\R,\NS}{\braket{B|B}}{\R,\NS}$ are approximately equal
(we note that $K(0)=K(\pi)=0$) 
\be
\lrsub{\R}{\braket{B|B}}{\R}\approx
\lrsub{\NS}{\braket{B|B}}{\NS}\approx
\exp\Bigl(L\int_0^\pi\frac{\mathrm d k}{2\pi}\log(1+K^2(k)) \Bigr)\, .
\ee 
Importantly we have
\be
\frac{\lrsub{\R}{\braket{B|B}}{\R}}{\lrsub{\NS}{\braket{B|B}}{\NS}}=1+{\cal
  O}\big(e{^{-\alpha L}}\big),
\label{normalizationRNS}
\ee
where $\alpha$ is a positive constant. Eqn \fr{normalizationRNS}
allows us to simplify the expression \fr{eq:order} for the 1-point
function to 
\be\label{eq:order2}
\langle\Psi_0(t)|\sigma^x_\ell|\Psi_0(t)\rangle={\rm Re}\frac{
{}_{\R}\langle B(t)|\sigma^x_\ell|B(t)\rangle_{\NS}}
{{}_{\rm R}\langle B|B\rangle_{\rm R}}+{\cal O}\big(L^{-2}\big).
\ee
As we are interested in the thermodynamic limit this representation is
most convenient and we will use it in the following. The normalization
can be expanded in powers of $K$ as
\begin{eqnarray}
\label{eq:normalization}
\fl\qquad
\lrsub{\tt R}{\braket{B|B}}{\tt R}=
1+\sum_{0<k\in{\R }}K^2(k)+\frac{1}{2}
\left[\sum_{0<k\in{\R }}K^2(k)\right]^2
-\frac{1}{2}\sum_{0<k\in{\R }}K^4(k)+\ldots\nn
\fl\qquad\quad
 \equiv 1+\sum_{n=1}^\infty \Upsilon_{2n}\, ,
\end{eqnarray}
where $\Upsilon_{2n}$ collects all the terms in which $2n$ functions
$K(k)$ are multiplied together. The following representation of
$\Upsilon_{2n}$ turns out to be particularly useful
\be
\Upsilon_{2n}=\frac{1}{n!}
\sideset{}{'}\sum_{0<k_1,\ldots,k_n\in\R}
%\left[\prod_{l=1}^n\prod_{m=l}^n(1-\delta_{k_l,k_m})\right]
\prod_{j=1}^nK^2(k_j).
\label{ups2n}
\ee
Here $\sum'$ indicates that the sum is only over terms with $k_l\neq
k_m$ for $l\neq m=1,\dots,n$.
Eqn \fr{eq:normalization} is a
formal series expansion as each term grows as $L^{n}$, \emph{i.e.}
diverges in the thermodynamic limit. The divergences in the
normalization \fr{eq:normalization} are mirrored by infinite volume
divergences in the numerator of \fr{eq:order}. 
Using \fr{eq:boundary} to expand
$\lrsub{\NS}{\braket{B(t)|\sigma^x_\ell|B(t)}}{\R}$ in powers of the
function $K(k)$ gives 
\begin{eqnarray}
\label{eq:formalexp}
\fl\quad
{\rm Re}\left[\lrsub{\NS}{\braket{B(t)|\sigma^x_\ell|B(t)}}{\R}\right]
=2m_0^x\biggl\{1+R_1(t)+\Bigl[R_2(t)+\Upsilon_2\Bigr]
+\Bigl[R_3(t)+\Upsilon_2 R_1(t)\Bigr]\nn
\qquad\qquad\quad+\Bigl[R_4(t)+\Upsilon_2
R_2(t)+\Upsilon_4\Bigr]+\dots\biggr\}\, , 
\end{eqnarray}
where $m_0^x\equiv \braket{\sigma^x_\ell}_0/2$ is the magnetization at the
initial time and the $R_n(t)$ are finite in the thermodynamic limit.
In terms of the ``connected'' contributions $R_n(t)$ the order
parameter expectation value \fr{eq:order2} is expressed as
\bea
\langle\Psi_0(t)|\sigma^x_\ell|\Psi_0(t)\rangle=2m_0^x
\left[1+\sum_{n=1}^\infty R_n(t)\right].
\label{linked_clusters}
\eea
Eqn \fr{linked_clusters} constitutes a linked cluster expansion, where
the contribution $R_n(t)$ is of order ${\cal O}\big(K^{n}\big)$.
As is shown in \ref{app:initial} (see Eqn \fr{densityofexc})
physically 
the formal expansion in powers of $K(q)$
corresponds to a 
low-density expansion, where the small parameter is the density of
excitations of the post-quench Hamiltonian in the initial state.

The next step is to determine the functions $R_n(t)$. Expanding
the boundary states in
$\lrsub{\NS}{\braket{B(t)|\sigma^x_\ell|B(t)}}{\R}$ we obtain 
\begin{eqnarray}\label{eq:Lehmann}
\fl\qquad
\lrsub{\NS}{\braket{B(t)|\sigma^x_\ell|B(t)}}{\R}&=
\sum_{l,n=0}^\infty\frac{i^{n-l}}{n!\ l!}
\sum_{q_1,\dots,q_n\in\NS\atop
p_1,\dots,p_l\in\R}\prod_{j=1}^nK(q_j)
\prod_{i=1}^lK(p_i)e^{2i t (\sum_{i=1}^n\varepsilon_{q_i}-\sum_{j=1}^l \varepsilon_{p_j})}\nn
&\qquad\qquad\times
\lrsub{\NS}{\braket{-q_1,q_1,\cdots,-q_n,q_n|\sigma^x_\ell|p_1,-p_1,\cdots,p_l,-p_l}}{\R}\,
\nn\fl
&\equiv\sum_{l,n=0}^\infty C_{(n,l)}.
\end{eqnarray}
In the following we consider the first few terms in the expansion
\fr{eq:Lehmann}. 
We will find that most terms exhibit long-time (infrared) divergences,
the strongest of which occur in ``diagonal'' contributions $n=l$. We
will determine and then sum these leading singularities to all orders
in the expansion \fr{eq:Lehmann}.

%%%%%%%%%%%%%%%%%%%%%%%%%%%%%%%%%%%%%%%%%%%%%%%%%%%%%%%%%%%%
\subsubsection{Order ${\cal O}\big(K^0\big)$ Contribution ($n=l=0$).}
%%%%%%%%%%%%%%%%%%%%%%%%%%%%%%%%%%%%%%%%%%%%%%%%%%%%%%%%%%%%

The zero-particle contribution in \fr{eq:Lehmann} is the ground state
(of the post-quench Hamiltonian $H(h)$) expectation value
\be
C_{(0|0)}=2 m_0^x=\sqrt{\xi}=(1-h^2)^{1/8}\, .
\ee

%%%%%%%%%%%%%%%%%%%%%%%%%%%%%%%%%%%%%%%%%%%%%%%%%%%%%%%%%%%%
\subsubsection{Order ${\cal O}\big(K\big)$ Contributions.}
%%%%%%%%%%%%%%%%%%%%%%%%%%%%%%%%%%%%%%%%%%%%%%%%%%%%%%%%%%%%

At first order in $K$ there are two contributions
\bea
\fl
C_{(1|0)}=i\sum_{0<q\in{\rm NS}}K(q)e^{2it\veps_q}\ {}_{\rm
  NS}\langle -q,q|\sigma^x_\ell|0\rangle_{\rm R}=
4J^2h\sqrt{\xi}\frac{1}{L}\sum_{0<q\in{\rm NS}}\frac{K(q)\sin
  q}{\veps_q^2}e^{2it\veps_q},\nn
\fl
C_{(0|1)}=-i\sum_{0<p\in{\rm R}}K(p)e^{-2it\veps_p}\ {}_{\rm
  NS}\langle 0|\sigma^x_\ell|p,-p\rangle_{\rm R}=
4J^2h\sqrt{\xi}\frac{1}{L}\sum_{0<p\in{\rm R}}\frac{K(p)\sin
  p}{\veps_p^2}e^{-2it\veps_p}.
\label{C10C01}
\eea
In the limit $L\to\infty$ we can turn the momentum sums
in \fr{C10C01} into integrals, which gives
\bea\fl\qquad
I(t)=\frac{1}{\sqrt{\xi}}\lim_{L\to\infty} \left[C_{(1|0)}+C_{(0|1)}\right]
=8J^2h\int_{0}^\pi\frac{dk}{2\pi}
\frac{K(k)\sin  k}{\veps_k^2}\cos\big(2t\veps_k\big).
\eea
For late times we can evaluate this integral by a stationary phase
approximation 
\be
I(t)= A_0\frac{\cos\big(2\varepsilon_0t+\frac{3\pi}{4}\big)}{t^{3/2}}
-A_\pi\frac{\cos\big(2\varepsilon_\pi
  t-\frac{3\pi}{4}\big)}{t^{3/2}}+o(t^{-3/2})\ ,
\label{ioft}
\ee
where
\be
%
% factor of two removed
%
A_k=\frac{hJ^2K'(k)}{\sqrt{\pi}\varepsilon^2_k|\varepsilon''_k|^{3/2}}\
,\qquad
k=0,\pi\, .
\ee
Comparison of \fr{eq:Lehmann} and \fr{eq:formalexp} then allows us to
identify the function $R_1(t)$ in the thermodynamic limit
\be
R_1(t)=I(t).
\ee
%%%%%%%%%%%%%%%%%%%%%%%%%%%%%%%%%%%%%%%%%%%%%%%%%%%%
\subsubsection{Order ${\cal O}\big(K^2\big)$ Contributions.}
%%%%%%%%%%%%%%%%%%%%%%%%%%%%%%%%%%%%%%%%%%%%%%%%%%%%

To order ${\cal O}(K^2)$ there are two types of contributions.
$C_{(2|0)}$ and $C_{(0|2)}$ are finite in the thermodynamic limit and
well behaved at late times. They don't play a significant role in the
following and we therefore refrain from presenting explicit 
expressions. The most important contribution to order
${\cal O}(K^2)$ is given by
\be\label{eq:2parmag}
C_{(1|1)}=\sum_{0<q\in\NS\atop 0<p\in \R}K(q)K(p)\ e^{-2it (\varepsilon_p-\varepsilon_q)}
\lrsub{\NS}{\braket{-q,q|\sigma^x_\ell|p,-p}}{\R}\, .
\ee
The relevant form factor is, \emph{cf.} eqn \fr{eq:formfactors},
\be
\lrsub{\NS}{\braket{-q,q|\sigma^x_\ell|p,-p}}{\R}\simeq
\frac{4 \sqrt{\xi}}{L^2}
\frac{\epsilon_{p,q}^4}{\varepsilon^2_q\varepsilon^2_p} \sin q\sin p\ c_{q p}^2\, ,
\ee
where we have defined
\be
c_{k k^\prime}\equiv\frac{1}{\cos k-\cos k^\prime}\, .
\ee
For large $t$ and $L$ the momentum sum in the NS-sector can be evaluated using
Lemma 1 \fr{id1} and retaining only the leading terms in $t$ and $L$. 
This results in
\be
C_{(1|1)}= \sqrt{\xi}\sum_{0<p\in
  \R}K^2(p)\left[1-\frac{4t\varepsilon^\prime_p}{L}\right]+\ldots
\label{eq:C11}
\ee
Crucially, this contribution exhibits an \emph{infinite volume}
divergence (for $L\to\infty$) as well as an \emph{infrared} divergence
(for $t\to\infty$). To ${\cal O}(1)$ (in both $L$ and $t$) \fr{eq:C11}
can be expressed as 
\be
C_{(1|1)}= \sqrt{\xi}\Upsilon_2-\sqrt{\xi}\int_{0}^{\pi}\frac{\mathrm
  d k}{\pi }K^2(k)\ 2 t\varepsilon^\prime_k+\ldots\, , 
\label{c11}
\ee
where $\Upsilon_2$ is defined in \fr{eq:normalization}.
By comparing \fr{c11} to \fr{eq:formalexp} we can determine
the function $R_2(t)$ for large $t$
\be
\label{eq:R2tK}
R_2(t)=-\int_{0}^{\pi}\frac{\mathrm d k}{\pi }K^2(k)2
t\varepsilon^\prime_k
+\ldots\, ,
\ee
where the dots denote terms that are subleading in $t$ at late times.
We note the leading term \fr{eq:R2tK} arises from the region $p\approx
q$ in \fr{eq:2parmag}. This observation will be important in what follows.
%%%%%%%%%%%%%%%%%%%%%%%%%%%%%%%%%%%%%%%%%%%%%%%%%%%%%
\subsubsection{Order ${\cal O}\big(K^3\big)$ Contributions.}
%%%%%%%%%%%%%%%%%%%%%%%%%%%%%%%%%%%%%%%%%%%%%%%%%%%%%
To this order there are several contributions. $C_{(3|0)}$ and
$C_{(0|3)}$ are completely regular and do not play an important role
in the following. The evaluation of the other contributions follows
closely the calculation for $C_{(1|1)}$ and results in
\bea
C_{(1|2)}+{\rm h.c.}&={\rm Re}\sum_{0<k_{1,2}\in\NS}\ \sum_{0<p\in\R}
iK(k_1)K(k_2)K(p)e^{2it[\veps_{k_1}+\veps_{k_2}-\veps_p]}\nn
&\qquad\qquad\qquad
\times\ {}_{\rm NS}\langle   -k_1,k_1,-k_2,k_2|\sigma_m^x|p,-p\rangle_{\rm R}\nn
&={\rm Re}\big(C_{(0|1)}\big)
\left[\Upsilon_2-\frac{1}{L}\sum_{0<p\in \R}4t\veps'_pK^2(p)
\right]+\ldots\ ,\nn
C_{(2|1)}+{\rm h.c.}&={\rm Re}\sum_{0<k_{1,2}\in\R}\ \sum_{0<p\in\NS}
iK(k_1)K(k_2)K(p)e^{2it[\veps_{k_1}+\veps_{k_2}-\veps_p]}\nn
&\qquad\qquad\qquad
\times\ {}_{\rm R}\langle   -k_1,k_1,-k_2,k_2|\sigma_m^x|p,-p\rangle_{\rm NS}\nn
&={\rm Re}\big(C_{(1|0)}\big)
\left[\Upsilon_2-\frac{1}{L}\sum_{0<p\in\NS}4t\veps'_pK^2(p)
\right]+\ldots
\label{k3lehmann}
\eea
Comparison with \fr{eq:formalexp}
then gives
\be
R_3(t)=-I(t)\int_0^\pi\frac{dp}{\pi}2t\veps'_pK^2(p)+\ldots,
\label{k3leading}
\ee
where the dots indicate contributions that are subleading in $t$. We
note that \fr{k3leading} arises from the regions
$k_{1,2}\approx p$ in \fr{k3lehmann}. 
%%%%%%%%%%%%%%%%%%%%%%%%%%%%%%%%%%%%%%%%%%%%%%%%%%%%%
\subsubsection{Order ${\cal O}\big(K^4\big)$ Contribution $C_{(2|2)}$.}
%%%%%%%%%%%%%%%%%%%%%%%%%%%%%%%%%%%%%%%%%%%%%%%%%%%%%
The most important contribution at order ${\cal O}(K^4)$ is
$C_{(2|2)}$ and we now discuss its evaluation in some detail. From
\fr{eq:Lehmann} we have
\begin{eqnarray}\label{eq:4cont}
\fl \qquad C_{(2|2)}=\frac{1}{2!^2}
\sum_{0<q\neq q^\prime\in \NS\atop 0<p\neq p^\prime\in\R}K(q)K(q^\prime)K(p)K(p^\prime)e^{-2it(\varepsilon_{p}+\varepsilon_{p^\prime}-\varepsilon_{q}-\varepsilon_{q^\prime})}\nonumber\\
\qquad\qquad\times\ \lrsub{\NS}{\braket{-q,q,-q^\prime,q^\prime|\sigma^x_\ell|p,-p,p^\prime,-p^\prime}}{\R}\, ,
\end{eqnarray}
where the form factor is given by
\begin{eqnarray}\label{eq:4matel}
\fl\qquad 
\lrsub{\NS}{\braket{-q,q,-q^\prime,q^\prime|\sigma^x_\ell
|p,-p,p^\prime,-p^\prime}}{\R}=\frac{16\sqrt{\xi}}{L^4}
\frac{\epsilon^4_{q, p}\epsilon^4_{q,p^\prime}
\epsilon^4_{q^\prime, p}\epsilon^4_{ q^\prime, p^\prime}
}{\varepsilon_q^2\varepsilon_{q^\prime}^2\varepsilon_p^2\varepsilon_{p^\prime}^2\epsilon^4_{q, q^\prime}\epsilon^4_{p, p^\prime}} \nn
\qquad\qquad\qquad\qquad\qquad\times \
\sin q \sin q^\prime \sin p\sin p^\prime \frac{c_{q p}^2 c_{q p^\prime}^2c_{q^\prime p}^2c_{q^\prime p^\prime}^2}{c_{q q^\prime}^2 c_{p p^\prime}^2}\, .
\end{eqnarray}
Each factor $c_{q q^\prime}$ is associated with a singularity in the
thermodynamic limit and it is useful to isolate these poles as
functions of the NS-sector momenta, e.g.
\begin{eqnarray}
\label{eq:trick1}
c_{p q}c_{p^\prime q}=c_{p p^\prime}(c_{p^\prime q}-c_{p q})\ ,\nn
c_{p q}^2c_{p^\prime q}^2=c_{p p^\prime}^2\Bigl[c_{p q}^2+c_{p^\prime
    q}^2-2 c_{p p^\prime}(c_{p^\prime q}-c_{p q})\Bigr]\ , \nn
\frac{c_{q p}^2 c_{q p^\prime}^2c_{q^\prime p}^2c_{q^\prime
    p^\prime}^2}{c_{q q^\prime}^2 c_{p p^\prime}^2}=c_{q p}^2
c^2_{q^\prime p^\prime}+c^2_{q p^\prime}c^2_{q^\prime p}-2c_{p
  p^\prime}^2(c_{p^\prime q}-c_{p q})(c_{p^\prime q^\prime}-c_{p
  q^\prime})\, .
\end{eqnarray}
Using the symmetry of \fr{eq:4cont} under exchange of $p$ and $p^\prime$
the contribution of the last term in \fr{eq:trick1} can be reexpressed
by substituting
\be\label{eq:4csimp2}
\frac{c_{q p}^2 c_{q p^\prime}^2c_{q^\prime p}^2c_{q^\prime p^\prime}^2}{c_{q q^\prime}^2 c_{p p^\prime}^2}\rightarrow 2 c_{q p}^2 c_{q^\prime p^\prime}^2-4c_{p p^\prime}^2 c_{p q}c_{p q^\prime}+4c^2_{p p^\prime}c_{p q}c_{p^\prime q^\prime}\, .
\ee
The sums over the NS momenta can then be carried out using 
Lemmas 1 \fr{id1} and 5 \fr{id5}. We begin with the contribution due
to the first term in \fr{eq:4csimp2}, which we denote by
$C_{(2|2)}^{[2,2]}$. Here the superscript indicates the pole
structure, namely two double poles.
The NS sector momentum sums in \fr{eq:4cont} are performed by using
(\ref{id1}) and retaining only the leading terms in $L$ and $t$, which gives
\bea
\label{eq:C22[22]}
C_{(2|2)}^{[2,2]}&=\frac{\!\sqrt{\xi}}{2}\sum_{0<p\neq p^\prime\in
  \R}K^2(p)\left[1-\frac{4t\varepsilon^\prime_p}{L}
\right]K^2(p^\prime)\left[1-\frac{4t\varepsilon^\prime_{p^\prime}}{L}
\right]+\ldots\nn
&=\frac{\sqrt{\xi}}{2}\bigl[R_2(t)+\Upsilon_2\bigr]^2
-\frac{\sqrt{\xi}}{2}\sum_{0<p\in \R}K^4(p)
\left[1-\frac{8t\varepsilon^\prime_p}{L}\right]+\ldots .
\eea
Working out the contributions arising from the other terms in
\fr{eq:4csimp2} is more involved. Carrying out the sums over $q$ and
$q'$ using (\ref{id5}) and retaining only the leading contributions in
$L$ and $t$ gives
\begin{eqnarray}
\label{eq:4C1}
\fl\qquad 
-4c_{p p^\prime}^2 c_{p q}c_{p q^\prime}&\rightarrow
\frac{4\sqrt{\xi}}{L^2}\sum_{0<p\neq p^\prime\in\R}K^3(p)K(p^\prime)
e^{-2it(\varepsilon_{p^\prime}-\varepsilon_p)}
\frac{\epsilon_{p,p^\prime}^4}{\varepsilon_p^2\varepsilon_{p^\prime}^2}\sin p\sin
p^\prime c_{p p^\prime}^2\ ,\nn
\fl\qquad 4c^2_{p p^\prime}c_{p q}c_{p^\prime q^\prime}&\rightarrow
-\frac{4
  \sqrt{\xi}}{L^2}\sum_{0<p\neq p^\prime\in\R}K^2(p)K^2(p^\prime)\sin p
\sin p^\prime c_{p p^\prime}^2\, .
\end{eqnarray}
The leading contribution at late times and large $L$ can then be
extracted by using Lemma 2a \fr{id1b}
\be
C_{(2|2)}^{[1,1]}=-\frac{4\sqrt{\xi}}{ L}\sum_{0<p\in \R}K^4(p)
t\varepsilon^\prime_p+\ldots\ .
\ee
Here the superscript $[1,1]$ indicates that we are considering the
contribution from two single poles. Putting everything together we
arrive at the following result for $C_{(2|2)}$
\be
C_{(2|2)}=C_{(2|2)}^{[2,2]}+C_{(2|2)}^{[1,1]}
=\sqrt{\xi}\Bigl[\frac{1}{2}\big(R_2(t)\big)^2
+\Upsilon_2 R_2(t)+\Upsilon_4\Bigr]+\ldots\, .
\ee
This allows us to identify the leading contribution to
the term $R_4(t)$ in \fr{eq:formalexp} as
\be
\label{eq:R4tK}
R_4(t)=\frac{1}{2}\big(R_2(t)\big)^2+\ldots\, .
\ee

%%%%%%%%%%%%%%%%%%%%%%%%%%%%%%%%%%%%%%%%%%%%%%%%%%%%%%%%%%%%%%%%%%%%%%%%%%%
\subsubsection{Exponentiation of the Contributions $C_{(n|n)}$.}%
%%%%%%%%%%%%%%%%%%%%%%%%%%%%%%%%%%%%%%%%%%%%%%%%%%%%%%%%%%%%%%%%%%%%%%%%%%%
Our analysis of the first few orders in powers of $K$ reveals the
general structure of the expansion \fr{eq:Lehmann}: at each order
(except the very lowest ones) there are late time divergences that
become stronger at higher orders. Moreover, the leading singularities
are found in the ``diagonal'' contributions $C_{(n|n)}$ and we will
now isolate these singular terms. At order ${\cal O}(K^{2n})$ we have
\begin{eqnarray}
C_{(n|n)}=
\frac{1}{n!^2}
\sideset{}{'}
\sum_{q_1,\dots,q_n\in\NS\atop
p_1,\dots,p_n\in\R}\prod_{j=1}^nK(q_j) K(p_j)e^{2i t
  \sum_{i=1}^n(\varepsilon_{q_i}- \varepsilon_{p_i})}\nn
\qquad\quad\times\ 
\lrsub{\NS}{\braket{-q_1,q_1,\dots,-q_{n},q_{n}|\sigma^x_\ell|p_1,-p_1
\dots,p_n,-p_n}}{\R}\, , 
\end{eqnarray}
where the form factor in the limit of large $L$ is given by
\begin{eqnarray}
\label{eq:formfactorsdiag}
\fl
\qquad\lrsub{\NS}{\braket{-q_1,q_1,\dots,-q_{n},q_{n}|\sigma^x_\ell
|p_1,-p_1\dots,p_n,-p_n}}{\R}=\nonumber\\
\fl \qquad\qquad \frac{4^n \sqrt{\xi}}{L^{2n}}
\left[\prod_{j,l=1}^{n}\epsilon_{q_j, p_l}^4c^2_{q_jp_l}
\right]
\left[\prod_{j=1}^{n} \frac{\sin(q_j)\sin(p_j)}{
\varepsilon_{q_j}^{2}\varepsilon_{p_j}^{2}}
\right]\left[\prod_{j<j'=1}^{n}
\frac{1}{\epsilon_{q_j, q_{j^\prime}}^{4} \epsilon_{p_j,
    p_{j^\prime}}^{4}
c^2_{q_j q_{j^\prime}}c^2_{p_j p_{j^\prime}}}\right]\, .
\end{eqnarray}
The most singular terms at late times arise from the regions in
momentum space
\be
p_j\approx q_{P(j)}\, ,
\ee
where $P$ is a permutation of $1, \dots, n$. There are $n!$ such
contributions with only double poles and they are all equal. They are
determined by replacing
\be
\label{eq:decdominant}
\frac{\prod_{j=1}^{n}\prod_{l=1}^n c_{q_j
    p_l}^2}{\prod_{j<j'=1}^{n}c^2_{q_j q_{j^\prime}}c^2_{p_j
    p_{j^\prime}}}\rightarrow n!\prod_{i=1}^n c^2_{q_i p_i}+\dots\, ,
\ee
and then carrying out the NS-sector momentum sums using \fr{id1}. This gives
\be
\label{eq:Cnn0}
C_{(n|n)}^{[2, \dots,
    2]}=\frac{1}{n!}\frac{4^n\sqrt{\xi}}{L^n}
\sideset{}{'}\sum_{0<p_1,\dots, p_n\in\R}\ \prod_{i=1}^n
K(p_i)^2\Bigl[\frac{L}{4}-t\varepsilon_{p_i}^\prime\Bigr]+\ldots\, . 
\ee
This is in agreement with \fr{eq:C11} and \fr{eq:C22[22]}. 
Using \fr{ups2n} we see that the $t$-independent part of \fr{eq:Cnn0}
equals $\Upsilon_{2n}$, so that
\be
\sum_{n=0}^{\infty}C_{(n|n)}^{[2,\dots, 2]}\Big|_{t=0}=\sqrt{\xi}
\exp\Bigl[\sum_{0<p\in\R}\log\Bigl(1+K^2(p)\Bigr)\Bigr]+\ldots\ .
% \lrsub{\R}{\braket{B|B}}{\R}+\ldots\, .
\ee
For $t>0$ we may invert the steps used to express \fr{eq:norm} in
terms of \fr{ups2n}, which gives
\begin{eqnarray}
\label{eq:exponentiation}
\fl\qquad\sum_{n=0}^{\infty}C_{(n|n)}^{[2,\dots,
    2]}&=\sqrt{\xi}\exp\Bigl[\sum_{0<p\in\R}\log\Bigl(1+K^2(p)\Bigl[1-\frac{4t\varepsilon_{p}^\prime}{L}\Bigr]\Bigr)\Bigr]+\ldots\nonumber\\ 
&=\sqrt\xi\
\lrsub{\R}{\braket{B|B}}{\R}\
\exp\Bigl[-t \int_0^\pi \frac{\mathrm d
    k}{\pi}\frac{K^2(k)}{1+K^2(k)}|2\varepsilon^\prime(k)|\Bigr]+\ldots\, . 
\end{eqnarray}
The
contribution of all these terms to the 1-point function of the order
parameter is thus
\bea
{\rm Re}\sum_{n=0}^\infty \frac{C_{(n|n)}}{\braket{B|B}_{\R}}
=\sqrt{\xi}\ \exp\Bigl[-t \int_0^\pi \frac{\mathrm d
    k}{\pi}\frac{K^2(k)}{1+K^2(k)}|2\varepsilon^\prime(k)|\Bigr]+\ldots\, .
\label{cnncontrib}
\eea

%%%%%%%%%%%%%%%%%%%%%%%%%%%%%%%%%%%%%%%%%%%%%%%%%%%%%%%%%%%%%%%%%%%%%%%%%%%
\subsubsection{Exponentiation of the Contributions $C_{(n\pm 1|n)}$.}%
%%%%%%%%%%%%%%%%%%%%%%%%%%%%%%%%%%%%%%%%%%%%%%%%%%%%%%%%%%%%%%%%%%%%%%%%%%%
The leading terms (at late times and large $L$) in $C_{(n\pm 1|n)}$
can be summed to all orders in a similar way to our treatment of
$C_{(n|n)}$. The contributions $C_{(n+1|n)}$ are given by
\begin{eqnarray}
\fl\qquad 
{\rm Re}\ C_{(n+1|n)}=\frac{1}{(n+1)!n!}\ {\rm Im}
\sideset{}{'}
\sum_{0<q_1,\dots, q_n,q\in \NS\atop 0<p_1,\dots, p_{n}\in \R}
K(q)\left[\prod_{j=1}^{n}K(q_j)K(p_j)\right]
e^{2i t \varepsilon_{q}+2i t \sum_{i=1}^{n}(\varepsilon_{q_i}- \varepsilon_{p_i})}\nn
\qquad\qquad\times \
\lrsub{\NS}{\braket{-q_1,q_1,\dots,-q_n,q_n,-q,q|\sigma^x_\ell
|p_1,-p_1, \dots p_{n},-p_{n}}}{\R}\, ,
\end{eqnarray}
where the form factors in the limit of large $L$ are
\bea
\fl
\lrsub{\NS}{\braket{-q_1,q_1,\dots,-q_n,q_n,-q,q|\sigma^x_\ell
|p_1,-p_1, \dots p_{n},-p_{n}}}{\R}=-4i hJ^2 \frac{4^{n} \sqrt{\xi}}{L^{2n+1}}
\prod_{j,l=1}^{n}\epsilon_{q_j, p_l}^4
\prod_{m=1}^{n} \varepsilon_{q_m}^{-2}\varepsilon_{p_m}^{-2}\nn
\fl\qquad\times\
\prod_{j<j'=1}^{n}\epsilon_{q_j, q_{j^\prime}}^{-4} \epsilon_{p_j, p_{j^\prime}}^{-4}
\prod_{i=1}^n\sin(q_i)\sin(p_i)
\frac{\prod_{j,l=1}^{n} c_{q_j p_l}^2}{\prod_{j<j'=1}^{n}c^2_{q_j
    q_{j^\prime}}c^2_{p_j p_{j^\prime}}}
\prod_{j=1}^n\frac{\epsilon_{q, p_j}^4}{\epsilon_{q, q_j}^4}\frac{\sin
  q}{\varepsilon_q^2}\prod_{j=1}^n\frac{ c_{q p_j}^2}{c^2_{q_j q}} 
\, .
\end{eqnarray}
Following the same reasoning that led to \fr{eq:decdominant}, we
focus on the contributions arising from the regions
\be
p_j\approx \tilde{q}_{j}\ ,\ j=1,\ldots,n\ ,
\ee
where $(\tilde{q}_1,\ldots,\tilde{q}_{n+1})$ is an arbitrary
permutation of $(q_1,\ldots,q_n,q)$. All of these contributions are
the same and hence the leading behaviour  at late times can be
extracted by the substitution
\be
\frac{\prod_{j=1}^{n}\prod_{l=1}^n c_{q_j p_l}^2}{\prod_{j<j'=1}^{n}c^2_{q_j q_{j^\prime}}c^2_{p_j p_{j^\prime}}}\prod_{j=1}^n\frac{ c_{q p_j}^2}{c^2_{q_j q}}\rightarrow (n+1)!\prod_{i=1}^n c^2_{q_i p_i}+\dots\, .
\ee
The NS-sector momentum sums can be carried out using (\ref{id1}),
which gives
\begin{eqnarray}
\label{eq:Cn+1n0}
\fl \quad 
{\rm Re}\ C_{(n+1|n)}^{[2, \dots, 2]}&=-\frac{1}{n!}\frac{4^{n+1}hJ^2
  \sqrt{\xi}}{L^{n+1}}\sum_{0<q\in NS\atop 0<p_1, \dots, p_n\in\R}K(q)
\cos(2t \varepsilon_q)\frac{\sin  q}{\varepsilon_q^2}\prod_{i=1}^n
K(p_i)^2\Bigl[\frac{L}{4}-t\varepsilon_{p_i}^\prime\Bigr]+\ldots\nn
\fl\quad
&= -4hJ^2 \ C_{(n|n)}^{[2,\dots,2]}\ 
\int_0^\pi\frac{\mathrm d k }{2\pi}K(k)\cos(2t
\varepsilon_k)\frac{\sin
  k}{\varepsilon_k^2}+\ldots\, .  
\end{eqnarray}
Summing over $n$ then results in
\begin{eqnarray}
\label{eq:exponentiation_n+1n}
\fl\qquad{\rm Re}\sum_{n=0}^{\infty}C_{(n+1|n)}^{[2,\dots,
    2]}&=\sqrt{\xi}\frac{I(t)}{2}\lrsub{\R}{\braket{B|B}}{\R}\
\exp\Bigl[-t \int_0^\pi \frac{\mathrm d
    k}{\pi}\frac{K^2(k)}{1+K^2(k)}|2\varepsilon^\prime(k)|\Bigr]+\ldots\, ,
\end{eqnarray}
where $I(t)$ is given by \fr{ioft}. The analysis for the contributions
$C_{(n|n+1)}$ is completely analogous and finally leads to the
following result
\bea
{\rm Re}\sum_{n=0}^\infty \frac{C_{(n+1|n)}+C_{(n|n+1)}}{\braket{B|B}_{\R}}
=\sqrt{\xi}I(t)\ \exp\Bigl[-t \int_0^\pi \frac{\mathrm d
    k}{\pi}\frac{K^2(k)}{1+K^2(k)}|2\varepsilon^\prime(k)|\Bigr]+\ldots\, ,
\label{resummed2}
\eea
We see that by virtue of the extra factor $I(t)$ this is always
subleading compared to the ``diagonal'' contribution \fr{cnncontrib}.

%%%%%%%%%%%%%%%%%%%%%%%%%%%%%%%%%%%%%%%%%%%%%%%%%%%%%%%%%%%%%%%%%%%%%%%%%%%
\subsubsection{Form Factor Result for the 1-point Function
  $\langle\Psi_0(t)|\sigma^x_\ell|\Psi_0(t)\rangle$} 
%%%%%%%%%%%%%%%%%%%%%%%%%%%%%%%%%%%%%%%%%%%%%%%%%%%%%%%%%%%%%%%%%%%%%%%%%%%
Combining the results in the previous subsections we conclude that
\bea\label{eq:1pointFF}
\langle\Psi_0(t)|\sigma^x_\ell|\Psi_0(t)\rangle\sim
\sqrt{\xi}\left[1+I(t)+{\cal O}\big(t^{-1}\big)\right]e^{-t/\tau}\ ,
\eea
where the decay time $\tau$ is given by
\be
\label{eq:ximform}
\tau^{-1}= \int_0^\pi \frac{\mathrm d k}{\pi}K^2(k)
\left[2\varepsilon^\prime(k)\right]+O(K^6)\, .
\ee
Here we have used \fr{eq:R4tK} to infer the absence of $O(K^4)$
corrections to the decay time. The corrections in the prefactor are
expected to be ${\cal O}\big(t^{-1}\big)$ as there are such
(subleading) contributions in the various terms $C_{(n|l)}$ we have
calculated above. We note that these terms are larger than $I(t)$ at
late times as $I(t)\sim t^{-3/2}$.

The result \fr{eq:ximform} is in agreement with
the one obtained by applying the cluster decomposition principle on
the asymptotic result for the two-point function obtained in the
determinant approach (\emph{cf.} \Eref{eq:prediction}), which gives
\be\label{eq:ximexact}
\tau^{-1}=\int_0^\pi \frac{\mathrm d
  k}{\pi}\log\Bigl|\frac{1+K^2(k)}{1-K^2(k)}\Bigr|
\varepsilon^\prime(k)\, . 
\ee
Indeed, expanding \fr{eq:ximexact} in the limit where $K^2(k)\ll 1$
gives precisely \fr{eq:ximform}. We note that \fr{eq:ximform} is a rather
good approximation for a wide range of magnetic fields $h_0$ and $h$.
For example, for a quench from $h_0=0.1$ to $h=0.8$ the relative error
is less than $1\%$. This shows the effectiveness of the form-factor approach.
%{\color{red} This still needs to be tied up with section 2}

%%%%%%%%%%%%%%%%%%%%%%%%%%%%%%%%%%%%%%%%%%%%%%%%%%%%%%%%
\subsection{Two-Point Function in the Ordered Phase}
\label{sec:2spoint}
%%%%%%%%%%%%%%%%%%%%%%%%%%%%%%%%%%%%%%%%%%%%%%%%%%%%%%%%
We now turn to the time evolution of the two-point function for a
quench within the ordered phase. The quantity we want to evaluate is
\bea
\rho^{xx}(t,\ell)&=&\langle\Psi_0(t)|\sigma^x_{m+\ell}\sigma^x_m|\Psi_0(t)
\rangle\nn
&=&\frac{{}_{\rm R}\langle
B(t)|\sigma^x_{m+\ell}\sigma^x_m|B(t)\rangle_{\rm R}}
{2\ {}_{\rm R}\langle B|B\rangle_{\rm R}}
+\frac{{}_{\rm NS}\langle B(t)|\sigma^x_{m+\ell}\sigma^x_m|B(t)\rangle_{\rm
  NS}}
{2\ {}_{\rm NS}\langle B|B\rangle_{\rm NS}}.
\label{2point0}
\eea
In the large-$L$ limit both terms contribute equally,
so that
\bea
\lim_{L\to\infty}\rho^{xx}(t,\ell)=
\lim_{L\to\infty}\frac{{}_{\rm R}\langle
B(t)|\sigma^x_{m+\ell}\sigma^x_m|B(t)\rangle_{\rm R}}
{{}_{\rm R}\langle B|B\rangle_{\rm R}}.
\label{2point}
\eea
We note that as a consequence of reflection symmetry the sign of $\ell$
does not matter, i.e. $\rho^{xx}(t,\ell)=\rho^{xx}(t,-\ell)$. We therefore
will assume from now on that
\be
\ell\geq 0.
\ee
The Lehmann representation for the numerator on the right hand
side of eqn \fr{2point} is
\bea
\fl
{}_{\rm R}\langle B(t)|\sigma^x_{m+\ell}\sigma^x_m|B(t)\rangle_{\rm R}=
\sum_{l,n=0}^\infty\frac{i^{n-l}}{n!\ l!}
\sumprime_{0<k_1,\dots,k_n\in\R\atop
0<p_1,\dots,p_l\in\R}\
\sum_{q_1,\dots,q_r\in\NS}\frac{1}{r!}
\left[\prod_{j=1}^nK(k_j)\right]
\left[\prod_{i=1}^lK(p_i)\right]\nn
\fl\hskip 4cm
\times\
e^{-2it[\sum_{j=1}^l\veps_{p_j}
-\sum_{s=1}^n\veps_{k_j}]}
{}_{\rm R}\langle -k_1,k_1,\ldots,-k_n,k_n|\sigma^x_{m+\ell}|
q_1,\ldots,q_r\rangle_{\rm NS}\nn
\fl\hskip 4cm
\times\
{}_{\rm NS}\langle q_r,\ldots,q_1|\sigma^x_m|
p_1,-p_1\ldots,p_l,-p_l\rangle_{\rm R}\nn
\fl\hskip 3.7cm\equiv\sum_{l,n,r=0}^\infty D_{(2n|r|2l)}.
\label{2point2}
\eea
Like in the case of the 1-point function we focus on the terms
with the strongest possible singularities in the form factors for a
given order in the formal expansion in powers of $K(p)$. These are
obtained by taking $2n=2l=r$. 
%%%%%%%%%%%%%%%%%%%%%%%%%%%%%%%%%%%%%%%%%%%%%%%%%%%%%%%%
\subsubsection{Contributions at ${\cal O}(K^0)$.}
%%%%%%%%%%%%%%%%%%%%%%%%%%%%%%%%%%%%%%%%%%%%%%%%%%%%%%%%
These are equal to the ground state two-point function
\be
{}_\R\langle0|\sigma^x_{m+\ell}\sigma^x_m|0\rangle_\R\ .
\label{gs2pt}
\ee
Using a Lehmann representation to evaluate \fr{gs2pt} gives rise to
contributions 
\be
D_{(0|2s|0)}=\sum_{0<q_1,\dots,q_{2s}\in\NS}\frac{1}{(2s)!}
{}_{\rm R}\langle 0|\sigma^x_{m+\ell}|
q_1,\ldots,q_{2s}\rangle_{\rm NS}\
{}_{\rm NS}\langle q_{2s},\ldots,q_1|\sigma^x_m|
0\rangle_{\rm R}.
\ee
For $s=0$ we have
\be
D_{(0|0|0)}=\big(2m_0^x\big)^2=\xi=(1-h^2)^{1/4},
\ee
while for $s=1$ we obtain
\bea
D_{(0|2|0)}
%&=\frac{1}{2}\sum_{q_1,q_2\in\NS}
%\Big|{}_{\rm R}\langle 0|\sigma^x_{m}|
%q_1,q_{2}\rangle_{\rm NS}\Big|^2e^{ij(q_1+q_2)}\nn
&=\frac{2h^2J^4\xi}{\pi^2}\int_{-\pi}^\pi dq_1dq_2\
\frac{\sin^2\big(\frac{q_1-q_2}{2}\big)}{\veps_{q_1}
\veps_{q_2}\eps^2_{q_1,q_2}}\
e^{i\ell(q_1+q_2)}+{\cal O}\big(L^{-1}\big)\propto h^{2\ell}.
%
% -\ell changed to +\ell
%
\eea
We see that at large $\ell$ $D_{(0|2|0)}$ is exponentially small
compared to $D_{(0|0|0)}$. The contributions $D_{(0|2s|0)}$ with $s>1$
are suppressed by additional exponential factors. We conclude that for
large distances $\ell$ we only need to retain $D_{(0|0|0)}$.
%%%%%%%%%%%%%%%%%%%%%%%%%%%%%%%%%%%%%%%%%%%%%%%%%%%%%%%%
%\subsubsection{Contributions at ${\cal O}(K^1)$.}
%%%%%%%%%%%%%%%%%%%%%%%%%%%%%%%%%%%%%%%%%%%%%%%%%%%%%%%%

%%%%%%%%%%%%%%%%%%%%%%%%%%%%%%%%%%%%%%%%%%%%%%%%%%%%%%%%
\subsubsection{Contributions at ${\cal O}(K^2)$.}
\label{ssec:D222}
%%%%%%%%%%%%%%%%%%%%%%%%%%%%%%%%%%%%%%%%%%%%%%%%%%%%%%%%

At order ${\cal O}(K^2)$ the largest contribution at large
distances and late times arises from
\bea
\fl\qquad
D_{(2|2|2)}&=\sum_{0<k,p\in \R}\frac{1}{2}\sum_{q_1,q_2\in\NS}
K(k)K(p)e^{-2it[\veps_p-\veps_k]}
e^{i\ell(q_1+q_2)}\ {}_{\rm R}\langle -k,k|\sigma^x_m|q_1,q_2\rangle_{\rm NS}\nn
&\qquad\qquad\times\
{}_{\rm NS}\langle q_2,q_1|\sigma^x_m|p,-p\rangle_{\rm R}
\equiv D_{(2|2|2)}^{(1)}+D_{(2|2|2)}^{(2)},
\label{2point2part}
\eea
where $D_{(2|2|2)}^{(1,2)}$ denote the contributions arising from terms
with $k\neq p$ and $k=p$ respectively. Using \fr{id3} to carry out the
summations over $q_{1,2}$ we arrive at
\bea
\fl\qquad
D_{(2|2|2)}^{(1)}=\frac{2\xi}{L^2}\sum_{0<k\neq
  p\in \R}e^{-2it[\veps_p-\veps_k]}
\frac{(\veps_k+\veps_p)^2}{\veps_k^2\veps_p^2}
\frac{K(k)K(p)}{(\cos k-\cos p)^2}\Bigg[
\sin k\sin p\ \veps_{kp}^2\nn
\fl\qquad\qquad\qquad
+\veps_k\veps_p\Big[\sin^2\Big(\frac{k-p}{2}\Big)
\cos\Big(\ell(k+p)\Big)-\sin^2\Big(\frac{k+p}{2}\Big)
\cos\Big(\ell(k-p)\Big)\Bigg]+\ldots
\eea
We note that in this expression there is no singularity if we consider
the limit $k- p\to 0$. Nevertheless for large $\ell$ and $t$ the
leading contribution to the double sum arises from the vicinity of
$k=p$. In order to isolate this contribution we turn the sum over $k$
into an integral using the Euler-Maclaurin sum formula and then deform
the integration contour into the upper half plane. As $\veps_p'>0$ for
all $p>0$ the contribution of the first term in square brackets will
be negligible for large $t$. The same holds true for the second
contribution as long 
as $2\veps_p't\pm \ell >0$. On the other hand, if $2\veps_p't\pm \ell <0$ we
need to deform the contour into the lower half-plane (in the variable
$k$). In doing so acquire a contribution from the double pole at
$p=k$. The residue is dominated by the factors involving $t$ and
$\ell$ as both of these are assumed to be large, so that we end up
with
\be
D_{(2|2|2)}^{(1)}=-\frac{4\xi}{L}\sum_{0<p\in\R} K^2(p)\
\theta_H(\ell-2\veps'(p)t)\left[2\veps_p't-\ell\right]+\ldots\ ,
\label{d222_1}
\ee
where $\theta_H(x)$ denotes the Heaviside step function.
The second contribution $D_{(2|2|2)}^{(2)}$ is given by
\bea
\fl\qquad
D_{(2|2|2)}^{(2)}&=&\frac{8\xi}{L^4}
\sum_{0<k\in\R}\frac{\sin^2(k)\
  K^2(k)}{\veps^4(k)}\sum_{q_{1,2}\in\NS}\frac{e^{i\ell(q_1+q_2)}}{\veps_{q_1}\veps_{q_2}}
\frac{\sin^2\Big(\frac{q_1-q_2}{2}\Big)
\ \eps_{q_1,k}^4\eps_{q_2,k}^4}{\eps_{q_1,q_2}^2}\ c_{kq_1}^2c_{kq_2}^2.
\label{d22}
\eea
Using \fr{id4a} to carry out the summations over $q_{1,2}$
we arrive at
\bea
D_{(2|2|2)}^{(2)}=\xi\sum_{0<k\in\R}K^2(k)\left[1-\frac{4\ell}{L}\right]
+\ldots
\label{d222_2}
\eea
where we again have only retained the leading terms at large $L$ and
$\ell$. Adding the two contributions \fr{d222_1} and \fr{d222_2} we
obtain 
\bea
\fl
D_{(2|2|2)}(t,\ell)&=\xi\left[\Upsilon_2-\frac{4}{L}\sum_{0<k\in\R}K^2(k)
\left(\theta_H\big(\ell-2t\varepsilon_k'\big)\left[
2t\varepsilon_k'-\ell\right]+\ell\right)\right]+\ldots\nn
\fl
&=\xi\left[\Upsilon_2-\frac{4}{L}\sum_{k>0}K^2(k)
\left(2t\varepsilon_k'\ \theta_H\big(\ell-2t\varepsilon'(k)\big)+
\ell\ \theta_H\big(2t\varepsilon_k'-\ell\big)\right)\right]+\ldots
\label{2partfinal}
\eea
This contains a part $\Upsilon_2$ that diverges in the thermodynamic
limit, but which will again be compensated by the denominator in
\fr{2point}. 
%%%%%%%%%%%%%%%%%%%%%%%%%%%%%%%%%%%%%%%%%%%%%%%%%%%%%%%%
\subsection{Exponentiation of $D_{(2n|2n|2n)}$}
\label{Exp1}
%%%%%%%%%%%%%%%%%%%%%%%%%%%%%%%%%%%%%%%%%%%%%%%%%%%%%%%%
The dominant contribution at order ${\cal O}\big(K^{2n}\big)$ is again
the ``diagonal'' one
\bea
\fl\quad
D_{(2n|2n|2n)}=&\frac{\xi}{(n!)^2(2n)!}\frac{1}{L^{4n}}
\sumprime_{0<k_1,\ldots,k_n\in\R\atop
0<p_1,\ldots,p_n\in\R}\prod_{r=1}^{n}\left[\frac{\sin(k_r)K(k_r)
e^{2it\veps_{k_r}}}{\veps^2_{k_r}}
\frac{\sin(p_r)K(p_r)e^{-2it\veps_{p_r}}}{\veps^2_{p_r}}\right]\nn
\fl
&\times\ \prod_{l<l'}^n\frac{1}{4c^2_{k_lk_{l'}}\eps^4_{k_l,k_{l'}}}
\frac{1}{4c^2_{p_lp_{l'}}\eps^4_{p_l,p_{l'}}}
\sum_{q_1,\ldots,q_{2n}\in\NS}
\left[\prod_{r=1}^{2n}\frac{e^{i\ell q_r}}{\veps_{q_r}}\right]\nn
\fl
&\times\ \prod_{l<l'}^{2n}\frac{\sin^2\big(\frac{q_l-q_{l'}}{2}\big)}{\eps^2_{q_l,q_{l'}}}
\prod_{r=1}^n\prod_{s=1}^{2n}4\eps^2_{k_r,q_s}\eps^2_{p_r,q_s}c_{k_rq_s}c_{p_rq_s}.\nn
\label{D2n2n2n}
\eea
In order to extract the large time and distance behaviour of
\fr{D2n2n2n} we need to focus on the regions where the form factors
exhibit the strongest singularities. For the case $n=1$ discussed in 
subsection \ref{ssec:D222} these regions were
\bea
\bullet\quad q_1&\approx \pm k\ ,\quad
q_2\approx\mp p\ {\rm and}\ p\approx k\ ,\nn
\bullet\quad q_2&\approx\pm k\ ,\quad
q_1\approx \mp p\ {\rm and}\ p\approx k.
\eea
For general $n$ we need to focus on the regions
\bea
\pmatrix{k_{Q_1}\cr\vdots\cr k_{Q_n}}\approx 
\pmatrix{p_{1}\cr\vdots\cr p_{n}}\ ,\quad
\pmatrix{q_{S_1}\cr \vdots\cr q_{S_{2n}}}
\approx\pmatrix{\sigma_1 k_{Q_1}\cr -\sigma_1 p_1\cr \vdots\cr
\sigma_n k_{Q_n}\cr -\sigma_n p_n }\ ,
\label{regions_ord}
\eea
where $\sigma_j=\pm$ and $(Q_1,\ldots,Q_{n})$,
$(S_1,\ldots,S_{2n})$ are permutations of $(1,2\ldots,n)$ and
$(1,2\ldots,2n)$ respectively. By symmetry all these regions
contribute equally, which gives rise to a combinatorial factor of
$n!(2n)!$. We therefore focus on the single region where
$(Q_1,\ldots,Q_n)=(1,\ldots,n)$ and
$(S_1,\ldots,S_{2n})=(1,\ldots,2n)$, together will all regions
obtained by exchanging any pair $S_{2r-1}\leftrightarrow S_{2r}$. 
We start by expressing \fr{D2n2n2n} in the form
\bea
\fl
D_{(2n|2n|2n)}&=\frac{1}{L^{4n}n!2^n}\sumprime_{0<k_1,\ldots,k_n\in\R\atop
0<p_1,\ldots,p_n\in\R}e^{2it\sum_{l=1}^n(\veps_{k_l}-\veps_{p_l})}\sum_{q_1,\ldots,q_{2n}\in\NS}
f(k_1,\ldots,k_n;p_1,\ldots p_n|q_1,\ldots,q_{2n})\nn
\fl
&\qquad\qquad\qquad\times\
\prod_{r=1}^ne^{i(q_{2r-1}+q_{2r})\ell}c_{k_rq_{2r-1}}c_{k_rq_{2r}}c_{p_rq_{2r-1}}c_{p_rq_{2r}},
\eea
where it is understood that the summation only terms arising from the
region we consider are retained.
We then carry out the sums over the $q_s$'s using \fr{id3a} and
\fr{id4b}. When doing this we need to distinguish the cases $k_r\neq
p_r$ and $k_r=p_r$ as they give rise to different kinds of
singularities in the $q_s$ summations. For $k_r=p_r$ the summation
over both $q_{2r-1}$ and $q_{2r}$ gives a leading contribution
\bea
\fl
\frac{1}{L^3}\sum_{q_{2r-1},q_{2r}\in\NS}e^{i(q_{2r-1}+q_{2r})\ell}
c^2_{p_rq_{2r-1}}c^2_{p_rq_{2r}}
f(k_1,\ldots,k_n;p_1,\ldots ,p_n|q_1,\ldots,q_{2n})\bigg|_{p_r=k_r}\nn
\fl\qquad
=\frac{L}{16}\left(1-\frac{2\ell}{L}\right)^2
\frac{\tilde{f}(p_r,-p_r)+\tilde{f}(-p_r,p_r)}{\sin^4(p_r)}
\bigg|_{k_r=p_r}+\ldots\ ,
\label{D2n2n2n_1}
\eea
where we have defined
\be
\fl\qquad
\tilde{f}(p,q)=f(k_1,\ldots, k_n;p_1,\ldots,p_n
|q_1,\ldots,q_{2r-2},p,q,q_{2r+1}\ldots,q_{2n}).
\ee
The extra factor $1/L$ in \fr{D2n2n2n_1} is present as we are
considering a single term in the sum over $k_r$.
On the other hand, for $p_r\neq k_r$ we obtain
\bea
\fl
\frac{1}{L^3}\sum_{0<k_r\in\R}e^{2it[\veps_{k_r}-\veps_{p_r}]}
\sum_{q_{2r-1},q_{2r}\in\NS}e^{i(q_{2r-1}+q_{2r})\ell}
c_{k_rq_{2r-1}}c_{k_rq_{2r}}c_{p_rq_{2r-1}}c_{p_rq_{2r}}\nn
\qquad\qquad\qquad\times\ 
f(k_1,\ldots,k_n;p_1,\ldots,p_n|q_1,\ldots,q_{2n})\nn
\fl
=\frac{1}{L}\sum_{0<k_r\in\R}e^{2it[\veps_{k_r}-\veps_{p_r}]}\
c^2_{p_rk_r}\bigg\{
-\frac{e^{-i\ell(p_r-k_r)}\tilde{f}(-p_r,k_r)+e^{i\ell(p_r-k_r)}\tilde{f}(p_r,-k_r)}
{4\sin k_r\sin p_r}\nn
\fl\qquad\qquad\qquad
+\frac{
\tilde{f}(p_r,-p_r)+\tilde{f}(-p_r,p_r)}{4\sin^4(p_r)}
+p_r\leftrightarrow k_r\bigg\}
+\ldots.
\eea
Similarly to our treatment of the $n=1$ case in subsection
\ref{ssec:D222} we can now carry out the summation over $k_r$ with the result
\bea
-\theta_H(\ell-2t\veps'_{p_r})[2t\veps'_{p_r}-\ell]
\frac{
\tilde{f}(p_r,-p_r)+\tilde{f}(-p_r,p_r)}{4\sin^4(p_r)}\bigg|_{k_r=p_r}+\ldots\ .
\label{D2n2n2n_2}
\eea
Combining the two contributions \fr{D2n2n2n_1} and \fr{D2n2n2n_2} we
find that the leading terms at large $\ell$ and $t$ are
\bea
\left[\frac{L}{4}-\ell-\theta_H(\ell-2t\veps'_{p_r})[2t\veps'_{p_r}-\ell]\right]
\frac{
\tilde{f}(p_r,-p_r)+\tilde{f}(-p_r,p_r)}{4\sin^4(p_r)}\bigg|_{k_r=p_r}+\ldots\ .
\label{D2n2n2n_3}
\eea
Carrying out all $q_s$ and $k_l$ summations in the same way gives
\bea
\fl\qquad
D_{(2n|2n|2n)}&=\frac{\xi}{n!2^nL^n}\sumprime_{0<p_1,\ldots,p_n\in\R}
\prod_{r=1}^n\left[L-4\ell-4\theta_H(\ell-2t\veps'_{p_r})
[2t\veps'_{p_r}-\ell]\right]\frac{1}{16\sin^4(p_r)} \nn
\fl&\qquad\times\sum_{\sigma_1,\ldots,\sigma_1=\pm} 
f(p_1,\ldots p_n;p_1,\ldots p_n|\sigma_1p_1,-\sigma_1p_1,\ldots,
\sigma_np_n,-\sigma_np_n)+\ldots\nn
\fl&=\frac{\xi}{n!}\sumprime_{0<p_1,\ldots,p_n\in\R}
\prod_{r=1}^nK^2(p_r)\left[1-\frac{4\ell}{L}-\frac{4}{L}
\theta_H(\ell-2t\veps'_{p_r})[2t\veps'_{p_r}-\ell]\right]+\ldots.
\eea
Like for the 1-point function the sum over $n$ can be taken by
inverting the steps used to express \fr{eq:norm} in terms of
\fr{ups2n}, which gives 
\bea
\fl\quad
\sum_{n=0}^\infty D_{(2n|2n|2n)}&=\xi \exp\bigg(\sum_{0<p\in\R}\ln\left[
1+K^2(p)\left(1-\frac{4}{L}\left(
\ell+\theta_H(\ell-2t\veps'_{p})[2t\veps'_{p}-\ell]\right)
\right)\right]\bigg)+\ldots\nn
&\simeq\xi\ \lrsub{\NS}{\braket{B|B}}{\NS}\
\exp\bigg[-\frac{4}{L}\sum_{0<p\in\R}
K^2(p)\left(
\ell+\theta_H(\ell-2t\veps'_{p})[2t\veps'_{p}-\ell]\right)
\bigg]+\ldots\nn
\eea
Here we have retained only the ${\cal O}(K^2)$ term in the exponent as
the higher orders are beyond the accuracy of our calculation.
This then gives the desired result for the two-point function
\fr{2point} in the ordered phase
\bea
\fl\quad
\lim_{L\to\infty}\rho^{xx}(t,\ell)
&\simeq(1-h^2)^{\frac{1}{4}}\ \exp\bigg[-2\int_0^\pi\frac{dp}{\pi}
K^2(p)\left(
\ell+\theta_H(\ell-2t\veps'_{p})[2t\veps'_{p}-\ell]\right)
\bigg]+\ldots
\label{2point_res}
\eea

Eqn \fr{2point_res}, including the prefactor, is expected to be
accurate at late times and large distances as long as we do not quench
too close to the critical point.

%%%%%%%%%%%%%%%%%%%%%%%%%%%%%%%%%%%%%%%%%%%%%%%%%%%%%%%%
\subsection{Two-Point Function for Quenches in the Disordered Phase
$h_0,h>1$}\label{ss:PtoP}
%%%%%%%%%%%%%%%%%%%%%%%%%%%%%%%%%%%%%%%%%%%%%%%%%%%%%%%%
As a consequence of the $\mathbb{Z}_2$ symmetry the 1-point function
of $\sigma^x_\ell$ is identically zero for quenches within the
disordered phase, i.e. $h_0,h>1$.
We therefore turn to the two-point function. As shown in
\ref{app:diag} the ground state in the paramagnetic phase is the NS
vacuum $|0;h_0\rangle_\NS$. Hence the 2-point function after the
quench is equal to
\bea
%\lim_{L\to\infty}
\rho^{xx}(t,\ell)=
%\lim_{L\to\infty}
\frac{{}_{\rm NS}\langle
B(t)|\sigma^x_{m+\ell}\sigma^x_m|B(t)\rangle_{\rm NS}}
{{}_{\rm NS}\langle B|B\rangle_{\rm NS}}.
\label{2pointDO}
\eea
In the basis underlying the expression \fr{eq:formfactors} for the
form factors, the boundary states are again
given by \fr{eq:boundary}, but now the function $K(k)$ is given by 
\be
\label{eq:k_disordered}
\fl\qquad
K(k)=-\frac{\sin(k)\ (h-h_0)}{\veps_{h_0}(k)
\veps_{h}(k)\big(2J\big)^{-2}+1+hh_0-(h+h_0)\cos(k)}\, . 
\ee
The Lehmann representation of the numerator is again given
\fr{2point2} if we replace R $\rightarrow$ NS. The difference to the
ordered phase is that now only form factors with \emph{odd} numbers
$r$ of particles in the intermediate states are non-vanishing.

%%%%%%%%%%%%%%%%%%%%%%%%%%%%%%%%%%%%%%%%%%%%%
\subsubsection{Order ${\cal O}\big(K^{0}\big)$ Contributions.}
\label{ssec:k0dis}
%%%%%%%%%%%%%%%%%%%%%%%%%%%%%%%%%%%%%%%%%%%%%
The ${\cal O}\big(K^{0}\big)$ contribution obtained upon expanding the
boundary states in powers of $K$ is
\be
{}_{\rm NS}\langle 0|\sigma^x_{m+\ell}(t)\sigma^x_m(t)|0\rangle_{\rm NS}
={}_{\rm NS}\langle 0|\sigma^x_{m+\ell}\sigma^x_m|0\rangle_{\rm NS}.
\ee
This is equal to the static zero temperature correlator in equilibrium.
Inserting a resolution of the identity we conclude that the large-$\ell$
behaviour in the $L\to\infty$ limit is determined by one-particle
intermediate states 
\bea
D_{(0|1|0)}(t,\ell)&=
\xi\sqrt{4J^2h}\frac{1}{L}\sum_{q\in {\rm R}}\frac{e^{iq|\ell|}}{\veps_q}
=2 J {\cal A}\int_{-\pi}^\pi\frac{dq}{2\pi}\frac{e^{iq\ell}}{\veps_q}
+{\cal O}\big(L^{-1}\big)\nn
%&=\frac{\xi}{\sqrt{4\pi}h^{j+1}}\sum_{k=0}^\infty
%\frac{\Gamma\big(j+2k+\frac{1}{2}\big)(2k-1)!!}{(j+2k)!(2k)!!\ h^{2k}}
%\propto e^{-j\ln(h)},
\eea
where we have defined
\be
{\cal A}=\xi\sqrt{h}.
\ee
The integral can be carried out approximately by bending the contour
around the branch cut of the energy in the upper half plane, which gives
\bea
D_{(0|1|0)}&\approx\xi\sqrt{\frac{h}{\pi(h^2-1)}}\
\frac{h^{-|\ell|}}{\sqrt{\ell}}
\equiv {\cal D}(\ell).
\label{d010}
\eea
We conclude that the leading contribution in $K(q)$ is exponentially
small, which means that we need to evaluate all subleading
terms with the same exponential accuracy as well.
The contributions due to 3,5,\dots particle intermediate states are
small for large $\ell$ when compared to \fr{d010}
\be
D_{(0|2s+1|0)}\propto h^{-(2s+1)|\ell|},
\ee
and can hence be ignored for our purposes.
%%%%%%%%%%%%%%%%%%%%%%%%%%%%%%%%%%
\subsubsection{${\cal O}(K)$ Contributions.}
%%%%%%%%%%%%%%%%%%%%%%%%%%%%%%%%%%
The leading contributions to order $K$ are obtained by taking
single-particle intermediate states into account only. This gives 
\be
D_{(2|1|0)}+D_{(0|1|2)}
=\frac{4 J{\cal A}}{L}\sum_{k}\frac{K(k)}{\veps_k}
\sin(2t\veps_k-k\ell)+\ldots .
\label{D11}
\ee
The corrections to the rhs of \fr{D11} can be evaluated using contour
techniques but are negligible at large distances. As a function of $t$
for fixed $n$ $D_{(2|1|0)}+D_{(0|1|2)}$ displays oscillatory behaviour
on top of a slow $t^{-3/2}$ power-law decay in time. This can be seen
by turning the sum in \fr{D11} into an integral and evaluating the
latter by means of a saddle-point approximation, which gives
\bea
D_{(2|1|0)}+D_{(0|1|2)}\simeq\frac{2 J{\cal A}}{\sqrt{\pi}}
\sum_{a=\pm}\frac{K(k_a)}{\veps_{k_a}}{\rm
  Im}\Bigg[\frac{e^{2it\veps_{k_a}-ik_a\ell}}{\sqrt{-i\veps''_{k_a}t}}\Bigg]\ ,
\eea
where 
\be
k_\pm={\rm
  arccos}\Big[\frac{x^2\pm\sqrt{x^4-x^2(1+h^2)+h^2}}{h}\Big]\ ,\quad
x=\frac{\ell}{4Jt}.
\ee
In the limit $\ell/t\to 0$ this can be simplified further
\bea
\fl
D_{(2|1|0)}+D_{(0|1|2)}&\simeq
-\frac{\ell}{\big(2Jt\big)^{3/2}}\left[\frac{h+1}{h-1}\right]^{1/4}
\frac{h-h_0}{4h(h_0-1)\sqrt{\pi}}
\sin\left(4Jt(h-1)+\frac{\pi}{4}-\frac{h-1}{4hJ}\frac{\ell^2}{t}\right)
\nn
&-\frac{\ell}{\big(2Jt\big)^{3/2}}\left[\frac{h-1}{h+1}\right]^{1/4}
\frac{h-h_0}{4h(h_0+1)\sqrt{\pi}}
\sin\left(4Jt(h+1)-\frac{\pi}{4}+\frac{h+1}{4hJ}\frac{\ell^2}{t}\right).
\eea
%%%%%%%%%%%%%%%%%%%%%%%%%%%%%%%%%%
\subsubsection{${\cal O}(K^2)$ contributions.}
\label{ssec:k2dis}
%%%%%%%%%%%%%%%%%%%%%%%%%%%%%%%%%%
The next contributions we need to consider are second order in $K(k)$
and arise from
\be
\sum_{0<k,p\in {\rm NS}}
K(k)K(p){}_{\rm NS}\langle
-k,k|\sigma^x_{m+\ell}(t)\sigma^x_m(t)|p,-p\rangle_{\rm NS}. 
\ee
Inserting a resolution of the identity between the spin operators we
see that for large $\ell$ and $t$ the dominant contributions are
generated by 1-particle and 3-particle intermediate states. The former
is given by 
\bea
D_{(2|1|2)}&=&
\sum_{k,p>0}\sum_{q}
K(k)K(p)\ {}_{\rm NS}\langle
-k,k|\sigma^x_{m+\ell}(t)|q\rangle_{\rm R}\nn
&&\qquad\times\ {}_{\rm R}\langle q|\sigma^x_m(t)|p,-p\rangle_{\rm NS}. 
\eea
Like in our analysis of the 2-point function for quenches within the
ordered phase we again have to consider the two cases $k\neq p$ and
$k=p$ separately. Denoting the corresponding contributions by
$D_{(2|1|2)}^{(1)}$ and $D_{(2|1|2)}^{(2)}$ respectively, we find in the limit
of large $\ell$, $t$ and $L\to\infty$
\bea
\fl\quad
\label{D212_2}
D_{(2|1|2)}^{(2)}&=%\xi\sqrt{4J^2h}
\frac{2 J\cal A}{L}\sum_{k\in{\rm NS}}K^2(k)
\frac{e^{ik|\ell|}}{\veps_k}+\ldots\ ,\\
%\ee
%and
%\bea
\fl\quad
D_{(2|1|2)}^{(1)}&=%\xi\sqrt{4J^2h}
\frac{8 J{\cal A}}{L^2}\sum_{0<k\neq p\in{\rm NS}}
\frac{K(k)K(p)}{\veps^2_k\veps^2_p}\eps_{k,p}^2
\frac{\sin(k)\sin(p)}{\cos(k)-\cos(p)}
\left[\frac{\veps_k\sin(k\ell)}{\sin(k)}
-\frac{\veps_p\sin(p\ell)}{\sin(p)}
\right]
e^{2it[\veps_k-\veps_p]}\nn
\fl\quad&\quad+\ldots
\label{D21_1}
\eea
In the thermodynamic limit \fr{D21_1} can be written as a double
integral, but we have not succeeded in simplifying %the latter 
it in a useful way.
The contribution $D_{(2|1|2)}^{(2)}$ is time independent and
exponentially small in $\ell$. In contrast to this,
$D_{(2|1|2)}^{(1)}$ displays power-law decay in $t$ for
fixed $\ell$
\be
D_{(2|1|2)}^{(1)}\propto t^{-3}.
\ee

We now turn to the ${\cal O}\big(K^2\big)$ contribution involving
3-particle intermediate states
\bea
D_{(2|3|2)}&=&
\frac{1}{6}
\sum_{0<k,p\in\NS}\ \sum_{q_1,q_2,q_3\in\R}
K(k)K(p)\
{}_{\rm NS}\langle
-k,k|\sigma^x_{m+\ell}(t)|q_1,q_2,q_3\rangle_{\rm R}\nn
&&\qquad\times\ {}_{\rm R}\langle q_3,q_2,q_1|\sigma^x_m(t)|p,-p\rangle_{\rm NS}. 
\eea
We denote the contributions from $k\neq p$ and $k=p$ by
$D_{(2|3|2)}^{(1)}$ and $D_{(2|3|2)}^{(2)}$ respectively. After some lengthy
calculations we find
\bea
\fl\qquad
D_{(2|3|2)}^{(2)}&=D_{(0|1|0)}\ \sum_{0<k\in{\rm
    NS}}K^2(k)\left[1-\frac{4\ell}{L}\right]\nn
\fl
&\quad+\frac{4 J\cal A}{L}\sum_{0<k\in\NS}\frac{K^2(k)}{\veps_k}\cos(\ell k)
-4(h^2-1){\cal D}(\ell)\frac{1}{L}\sum_{0<k\in\NS}\frac{4 J^2K^2(k)}{\veps^2_k}+
\ldots
\label{D232}
\eea
This is time independent and contains a piece that diverges with the
volume as expected.

The contribution $D_{(2|3|2)}^{(1)}$ is given by
\bea
\fl\qquad
D_{(2|3|2)}^{(1)}&=\frac{2 J\cal A}{L^5}\sum_{0<k\neq p\in\NS}
K(k)K(p)\frac{\sin(k)\sin(p)}
{\veps^2_k\veps^2_p}\cos\big(2t[\veps_k-\veps_p]\big)\nn
\fl
&\times\ \frac{1}{6}
\sum_{q_1,q_2,q_3\in\R}\left[\prod_{j<l}
\frac{\sin^2\big(\frac{q_j-q_l}{2}\big)}{\eps_{q_j,q_l}^2}
%g^2(q_j,q_l)
\right]
\left[\prod_{m=1}^3\frac{\eps^2_{k,q_m}\eps^2_{p,q_m}e^{iq_mn}}{\veps_{q_m}}
4c_{q_mk}c_{q_mp}
%\frac{4}{\big(\cos(q_m)-\cos(k)\big)\big(\cos(q_m)-\cos(p)\big)}.
\right].
\eea
Carrying out the sums over $q_j$ using the same techniques as for the
other contributions we eventually arrive at
\bea
\fl\quad
D_{(2|3|2)}^{(1)}&=%\xi\sqrt{4J^2h}
\frac{8 J {\cal A}}{L^2}\sum_{0<k\neq p\in{\rm NS}}
\frac{K(k)K(p)}{\veps^2_k\veps^2_p}\eps_{k,p}^2
\frac{\sin(k)\sin(p)}{\cos(k)-\cos(p)}
\left[\frac{\veps_k\sin(k\ell)}{\sin(k)}
-\frac{\veps_p\sin(p\ell)}{\sin(p)}
\right]
e^{2it[\veps_k-\veps_p]}\nn
\fl\quad&\quad+\ldots
\label{D232_1}
\eea
The large time and distance behaviour is thus the same as for
$D_{(2|1|2)}$. 

In summary, the combined ${\cal O}(K^2)$ contributions can be divided
into two categories:
\begin{enumerate}
\item{Time-independent contributions}
\bea
\fl\quad
4 J{\cal A}\int_{-\pi}^\pi\frac{dk}{2\pi}\frac{K^2(k)}{\veps_k}
e^{ik\ell}
+D_{(0|1|0)}\left[\Upsilon_2-4\ell
  \int_0^\pi\frac{dk}{2\pi}K^2(k)\right]\nn
\fl\qquad 
-4(h^2-1){\cal
  D}(\ell)\ \int_0^\pi\frac{dk}{2\pi}\frac{4J^2K^2(k)}{\veps^2(k)}\ . 
\eea
\item{Time-dependent oscillatory contributions}
\be
\fl\quad
4 J {\cal A}\int_0^\pi\frac{dk\ dp}{\pi^2}
\frac{K(k)K(p)}{\veps^2(k)\veps^2(p)}\eps_{k,p}^2
\frac{\sin(k)\sin(p)}{\cos(k)-\cos(p)}
\left[\frac{\veps_k\sin(k\ell)}{\sin(k)}
-\frac{\veps_p\sin(p\ell)}{\sin(p)}
\right]
e^{2it[\veps_k-\veps_p]}.
\ee
\end{enumerate}
%%%%%%%%%%%%%%%%%%%%%%%%%%%%%%%%%%
\subsubsection{${\cal O}(K^3)$ contributions.}
%%%%%%%%%%%%%%%%%%%%%%%%%%%%%%%%%%
In order to infer the structure of higher-order oscillatory
contributions we determine the following ${\cal O}(K^3)$ term
\bea
\fl
D_{(4|3|2)}+D_{(2|3|4)}
%&=&\sum_{{0<k_{1,2},p\in{\rm NS}}\atop
%{q_{1,2,3}\in{\rm R}}}
%-\frac{i}{12}K(k_1)K(k_2)K(p)\
%{}_{\rm NS}\langle
%-k_1,k_1,-k_2,k_2|\sigma^x_{m+j}(t)|q_1q_2q_3\rangle_{\rm R}\ 
%{}_{\rm R}\langle q_3,q_2,q_1|\sigma^x_m(t)|p,-p\rangle_{\rm NS}
%+{\rm h.c.}\nn
&=\frac{J{\cal A}}{3L^6}{\rm Re}\ \sum_{{0<k_{1,2},p\in{\rm NS}}\atop
{q_{1,2,3}\in{\rm R}}}K(k_1)K(k_2)K(p)\
\frac{\sin k_1\sin k_2\sin p}{\veps^2_{k_1}\veps^2_{k_2}\veps^2_{p}}
g^2_{k_1,k_2}g^2_{k_1,-k_2}
\nn
\fl
&\times 
\frac{g^2_{q_1,q_2}g^2_{q_1,q_3}g^2_{q_2,q_3}}{\veps_{q_1}\veps_{q_2}\veps_{q_3}}
\ e^{2it[\veps_{k_1}+\veps_{k_2}-\veps_p]}
\prod_{j=1}^32\eps^2_{p,q_j}c_{pq_j}e^{i\ell q_j}
\prod_{l=1}^22\eps^2_{k_l,q_j}c_{k_lq_j}\ ,
\label{D432+D234}
\eea
where we have defined
\bea
g_{p,k}&=&\frac{2\sin\Big(\frac{p-k}{2}\Big)}{\veps_p+\veps_k}\ .
\label{eg}
\eea

%%%%%%%%%%%%%%%%%%%%%%%%%%%%%%%%%%%%%%%%%%%%%%
\noindent\emph{(i) Contributions for $p\neq k_{1,2}$.}
%%%%%%%%%%%%%%%%%%%%%%%%%%%%%%%%%%%%%%%%%%%%%%
We first consider only the contributions with $p\neq k_{1,2}$ and
denote them by $D^{(1)}_{(4|3|2)}+D^{(1)}_{(2|3|4)}$. In order to
carry out the sums over $q_j$ it is useful to rewrite the product of
pole factors $\prod_{l=1}^3\prod_{j=1}^32c_{k_lq_j}$ using the identity
\bea
c_{k_1q}c_{k_2q}c_{p q}&=&
c_{k_2k_1}\Big\{c_{pk_1}\left[c_{k_1q}-c_{pq}\right]
-c_{pk_2}\left[c_{k_2q}-c_{pq}\right]\Big\}.
\eea
The fully decomposed expression reads
\bea
\fl\quad
\prod_{j=1}^38c_{k_1q_j}c_{k_2q_j}c_{pq_j}=
\left[8c_{k_2k_1}\right]^3\Biggl\{
-c^3_{pk_1}\left[
\frac{1}{2}\widetilde{\cal C}_{k_1,k_1,p}
-{\cal C}_{k_1,k_1,k_1}
-\frac{1}{2}\widetilde{\cal C}_{p,p,k_1}
+{\cal C}_{p,p,p}\right]\nn
\fl
\quad+c^2_{pk_1}c_{pk_2}\Bigg[
\widetilde{\cal C}_{k_1,k_2,p}
-\widetilde{\cal C}_{k_1,p,p}
+3{\cal C}_{p,p,p}
-\frac{1}{2}\widetilde{\cal C}_{p,p,k_2}
+\frac{1}{2}\widetilde{\cal C}_{k_1,p,k_1}
-\frac{1}{2}\widetilde{\cal C}_{k_1,k_1,k_2}
\Bigg]
-\{k_1\leftrightarrow k_2\}\Bigg\},
\label{decomposition}
\eea
where we have defined
\bea
{\cal C}_{k_1,k_2,k_3}=\prod_{j=1}^3c_{k_j,q_j}\ ,\quad
\widetilde{\cal C}_{k_1,k_2,k_3}=\sum_{P\in S_3}{\cal C}_{k_{P_1},k_{P_2},k_{P_3}}.
\eea
We then can carry out the sums over $q_j$ using Lemma 4
\fr{id4a} and retaining only the pole contributions.
\begin{itemize}
\item[A.]{} 
The combined contributions to \fr{D432+D234} arising from
$c^3_{pk_1}\widetilde{\cal C}_{k_1,k_1,p}$, $c^3_{pk_2}\widetilde{\cal C}_{k_2,k_2,p}$,
$c^2_{pk_1}c_{pk_2}\widetilde{\cal C}_{k_1,p,k_1}$ and $c^2_{pk_2}c_{pk_1}\widetilde{\cal  C}_{k_2,p,k_2}$ 
in the decomposition \fr{decomposition} are of the form
\bea
\fl
%$D^{(1,1)}_{(2,1),3}
D_{(4|3|2)}^{(1,1)}+D_{(2|3|4)}^{(1,1)}
&=\frac{16 J{\cal A}
%\xi\sqrt{4J^2h}
}{L^3}\sum_{0<k_{1,2}\neq
  p\in{\rm NS}}
K(k_1)K(k_2)K(p)\ e^{2it[\veps(k_1)+\veps(k_2)-\veps_p]}\nn\fl
&\quad\times\ \frac{\eps^2_{k_1,p}\eps^2_{k_2,p}c_{k_2k_1}\sin k_1\sin k_2}
{\veps^2_{k_1}\veps^2_{k_2}\veps_p}
%&\hskip 4cm\times\
\left[c_{pk_1}-c_{pk_2}\right]\sin\big(\ell p\big)+{\rm h.c.}+\ldots\ .
\eea
\item[B.]{}
The contributions due to $c^3_{pk_1}\widetilde{\cal C}_{p,p,k_1}$ and 
$c^3_{pk_2}\widetilde{\cal C}_{p,p,k_2}$ in \fr{decomposition} are
\bea
\fl
D_{(4|3|2)}^{(1,2)}+D_{(2|3|4)}^{(1,2)}
&=-\frac{8 J{\cal A}
%\xi\sqrt{4J^2h}
}{L^3}
\sum_{0<k_{1,2}\neq  p\in{\rm NS}}
\frac{K(k_1)K(k_2)K(p)}{\veps^2_{k_1}\veps^2_{k_2}\veps^2_p}\ e^{2it[\veps(k_1)+\veps(k_2)-\veps_p]}
\frac{\eps^2_{k_1,p}\eps^2_{k_2,p}}{\eps^2_{k_1,k_2}}c_{k_2k_1}
\nn
&\quad\times\
\sin p
\left[\sin(\ell k_1)\sin k_2\veps_{k_1}\eps^2_{k_2,p}c_{pk_1}-\{k_1\leftrightarrow
k_2\}\right]+{\rm h.c.}+\ldots\ .
\label{342-12}
\eea
\item[C.]{}
The contributions of the $c^2_{pk_1}c_{pk_2}\widetilde{\cal C}_{p,p,k_1}$ and
$c_{pk_1}c^2_{pk_2}\widetilde{\cal C}_{p,p,k_2}$  terms are
\bea
\fl
D_{(4|3|2)}^{(1,3)}+D_{(2|3|4)}^{(1,3)}
&=\frac{16 J {\cal A}}{L^3}\sum_{0<k_{1,2}\neq  p\in{\rm NS}}
\frac{K(k_1)K(k_2)K(p)}{\veps_{k_1}\veps_{k_2}\veps^2_p}
\ e^{2it[\veps(k_1)+\veps(k_2)-\veps_p]}\frac{\eps^2_{k_1,p}\eps^2_{k_2,p}}
{\eps^2_{k_1,k_2}}\nn 
&\quad\times c_{k_2k_1}\sin p
\left[\sin(\ell k_1)\sin k_2\frac{\eps^2_{k_2,p}}{\veps_{k_2}}
c_{pk_2}-\{k_1\leftrightarrow k_2\}\right]+{\rm h.c.}+\ldots\ .
\label{342-13}
\eea
\item[D.]{}
The contributions of the $c^2_{pk_1}c_{pk_2}\widetilde{\cal C}_{p,p,k_2}$ and
$c_{pk_1}c^2_{pk_2}\widetilde{\cal C}_{p,p,k_1}$  terms are
\bea
\fl
D_{(4|3|2)}^{(1,4)}+D_{(2|3|4)}^{(1,4)}&=
-\frac{8 J {\cal A}}{L^3}\sum_{0<k_{1,2}\neq p\in{\rm NS}}
\frac{K(k_1)K(k_2)K(p)}{\veps_{k_1}\veps_{k_2}\veps^2_p}
\ e^{2it[\veps(k_1)+\veps(k_2)-\veps_p]}
\frac{\eps^2_{k_1,p}\eps^2_{k_2,p}}{\eps^2_{k_1,k_2}}\nn 
&\quad\times\ c_{k_2k_1}\sin p
\left[\sin(\ell k_1)\sin k_2\frac{\eps^2_{k_2,p}}{\veps_{k_2}}
\frac{c^2_{pk_2}}{c_{pk_1}}-\{k_1\leftrightarrow
k_2\}\right]+{\rm h.c.}+\ldots\ .
\label{342-14}
\eea
\item[E.]{}
The contributions of the $c^2_{pk_1}c_{pk_2}\widetilde{\cal C}_{k_1,k_1,k_2}$ and
$c_{pk_1}c^2_{pk_2}\widetilde{\cal C}_{k_2,k_2,k_1}$  terms are
\bea
\fl
D_{(4|3|2)}^{(1,5)}+D_{(2|3|4)}^{(1,5)}&=
\frac{8 J {\cal A}}{L^3}\sum_{0<k_{1,2}\neq
p\in{\rm NS}}
\frac{K(k_1)K(k_2)K(p)}{\veps_{k_1}\veps_{k_2}\veps^2_p}
\ e^{2it[\veps(k_1)+\veps(k_2)-\veps_p]}
\frac{\eps^2_{k_1,p}\eps^2_{k_2,p}}
{\eps^2_{k_1,k_2}}\nn 
\fl
&\quad\times\
\frac{c_{pk_1}c_{pk_2}}{c_{k_1k_2}}
\sin p\left[\sin(\ell k_2)\sin k_1\frac{\eps^2_{k_1,p}}{\veps_{k_1}}
c_{pk_1}-\{k_1\leftrightarrow
k_2\}\right]+{\rm h.c.}+\ldots\ 
\label{342-15}
\eea
\item[F.]{}
Finally, the contributions of the $c^2_{pk_1}c_{pk_2}
\widetilde{\cal C}_{k_1,k_2,p}$
and $c^2_{pk_2}c_{pk_1}\widetilde{\cal  C}_{k_2,k_1,p}$ terms are 
\bea
\fl
D_{(4|3|2)}^{(1,6)}+D_{(2|3|4)}^{(1,6)}&=
-\frac{32 J {\cal A}}{L^3}\sum_{k_{1,2}\neq
  p\in{\rm NS}} 
\frac{K(k_1)K(k_2)K(p)}{\veps_{k_1}\veps_{k_2}\veps_p}
\ \cos\Big(2t[\veps_{k_1}+\veps_{k_2}-\veps_p]\Big)\nn
&\quad\times\ \eps^4_{k_1,p}\eps^4_{k_2,p}c^2_{k_1p}c^2_{k_2p}
g^2_{k_1,k_2}g^2_{k_1,p}g^2_{k_2,p}\sin\big(\ell (k_1+k_2+p)\big),
%-g^2(-k_1,k_2)g^2(-k_1,p)g^2(k_2,p)\sin\big(j(-k_1+k_2+p)\big)\nn
%&&-g^2(k_1,-k_2)g^2(k_1,p)g^2(-k_2,p)\sin\big(j(k_1-k_2+p)\big)
%-g^2(k_1,k_2)g^2(k_1,-p)g^2(k_2,-p)\sin\big(j(k_1+k_2-p)\big)
%\Bigg]
\label{342-16}
\eea
where here the momentum sums are over the entire Brillouin zone. 
\end{itemize}
The contributions from all other terms are subleading.
We observe that the leading terms at large $\ell$ and $t$ in
\fr{342-12},\fr{342-13},\fr{342-14} and \fr{342-15} combine to
\bea
\fl
\sum_{a=2}^5D^{(1,a)}_{(4|3|2)}+D^{(1,a)}_{(2|3|4)}
&\sim
-\frac{64 J{\cal A}}{L^3}\sum_{0<k_{1,2}\neq p\in{\rm NS}}
\frac{K(k_1)K(k_2)K(p)}{\veps_{k_1}\veps^2_{k_2}\veps^2_p}
\ \cos\Big(2t[\veps_{k_1}+\veps_{k_2}-\veps_p]\Big)\nn
&\quad\times
\frac{\eps^2_{k_1,p}\eps^4_{k_2,p}\sin k_2\sin p}
{\eps^2_{k_1,k_2}}
c_{pk_2}\left[c_{pk_2}-c_{pk_1}\right]\ \sin(\ell k_1)+\ldots\ .
\label{342-12-15}
\eea
Both \fr{342-12-15} and \fr{342-16} can be simplified further, because
for large $t$ and $\ell$ the dominant contributions arise from the
``double pole'' factors and can be extracted using Lemmas 2a and 2b of
\ref{a:useful}. This leaves us with
\bea
\fl
D_{(4|3|2)}^{(1,6)}+D_{(2|3|4)}^{(1,6)}&=
-\frac{16 J {\cal A}}{L^2}\sum_{0< p\in{\rm NS}}
\sum_{k \in{\rm NS}}
\frac{K(k)}{\varepsilon_k}\sin\big(2t\varepsilon_k-\ell k\big)\ K^2(p)\nn
&\qquad\times\
\left[\frac{L}{6}-\left[\ell +(2t\varepsilon'_p-\ell )
\theta_H(2t\varepsilon'_p-\ell )\right]\right]+\ldots,
\label{342_16_b}
\eea
\bea
\fl
\sum_{a=2}^5D^{(1,a)}_{(4|3|2)}+D^{(1,a)}_{(2|3|4)}&=
\frac{16 J{\cal A}}{L^2}\sum_{0< p\in{\rm NS}}
\sum_{k \in{\rm NS}}
\frac{K(k)}{\varepsilon_k}\sin\big(2t\varepsilon_k-\ell k\big)\ K^2(p)
\left[\frac{L}{6}-2t\varepsilon'_p\right]+\ldots\ .\nn
\label{342_12-15_b}
\eea
Combining \fr{342_16_b} and \fr{342_12-15_b} we arrive at our final
result for the leading asymptotics of the $p\neq k_{1,2}$
contributions to \fr{D432+D234}
\bea
\fl
D_{(4|3|2)}^{(1)}+D_{(2|3|4)}^{(1)}&=
\left[\frac{4 J {\cal A}}{L}\sum_{k \in{\rm NS}}
\frac{K(k)}{\varepsilon_k}\sin\big(2t\varepsilon_k-\ell k\big)\right]
\!
\left[-\frac{4}{L}\sum_{0< p\in{\rm NS}} K^2(p)
\left[2t\varepsilon'_p-\ell\right]
\theta_H\big(\ell-2t\varepsilon'_p\big)\right]\nn
&\qquad+\ldots\ \nn
&=\left[D_{(2|1|0)}+D_{(0|1|2)}\right]
\left[-\frac{4}{L}\sum_{0< p\in{\rm NS}}\ K^2(p)
\left[2t\varepsilon'_p-\ell\right]
\theta_H\big(\ell-2t\varepsilon'_p\big)\right]+\ldots\ .
\label{d432+d234_1}
\eea

\vskip .2cm
%%%%%%%%%%%%%%%%%%%%%%%%%%%%%%%%%%%%%%%%%%%%%%
\noindent\emph{(ii) Contributions with $p=k_1$ or $p=k_2$.}
%%%%%%%%%%%%%%%%%%%%%%%%%%%%%%%%%%%%%%%%%%%%%%
Using the symmetry under $k_1\leftrightarrow k_2$ we can express these
contributions in the form
\bea
\fl
D_{(2|3|4)}^{(2)}+D_{(4|3|2)}^{(2)}
&=-\frac{i}{6}\sum_{0<k_{1,2}\in{\rm NS}}\sum_{q_{1,2,3}\in
  {\rm R}}K^2(k_1)K(k_2)\ e^{2it\veps(k_2)}\ \nn
&\times\ 
{}_{\rm NS}\langle -k_1,k_1,-k_2,k_2|\sigma^x_{m+\ell }|q_1,q_2,q_3\rangle_{\rm R}\
{}_{\rm
R}\langle q_3,q_2,q_1|\sigma^x_m|k_1,-k_1\rangle_{\rm NS}+{\rm h.c.}
\eea
In order to carry out the $q_j$ sums we rewrite the pole factors as
\bea
\fl
\prod_{j=1}^38 c^2_{k_1q_j}c_{k_2q_j}
&=\left(8c_{k_2k_1}\right)^3\Big\{{\cal C}^2(k_1,k_1,k_1)
-c_{k_2k_1}b(k_1,k_2|q_1)c^2_{k_1q_2}c^2_{k_1q_3}
-c_{k_2k_1}c^2_{k_1q_1}b(k_1,k_2|q_2)c^2_{k_1q_3}\nn
\fl
&-c_{k_2k_1}c^2_{k_1q_1}c^2_{k_1,q_2}b(k_1,k_2,|q_3)
+c^2_{k_1k_2}c^2_{k_1q_1}b(k_1,k_2|q_2)b(k_1,k_2|q_3)\nn\fl
&+c^2_{k_1k_2}c^2_{k_1q_2}b(k_1,k_2|q_1)b(k_1,k_2|q_3)
+c^2_{k_1k_2}c^2_{k_1q_3}b(k_1,k_2|q_1)b(k_1,k_2|q_2)\nn\fl
&-c^3_{k_2k_1}b(k_1,k_2|q_1)b(k_1,k_2|q_2)b(k_1,k_2|q_3)
\Big\},
\label{decomp2}
\eea
where we have defined
\be
b(k_1,k_2|q)=c_{k_1q}-c_{k_2q}.
\ee
The various terms in \fr{decomp2} contribute in qualitatively
different ways, depending on their structure when viewed as functions
of (the complex variables) $q_1$, $q_2$ and $q_3$.
\begin{enumerate}
%%%%%%%%%%%%%%%%%%%%%%%%%%%%%
\item[1.]{{\it Terms with only simple poles}} give rise
%%%%%%%%%%%%%%%%%%%%%%%%%%%%%
to contributions of order ${\cal O}(L^{-1})$ and can be ignored.

%%%%%%%%%%%%%%%%%%%%%%%%%%%%%
\item[2.]{{\it Terms with only double poles}}
%%%%%%%%%%%%%%%%%%%%%%%%%%%%% 
arise from the first contribution on the r.h.s. of \fr{decomp2}.
Using \fr{id4a} to carry out the sums over $q_j$ we obtain after some
calculations 
\bea\fl\quad
D_{(4|3|2)}^{(2,1)}+D_{(2|3|4)}^{(2,1)}
=&\frac{1}{L}\left[1-\frac{4\ell }{L}\right]{\cal
  D}(\ell )\sum_{0<k_{1,2}\in{\rm
    NS}}K^2(k_1)K(k_2)\cos(2t\veps_{k_2})c_{k_2,k_1}\nn
\fl&
+\frac{4 J {\cal
    A}}{L}\sum_{k\in\NS}\frac{K^3(k)}{\veps_k}\sin(2t\veps_k-\ell k)
+\ldots,
\eea
which can be simplified further if required.

%%%%%%%%%%%%%%%%%%%%%%%%%%%%%
\item[3.]{\it Terms with only one double pole}
%%%%%%%%%%%%%%%%%%%%%%%%%%%%%
arise from the fifth, sixth and seventh contributions on the r.h.s. of
\fr{decomp2}. Using \fr{id4a} and \fr{id3} to carry out the sums over
$q_j$ we obtain after some calculations 
\bea
\fl
\quad
D_{(4|3|2)}^{(2,2)}+D_{(2|3|4)}^{(2,2)}
&=\frac{16 J {\cal A}}{L^2}
\sum_{0<k_1\neq k_2\in{\rm NS}}K^2(k_1)K(k_2)
\cos\big(2t\veps_{k_2}\big)\sin(k_2)\cos(k_1\ell)\nn
\fl
&\qquad\qquad\qquad\times\ 
c_{k_2k_1}\frac{\eps^2_{k_1,k_2}}{\veps_{k_1}
\veps^2_{k_2}}+\ldots\ .
\eea
Here we can carry out the $k_1$ sum using Lemma 3 \ref{id3}, which gives
\bea
D_{(4|3|2)}^{(2,2)}+D_{(2|3|4)}^{(2,2)}
=\frac{4 J {\cal A}}{L}
\sum_{k\in{\rm NS}}\frac{K^3(k)}{\veps_k}
\sin\big(2t\veps_k-\ell k\big)+\ldots\ .
\eea

%%%%%%%%%%%%%%%%%%%%%%%%%%%%%
\item[4.]{\it Terms with two double poles}
%%%%%%%%%%%%%%%%%%%%%%%%%%%%%
arise from the second, third and fourth contributions on the
r.h.s. of \fr{decomp2}. Using \fr{id4a} 
and \fr{id3} to carry out the sums over
$q_j$ we obtain after some calculations 
\bea\fl
D_{(4|3|2)}^{(2,3)}+D_{(2|3|4)}^{(2,3)}
=\left[\frac{4 J {\cal A}}{L}
\sum_{k_2\in{\rm NS}}\frac{K(k_2)}{\veps_{k_2}}\sin(2t\veps(k_2)-\ell k_2)\right]
\sum_{0<k_1\in{\rm NS}}K^2(k_1)\left[1-\frac{4\ell }{L}\right]+\ldots\nn
\label{d432+d234_2}
\eea
This is in fact the leading contribution.
\end{enumerate}

Combining \fr{d432+d234_1} and \fr{d432+d234_2} we arrive at the
following result for our ${\cal O}(K^3)$ contributions
\bea
\fl
D_{(4|3|2)}+D_{(2|3|4)}&=\left[D_{(2|1|0)}+D_{(0|1|2)}\right]
\ \frac{4}{L}\sum_{0<p\in\NS}K^2(p)\
\left[\frac{L}{4}-\ell -\big[2t\varepsilon'(p)-\ell\big]
\theta_H\big(\ell -2t\varepsilon'(p)\big) \right]\nn
&\qquad+\frac{8 J{\cal A}}{L}
\sum_{k\in{\rm NS}}\frac{K^3(k)}{\veps_k}
\sin\big(2t\veps_k-\ell k\big)+\ldots\ .
\eea
We see that as expected there is a contribution that diverges in the
thermodynamic limit. However, in addition there are terms that become
very large for $\ell\gg 1$

%%%%%%%%%%%%%%%%%%%%%%%%%%%%%%%%%%%%%%%%%%%%%%%%%%%%%%%%%%%%%%%%%%%%
\subsubsection{Resummation of Leading $D_{(2n+2|2n+1|2n)}+D_{(2n|2n+1|2n+2)}$ 
Contributions.}
%%%%%%%%%%%%%%%%%%%%%%%%%%%%%%%%%%%%%%%%%%%%%%%%%%%%%%%%%%%%%%%%%%%%
\hfill\break
The above calculation of the ${\cal O}(K^3)$ contribution shows that
the dominant contributions arise from the regions
\bea
k_{Q_1}\approx p\ ,\quad
\pmatrix{q_{S_1}\cr q_{S_2}\cr q_{S_3}}
\approx\pmatrix{\sigma_1 k_{Q_1}\cr -\sigma_1 p\cr
  \sigma_2k_{Q_2}}\ ,\quad
\sigma_j=\pm,
\eea
where $(Q_1,Q_2)$ and $(S_1,S_2,S_3)$ are permutations of $(1,2)$ and
$(1,2,3)$ respectively. Motivated by this observation we therefore
consider the analogous regions for the ${\cal O}(K^{2n+1})$
contributions $D_{(2n+2|2n+1|2n)}$ and $D_{(2n|2n+1|2n+2)}$. They are
\bea
\pmatrix{k_{Q_1}\cr\vdots\cr k_{Q_n}}\approx 
\pmatrix{p_{1}\cr\vdots\cr p_{n}}\ ,\quad
\pmatrix{q_{S_1}\cr q_{S_2}\cr \vdots\cr q_{S_{2n-1}}\cr q_{S_{2n}}\cr q_{S_{2n+1}}}
\approx\pmatrix{\sigma_1 k_{Q_1}\cr -\sigma_1 p_1\cr \vdots\cr
\sigma_n k_{Q_n}\cr -\sigma_n p_n \cr \sigma_{n+1}k_{Q_{n+1}}}\ ,
\label{regions_dis}
\eea
where $\sigma_j=\pm$ and $(Q_1,\ldots,Q_{n+1})$,
$(S_1,\ldots,S_{2n+1})$ are permutations of $(1,2\ldots,{n+1})$ and
$(1,2\ldots,{2n+1})$ respectively. Our goal is to determine the
contribution of the regions \fr{regions_dis} to
\bea
\fl
D_{(2n+2|2n+1|2n)}+{\rm h.c.}=&{\cal C}\ {\rm Re}
\sumprime_{k_1,\ldots,k_{n+1}\in\NS\atop
p_1,\ldots,p_n\in\NS}\left[\prod_{r=1}^{n+1}
\frac{\sin(k_r)K(k_r)
e^{2it\veps_{k_r}}}{\veps^2_{k_r}}\right]
\left[\prod_{s=1}^{n}
\frac{\sin(p_s)K(p_s)e^{-2it\veps_{p_s}}}{\veps^2_{p_s}}\right]\nn
\fl
&\times\ 
\prod_{l<l'}^{n+1}\frac{4}{c^2_{k_lk_{l'}}\eps^4_{k_l,k_{l'}}}
\prod_{m<m'}^{n}\frac{4}{c^2_{p_mp_{m'}}\eps^4_{p_m,p_{m'}}}
\sum_{q_1,\ldots,q_{2n}\in\NS}
\left[\prod_{r=1}^{{2n+1}}\frac{e^{i\ell q_r}}{\veps_{q_r}}\right]
\prod_{l<l'}^{2n+1}g^2_{q_lq_{l'}}
%\frac{\sin^2\big(\frac{q_l-q_{l'}}{2}\big)}{\eps^2_{q_lq_{l'}}}
\nn
\fl
&\times\
\prod_{s=1}^{2n+1}2\eps^2_{k_{n+1},q_s}c_{k_{n+1}q_2}
\left[\prod_{r=1}^n4\eps^2_{k_r,q_s}\eps^2_{p_r,q_s}c_{k_rq_s}c_{p_rq_s}
\right],
\label{D2n+22n+12n}
\eea
where the constant ${\cal C}$ is
\be
{\cal C}=\frac{4 J \cal A}{L^{4n+2}(n+1)!n!(2n+1)!}.
\ee
As all regions \fr{regions_dis} contribute equally, we focus on the
case $(Q_1,\ldots,Q_{n+1})=(1,\ldots,n+1)$,
$(S_1,\ldots,S_{2n+1})=(1,\ldots,2n+1)$ and multiply the result by a
combinatorial factor $(n+1)!(2n+1)!$. We first carry out the
summations over $q_1,\ldots, q_{2n}$ and $k_1,\ldots, k_n$ by following
the analogous calculation for the ordered phase, see section
\ref{Exp1}. Finally, we carry out the sum over $q_{2n+1}$ using
\fr{id3}. This results in
\bea
\fl
D_{(2n+2|2n+1|2n)}+{\rm h.c.}=&
-\frac{2{\cal A}}{n!}
\sumprime_{p_1,\ldots,p_n\in\NS}
\prod_{r=1}^nK^2(p_r)\left[1-\frac{4\ell}{L}-\frac{4}{L}
\theta_H(\ell-2t\veps'_{p_r})[2t\veps'_{p_r}-\ell]\right]\nn
\fl
&\times
\frac{4 J}{L}\sum_{0<k_{n+1}\in\NS}\frac{K(k_{n+1})}{\veps_{k_{n+1}}}
\sin(k_{n+1}\ell) \cos(2t\veps_{k_{n+1}})+\ldots\nn
&=\frac{1}{n!}
\sumprime_{p_1,\ldots,p_n\in\NS}
\prod_{r=1}^nK^2(p_r)\left[1-\frac{4\ell}{L}-\frac{4}{L}
\theta_H(\ell-2t\veps'_{p_r})[2t\veps'_{p_r}-\ell]\right]\nn
&\quad\times\left[D_{(2|1|0)}+D_{(0|1|2)}\right]+\ldots\ .
\eea
The sum over $n$ can be again taken by inverting the steps used to
express \fr{eq:norm} in terms of \fr{ups2n}, which gives 
\bea
\fl
\sum_{n=0}^\infty D_{(2n+2|2n+1|2n)}+{\rm h.c.}\nn
\fl\quad
=\left[D_{(2|1|0)}+D_{(0|1|2)}\right]
\exp\bigg(\sum_{0<p\in\NS}\ln\left[
1+K^2(p)\left(1-\frac{4}{L}\left(
\ell+\theta_H(\ell-2t\veps'_{p})[2t\veps'_{p}-\ell]\right)
\right)\right]\bigg)+\ldots\nn
\fl\quad
\approx\left[D_{(2|1|0)}+D_{(0|1|2)}\right]
\lrsub{\NS}{\braket{B|B}}{\NS}\
\exp\bigg[-\frac{4}{L}\sum_{0<p\in\NS}
{K^2(p)}\left(
\ell+\theta_H(\ell-2t\veps'_{p})[2t\veps'_{p}-\ell]\right)
\bigg].\nn
\eea
In the last step we have only retained the ${\cal O}(K^2)$ term in the
exponent in order not to exceed the accuracy of our calculation.
%%%%%%%%%%%%%%%%%%%%%%%%%%%%%%%%%%%%%%%%%%%%%%%%
\subsubsection{Full Answer For the 2-Point 
Function in the Disordered Phase}
%%%%%%%%%%%%%%%%%%%%%%%%%%%%%%%%%%%%%%%%%%%%%%%%
\hfill\break
The above results suggests that the two-point function consists of two
parts 
\be
\lim_{L\to\infty}\rho^{xx}(t,\ell)=F_1(\ell )+F_2(t,\ell ).
\ee
The two kinds of contributions are
\begin{itemize}
\item{} An oscillating time dependent contribution 
arising from the terms $D_{(2n+2|2n+1|2n)}+{\rm h.c.}$
\bea
\fl\quad
F_2(t,\ell)&\simeq{2 J \cal
  A}\int_{-\pi}^\pi\frac{dk}{\pi}\frac{K(k)}{\veps_k}
\sin(2t\veps_k-k\ell)\nn
\fl
&\qquad\times\ \exp\bigg[-2\int_0^\pi\frac{dp}{\pi}
K^2(p)\left(
\ell+\theta_H(\ell-2t\veps'_{p})[2t\veps'_{p}-\ell]\right)
\bigg].
\label{2point_res_para}
\eea
\item{} An exponentially small, time-independent contribution arising
  from the ``diagonal'' terms
\bea
\quad\fl
F_1(\ell)=&\frac{1}{{}_{\NS}\langle B|B\rangle_{\NS}}\
\sum_{n=0}^\infty\frac{1}{n!}
\ \sumprime_{0<k_1,\dots,k_n\in\NS}\
\left[\prod_{j=1}^nK^2(k_j)\right]\nn
\fl
&\qquad\times\ {}_{\rm NS}\langle -k_n,k_n,\ldots,-k_1,k_1|\sigma^x_{m+\ell}
\sigma^x_m|k_1,-k_1\ldots,k_n,-k_n\rangle_{\rm NS}\ .
\label{F1}
\eea
The ${\cal O}(K^0)$ and ${\cal O}(K^2)$ have been evaluated in
sections
\ref{ssec:k0dis} and \ref{ssec:k2dis} respectively and are given by
\bea
\fl\quad
F_1(\ell )&\approx
2 J{\cal A}\int_{-\pi}^\pi\frac{dq}{2\pi}\frac{e^{iq\ell}}{\veps_q}
\left[1+2K^2(q)
-4\ell\int_0^\pi\frac{dk}{2\pi}K^2(k)\right]\nn
\fl
&-4(h^2-1)\int_0^\pi\frac{dk}{2\pi}\frac{4J^2K^2(k)}{\veps^2(k)}\ {\cal
  D}(\ell).
\eea
It is shown in \ref{app:PE} that the ``pair ensemble'' average \fr{F1}
is equal to the average in the generalized Gibbs ensemble, which was
previously calculated in \cite{CEF}. Hence we conclude that
\bea
F_1(\ell)&\simeq{\cal C}_{\rm PP}(\ell)\exp\left(-\frac{\ell}{\xi}\right)\ ,\nn
\xi^{-1}&=\ln\left( {\rm min}[h_0,h_1]\right)
-\ln\left[\frac{1+hh_0+\sqrt{(h^2-1)(h_0^2-1)}}{2hh_0}\right],
\eea
where $h_1=\frac{1+h h_0+\sqrt{(h^2-1)(h_0^2-1)}}{h+h_0}$ and the
large distance behaviour of ${\cal C}_{\rm PP}(\ell)$ is determined in
paper II. 
\end{itemize}

%%%%%%%%%%%%%%%%%%%%%%%%%
\section{Scaling Limit of the Ising Model}
\label{sec:scaling}
%%%%%%%%%%%%%%%%%%%%%%%%%
So far we have focussed on the quench dynamics in the transverse field
Ising lattice model. In the vicinity of the quantum critical point at
$h=1$ a quantum field theory description applies, see
e.g. \cite{ItzyksonDrouffe}. As quantum quenches in integrable fields
theories are of great current interest, we now present explicit 
expressions for the quench dynamics in the field theory limit.

The scaling limit of the transverse field Ising chain is ($a_0$ is the
lattice spacing) \cite{ItzyksonDrouffe}
\be
J\to\infty\ ,\quad h\to 1\ ,\quad a_0\to 0,
\label{scalingI}
\ee
while keeping fixed both the gap $\Delta$ and the velocity $v$
\be
2J|1-h|=\Delta\ ,\quad 2Ja_0=v.
\label{scalingII}
\ee
In this limit the dispersion and Bogoliubov angle become
\bea
\veps(q)&=&\sqrt{\Delta^2+v^2q^2}\ ,\\
\theta_h(q)&\rightarrow&{\rm arctan}\Big(\frac{vq}{\Delta}\Big).
\eea
In our quench problem both the initial and the final magnetic field
are scaled to the critical point, i.e. we need to take
\be
h_0\to 1\ ,\quad J(1-h_0)=\Delta_0=\ {\rm fixed}.
\ee
In this limit $K$-matrix turns into
\be
K(q)=\tan\left[\frac{{\rm arctan}\Big(\frac{vq}{\Delta}\Big)-
{\rm arctan}\Big(\frac{vq}{\Delta_0}\Big)}{2}\right].
\ee
Here the physical momentum is defined as
\be
q=\frac{k}{a_0}\ ,\quad -\infty<q<\infty.
\ee
The Hamiltonian describing the scaling limit is expressed in terms of
Majorana fermions as
\begin{equation}
H =\int_{-\infty}^\infty \frac{dx}{2\pi}\left[ \frac{iv}{2}({\bar\psi}
\partial_x\bar\psi - \psi\partial_x\psi) -
i\Delta\psi{\bar\psi}\right].
\label{Hscaling}
\end{equation} 

The order parameter in the scaling limit must be defined as
\be
\sigma(x)\propto(1-h^2)^{-\frac{1}{8}}\sigma^x_n.
\ee
where $x=na_0$. It is customary to choose the normalization of the
field $\sigma(x)$ such that 
\be
\lim_{x\to 0}\ \langle
0|\sigma(x)\sigma(0)|0\rangle=\frac{1}{|x|^\frac{1}{4}},
\ee
which implies that
\be
\sigma^x_j\rightarrow 2^{1/24}e^{1/8}{\cal
  A}^{-3/2}a_0^{1/8}\sigma(x)\ ,
\ee
where
\be
{\cal A}=1.28242712910062...
\ee

%%%%%%%%%%%%%%%%%%%%%%%%%%%%
\subsection{Ordered Phase}
%%%%%%%%%%%%%%%%%%%%%%%%%%%%
The result for the 2-point function after a quench within the ordered
phase in the scaling limit is
\bea
\fl
\frac{\langle\psi_0(t)|\sigma(x)\sigma(0)|\psi_0(t)\rangle}
{\langle\psi_0|\psi_0\rangle}\propto
\exp\left(\int_0^\infty\frac{dq}{\pi}\ \ln\left[\frac{1-K^2(q)}{1+K^2(q)}\right]
\left[x\theta(2\veps'(q)t-x)+2\veps'(q)t\theta(x-2\veps'(q)t)\right]\right)
,\nn
\label{rhoscaling}
\eea
where $x=na_0$. This expression is well-defined because at large $q$ we have
\be
K(q)\sim\frac{\Delta-\Delta_0}{2vq}.
\ee
In the stationary state we have
\be
\lim_{t\to\infty}\frac{\langle\psi_0(t)|\sigma(x)\sigma(0)|\psi_0(t)\rangle}
{\langle\psi_0|\psi_0\rangle}\propto\exp\left(-\frac{x}{\zeta}\right),
\ee
where
\bea
\zeta^{-1}&=&\frac{\Delta+\Delta_0}{2v}-\frac{\sqrt{\Delta\Delta_0}}{v}.
\label{scalingzeta}
\eea

%%%%%%%%%%%%%%%%%%%%%%%%%%%%%%
\subsection{Disordered Phase}
%%%%%%%%%%%%%%%%%%%%%%%%%%%%%%
For quenches within the disordered phase we can take the scaling limits
of the results \fr{2point_res_para} and \fr{F1}, which give
\be
\frac{\langle\psi_0(t)|\sigma(x)\sigma(0)|\psi_0(t)\rangle}
{\langle\psi_0|\psi_0\rangle}\propto{\cal F}_1(x)+{\cal F}_2(t,x)\ ,
\ee
where the time-dependent part dominates except at very late times and
is given by
\bea
\fl\qquad
F_2(t,x)\approx&
2v\int_{-\infty}^\infty\frac{dq}{2\pi}\frac{K(q)}{\veps(q)}\
\sin\big(2t\veps(q)-qx\big)\nn
&\times
\exp\left(\int_0^\infty\frac{dq}{\pi}\ \ln\left[\frac{1-K^2(q)}{1+K^2(q)}\right]
\left[x\theta(2\veps'(q)t-x)+2\veps'(q)t\theta(x-2\veps'(q)t)\right]\right)
.
\eea
The stationary state component is
\be
{\cal F}_1(x)\propto\exp\left(-\frac{x}{\tilde{\zeta}}\right),
\ee
where now
\bea
\tilde{\zeta}^{-1}&=&\frac{\Delta+\Delta_0}{2v}-\frac{\sqrt{\Delta\Delta_0}}{v}
+\frac{{\rm min}\big(\Delta_0,\sqrt{\Delta\Delta_0}\big)}{v}.
\label{scalingzeta_para}
\eea

%%%%%%%%%%%%
\section{Conclusions} %
\label{sec:Conclusions}
%%%%%%%%%%%%
In this work we have derived analytic expressions for the time
evolution of one and two point functions in the transverse field Ising
chain after a sudden quench of the magnetic field. To do so we have
developed two novel methods based on determinants and form factor sums
respectively. The former is applicable to quenches in models with free
fermionic spectrum and our analysis generalizes straightforwardly
e.g. to the spin-1/2 XY chain in a magnetic field
\cite{fagotti_unpublished}. Results obtained by this method are
\emph{exact} for asymptotically large times and distances 
in what we call the space-time-scaling limit ($t,\ell\to\infty$
keeping their ratio fixed).
The form factor approach is applicable more generally to integrable
quenches \cite{fm-10} in integrable quantum field theories \cite{SE12}
such as the sine-Gordon model. It is furthermore straightforwardly
extended to the study of non-equal time correlation functions \cite{EEF12}.
The form factor method provides approximate results that become exact
in the limit of small quenches, defined by the requirement that the
density of excitations (of the post-quench Hamiltonian) in the initial
state is low. We observe that the difference of the form factor and
exact results for quenches within either the ferromagnetic or
paramagnetic phase are generally very small, except for quenches
originating or terminating in the close vicinity of the quantum
critical point.

%%%%%%%%%%%%%%%%%%%%%%%%%%%%
\ack
We thank John Cardy, Dirk Schuricht and Alessandro Silva for helpful
discussions. This work was supported by the EPSRC under grant
EP/I032487/1 (FHLE and MF), the ERC under the Starting Grant n. 279391
EDEQS (PC), and by the ESF network INSTANS (PC and MF).
We thank the Galileo Galilei Institute for Theoretical Physics for the
hospitality and the INFN for partial support during the completion of
this work.

\appendix
\setcounter{section}{0}
\section{Diagonalization of the Transverse Field Ising Model}
\label{app:diag}
%%%%%%%%%%%%%%%%%%%%%%%%%%%%%%%%%%%%%%%%%%%%%%%%%%%%%%%%%%%%%%
In this Appendix we summarize the diagonalization of the TFIM \cite{n-67}
with periodic boundary conditions 
\be
H(h)=-J\sum_{j=1}^{L}[\sigma_j^x\sigma_{j+1}^x+h\sigma_j^z]\, ,
\ee 
where $l$ is even, $\sigma_j^\alpha$ are the Pauli matrices at site $j$ and
\be
\sigma^\alpha_{L+1}=\sigma^\alpha_1\ ,\quad \alpha=x,y,z.
\ee
The dimensionless constant $h$ describes the coupling with an external
magnetic field $J h$. The quantum Ising chain is mapped to a model of
spinless fermions by means of a Jordan-Wigner transformation. Defining
$\sigma^\pm_j=\big(\sigma^x_\ell\pm i\sigma^y_j\big)/2$ we construct
spinless fermion creation and annihilation operators by
\be
c_l^\dag=\prod_{j=1}^{l-1}\sigma_j^z \sigma_l^-\ ,\quad
\{c_j,c^\dagger_l\}=\delta_{j,l}.
\ee
The inverse transformation is
\be
\sigma^z_j=1-2c^\dagger_jc_j\ ,\quad
\sigma^x_j=\prod_{l=1}^{j-1}(1-2c^\dagger_lc_l)(c_j+c^\dagger_j)\ .
\ee
The Hamiltonian can be expressed in terms of the fermions as
\bea
H(h)&=&-J\sum_{j=1}^{L-1} [c^\dagger_j-c_j][c_{j+1}+c^\dagger_{j+1}]
-Jh\sum_{j=1}^Lc_jc^\dagger_j-c^\dagger_jc_j\nn
&-&Je^{i\pi {\hat N}}(c_L-c_L^\dagger)(c_1+c_1^\dagger),
\eea
where
\be
\hat{N}=\sum_{j=1}^Lc^\dagger_j c_j.
\label{fermionnumber}
\ee
As $[H,e^{i\pi {\hat N}}]=0$ we may diagonalize the two operators
simultaneously. The Hamiltonian is block diagonal $H=H_e\oplus H_o$,
where $H_{e/o}$ act on the subspaces of the Fock space with an even/odd
number of fermions respectively. 
%%%%%%%%%%%%%%%%%%%%%%%%%%%%%%%%%%%
\subsection{Even Fermion Number}
%%%%%%%%%%%%%%%%%%%%%%%%%%%%%%%%%%%
In the sector with an even number of fermions we have
$e^{i\pi{\hat{N}}}=1$ and the Hamiltonian can be written in the form
\bea
H_e(h)&=&-J\sum_{j=1}^{L} [c^\dagger_j-c_j][c_{j+1}+c^\dagger_{j+1}]
-Jh\sum_{j=1}^Lc_jc^\dagger_j-c^\dagger_jc_j\ ,
\label{Heven}
\eea
where we have imposed antiperiodic boundary conditions on the fermions
\be
c_{L+1}=-c_1.
\label{boundary conditions}
\ee
The Hamiltonian $H_e$ is diagonalized by going to Fourier space
\be
c(k_n)=\frac{1}{\sqrt{L}}\sum_{j=1}^Lc_j\ e^{ik_n j},
\ee
where $k_n$ are quantized according to (\ref{boundary conditions})
\be
k_{n}=\frac{2\pi (n+1/2)}{L}\ ,\quad n=-\frac{L}{2},\ldots \frac{L}{2}-1.
\label{NS}
\ee
The antiperiodic sector is commonly referred to as Neveu-Schwarz (NS)
sector. Following this nomenclature we introduce the notation
$k\in {\rm NS}$ to describe the set \fr{NS}.
Introducing Bogoliubov fermions by
\bea
c(k_n)&=&\cos(\theta_{k_n}/2) \alpha_{k_n}
+ i\sin(\theta_{k_n}/2)\alpha_{-k_n}^\dagger\ ,\nn
c^\dagger(-k_n)&=&i\sin(\theta_{k_n}/2) \alpha_{k_n}
+\cos(\theta_{k_n}/2)\alpha_{-k_n}^\dagger,
\label{Bogoliubovtrafo}
\eea
where the Bogoliubov angle fulfils
\bea
%\tan\theta_k=\left[\frac{\sin(k)}{\cos(k)-h}\right],
e^{i\theta_k}=\frac{h-e^{ik}}{\sqrt{1+h^2-2h\cos k}},
\label{angle<}
\eea
the Hamiltonian becomes diagonal
\be
H_e(h)=\sum_{n=-\frac{L}{2}}^{\frac{L}{2}-1}
\veps(k_n)\left[\alpha^\dagger_{k_{n}}\alpha_{k_{n}}-\frac{1}{2}\right].
\ee
Here the dispersion relation is
\bea
\veps_k&=&2J\sqrt{1+h^2-2h\cos(k)}.
\eea
A basis for the Fock space in the sector with even fermion number is
then given by
\bea
|k_1,\ldots,k_{2m};h\rangle_{\rm NS}&=&\prod_{j=1}^{2m}
\alpha^\dagger_{k_{j}}|0;h\rangle_{\rm NS}\ ,\quad
k_j\in{\rm NS},
\eea
where the fermion vacuum $|0;h\rangle_{\rm NS}$ is the state annihilated
by all $\alpha_{k_j}$ ($j=-\frac{L}{2},\ldots,\frac{L}{2}-1$).
%%%%%%%%%%%%%%%%%%%%%%%%%%%%%%%%%%
\subsection{Odd Fermion Number}
%%%%%%%%%%%%%%%%%%%%%%%%%%%%%%%%%%
In the sector with an odd number of fermions we have
$e^{i\pi{\hat{N}}}=-1$. The Hamiltonian can again be written in the form
\bea
H_o(h)&=&-J\sum_{j=1}^{L} [c^\dagger_j-c_j][c_{j+1}+c^\dagger_{j+1}]
-Jh\sum_{j=1}^Lc_jc^\dagger_j-c^\dagger_jc_j\ ,
\label{Hodd}
\eea
but now we have to impose periodic boundary conditions on the fermions
\be
c_{L+1}=c_1.
\label{periodic boundary conditions}
\ee
In Fourier space we therefore now have
\be
c(p_n)=\frac{1}{\sqrt{L}}\sum_{j=1}^Lc_j\ e^{ip_n j},
\ee
where $p_n$ are quantized according to (\ref{periodic boundary conditions})
\be
p_{n}=\frac{2\pi n}{L}\ ,\quad n=-\frac{L}{2},\ldots \frac{L}{2}-1.
\label{R}
\ee
The periodic sector is known as Ramond sector and we will denote the
set \fr{R} by $p_n\in {\rm R}$. Defining Bogoliubov fermions
${\alpha}_{p_n}$ for $p_n\neq 0$ by 
\bea
c(p_n)&=&\cos(\theta_{p_n}/2) \alpha_{p_n}
+ i\sin(\theta_{p_n}/2)\alpha_{-p_n}^\dagger\ ,\nn
c^\dagger(-p_n)&=&i\sin(\theta_{p_n}/2) \alpha_{p_n}
+\cos(\theta_{p_n}/2)\alpha_{-p_n}^\dagger,
\label{Bogoliubovtrafo2}
\eea
we can express the Hamiltonian as 
\bea
%\fl
H_o(h)=\sum_{{n=-\frac{L}{2}}\atop{n\neq0}}^{\frac{L}{2}-1}
\veps(p_n)\left[{\alpha}^\dagger_{p_{n}}{\alpha}_{p_{n}}-\frac{1}{2}\right]
-2J(1-h)\left[{\alpha}^\dagger_{0}
{\alpha}_{0}-\frac{1}{2}\right].
\eea
A basis of the subspace of the Fock space with odd fermion numbers is
then given by
\bea
|p_1,\ldots,p_{2m+1};h\rangle&=&\prod_{j=1}^{2m+1}
\alpha^\dagger_{p_{j}}|0;h\rangle_{\rm R}\ ,\quad p_j\in {\rm R},
\eea
where the fermion vacuum $|0\rangle_{\rm R}$ is the state annihilated by all
$\alpha_{p_j}$ ($j=-\frac{L}{2},\ldots,\frac{L}{2}-1$).

%%%%%%%%%%%%%%%%%%%%%%%%%%%%%%%%%%%%%%%
\subsection{Paramagnetic Phase $h>1$}
%%%%%%%%%%%%%%%%%%%%%%%%%%%%%%%%%%%%%%%%
Here the ground state is
\be
|0\rangle_{\rm NS}.
\ee
A complete set of states is then given by
\bea
|p_1,\ldots,p_{2m+1};h\rangle_{\rm R}&=&\prod_{k_j\in{\rm R}}^{2m+1}
\alpha^\dagger_{p_j}|0;h\rangle_{\rm R}\ 
,\nn 
|k_1,\ldots,k_{2m};h\rangle_{\rm NS}&=&\prod_{p_j\in{\rm NS}}^{2m}
\alpha^\dagger_{k_j}|0;h\rangle_{\rm NS}\ .
\eea
The Hamiltonians can be written as
\bea
H_e(h)&=&\sum_{k_n\in {\rm NS}}
\veps(k_n)\ {\alpha}^\dagger_{k_{n}}{\alpha}_{k_{n}}
+E_0^{\rm NS}(h)\ ,\nn
H_o(h)&=&\sum_{p_n\in {\rm R}}
\veps(p_n)\ {\alpha}^\dagger_{p_{n}}{\alpha}_{p_{n}}
+E_0^{\rm R}(h),
\label{hamils}
\eea
where $E_0^{\rm a}(h)=-\frac{1}{2}\sum_{q\in {\rm a}}\veps(q)$, ${\rm a}={\rm
  R}, {\rm NS}$.
%%%%%%%%%%%%%%%%%%%%%%%%%%%%%%%%%%%%%%%
\subsection{Ferromagnetic Phase $h<1$}
\label{ferro}
%%%%%%%%%%%%%%%%%%%%%%%%%%%%%%%%%%%%%%%%
As the zero momentum mode has negative energy it is useful to perform
a particle-hole transformation
\be
\alpha_0\longrightarrow{\alpha}^\dagger_0\ .
\ee
Redefining the Ramond vacuum as the state that is annihilated by
all $\alpha_{p_n}$ after the particle-hole transformation we can
construct a complete set of states as
\bea
|k_1,\ldots,k_{2m};h\rangle_{\rm R}&=&\prod_{k_j\in{\rm R}}^{2m}
\alpha^\dagger_{k_j}|0;h\rangle_{\rm R}\ 
,\nn 
|p_1,\ldots,p_{2m};h\rangle_{\rm NS}&=&\prod_{p_j\in{\rm NS}}^{2m}
\alpha^\dagger_{p_j}|0;h\rangle_{\rm NS}\ .
\eea
The Hamiltonians are then again given by \fr{hamils}.
%\bea
%H_e(h)&=&\sum_{k_n\in {\rm NS}}
%\veps(k_n)\ {\alpha}^\dagger_{k_{n}}{\alpha}_{k_{n}}
%+C_e(h)\ ,\nn
%H_o(h)&=&\sum_{p_n\in {\rm R}}
%\veps(p_n)\ {\alpha}^\dagger_{p_{n}}{\alpha}_{p_{n}}
%+C_o(h)-2J(1-h).
%\eea
%but now
%\be
%E_0^{\rm NS}(h)=-\frac{1}{2}\displaystyle\sum_{k\in{\rm NS}}\veps_k\ ,\quad
%E_0^{\rm R}(h)=-\frac{1}{2}\displaystyle\sum_{p\in{\rm R}}\veps_p.
%\ee
For large $L$ we have
$
E_0^{\rm NS}(h)-E_0^{\rm R}(h)={\cal O}\big(L^{-1}\big),
$
so that there are two low-energy states
\be
|0;h\rangle_{\rm R}\ ,\quad
|0;h\rangle_{\rm NS}.
\label{groundstatesL}
\ee
As long as $L$ is finite the ground state is $|0;h\rangle_{\rm NS}$.
On the other hand, in the thermodynamic limit the states
\fr{groundstatesL} become degenerate and by spontaneous symmetry
breaking one of the two combinations 
\be
\frac{1}{\sqrt{2}}\left[|0;h\rangle_{\rm R}\pm
|0;h\rangle_{\rm NS}\right]
\ee
is selected as the ground state.
%%%%%%%%%%%%%%%%%%%%%%%%%%%%%%%%%%%%%%%%%%%%%%%%%%%%%%%%%%
\section{Initial State}
\label{app:initial}
%%%%%%%%%%%%%%%%%%%%%%%%%%%%%%%%%%%%%%%%%%%%%%%%%%%%%%%%%%
As described in \ref{app:diag} the Hamiltonian $H(h_0)$ can be
diagonalized by a Bogoliubov transformation. Let us denote the
corresponding Bogoliubov fermions by $\tilde{\alpha}_k$, the
Bogoliubov angle by ${\theta}^0_k$ and the NS
vacuum by $|0;h_0\rangle_{\rm NS}$. Similarly the Hamiltonian
$H(h)$ is diagonalized by the Bogoliubov transformation
\fr{Bogoliubovtrafo} and its lowest energy state in the even fermion
sector is $|0;h\rangle_{\rm NS}$. As both sets of Bogoliubov fermions
are given in terms of the same spinless fermions $c_j$ and
$c^\dagger_j$ ($j=1,\ldots,L$), they can be expressed in terms of one
another by
\bea
\tilde{\alpha}_{k_n}=
\cos\Big(\frac{{\theta}_{k_n}-{\theta}^0_{k_n}}{2}\Big)
{\alpha}_{k_n}
+i\sin\Big(\frac{{\theta}_{k_n}-{\theta}^0_{k_n}}{2}\Big)
{\alpha}^\dagger_{-k_n}\ .
\label{relation}
\eea
As both sets of fermions can be used to construct a basis of the even
Fock space we can express $|{0};h_0\rangle_{\rm NS}$ in the form
\bea
|0;h_0\rangle_{\rm NS}=\sum_{n=0}^\infty\sum_{k^{(1)},\ldots,k^{(n)}\in{\rm NS}}
f_{k^{(1)},\ldots, k^{(n)}}
\prod_{j=1}^n\alpha^\dagger_{k^{(j)}}|0;h\rangle_{\rm NS}
\eea
Using the expression \fr{relation} in the condition
\be
\tilde{\alpha}_{k_n}|{0};h_0\rangle_{\rm NS}=0,
\ee
allows determination of the coefficients $f_{k^{(1)},\ldots,
  k^{(n)}}$. A simple calculation gives
\be
|{0};h_0\rangle_{\rm NS}=\frac{1}{{\cal N}_{\rm NS}}
\exp\left[i\sum_{p\in{\rm NS}}K(p)
\alpha^\dagger_{-p}\alpha^\dagger_{p}\right]
|{0};h\rangle_{\rm NS},
\ee
where ${\cal N}_{\rm NS}$ is a normalization constant and the function
$K(k)$ is given by 
\be
K(k)=\tan\Bigl(\frac{\theta_k-{\theta}^0_k}{2}\Bigr)\, .
\label{eq:k}
\ee
The equations of motion for $\alpha_k$ imply that
\be
\alpha_k(t)=e^{it\veps_k}\alpha_k(0),
\ee
so that \cite{rsms-08}
\be
e^{-itH_e(h)}|{0};h_0\rangle_{\rm NS}=\frac{|B(t)\rangle_{\rm NS}}
{\sqrt{{}_{\rm NS}\langle B|B\rangle}_{\rm NS}}\ ,
\ee
where
\bea
|B(t)\rangle_{\rm NS}= e^{-itE_0^{\rm NS}}
\exp\left[i\sum_{0<p\in{\rm NS}}
e^{-2it\veps_p}K(p)
\alpha^\dagger_{-p}\alpha^\dagger_p\right]
|{0};h\rangle_{\rm NS}.
\label{BS_NS}
\eea
Similarly one can show that
\be
e^{-itH_o(h)}|{0};h_0\rangle_{\rm R}=\frac{|B(t)\rangle_{\rm R}}
{\sqrt{{}_{\rm R}\langle B|B\rangle}_{\rm R}}\ ,
\ee
where
\bea
|B(t)\rangle_{\rm R}=
e^{-itE_0^{\rm R}}
\exp\left[i\sum_{0<p\in{\rm R}}
e^{-2it\veps_p}K(p)
\alpha^\dagger_{-p}\alpha^\dagger_p\right]
|{0};h\rangle_{\rm R}.
\label{BS_R}
\eea
The states \fr{BS_NS} and \fr{BS_R} are of the same form as
\emph{boundary states} in integrable scattering theories
\cite{boundarystate,cc-06}. 

A physical interpretation of the function $K(k)$ \fr{eq:k} is obtained
as follows. The density of post-quench Bogoliubov fermions
$\alpha^\dagger_k\alpha_k$ in the initial state is given by
\bea
\fl\qquad
%
% typo corrected h_0 -> h in expectation value on the rhs
%
{}_{\tt a}\langle 0;h_0|\alpha^\dagger_k\alpha_k|0;h_0\rangle_{\tt
  a}&=\frac{1}{{\cal N}_{\tt a}^2}\
{}_{\tt a}\langle 0;h|\prod_{k\in {\tt a}}
\left(1-iK(k)\alpha_k\alpha_{-k}\right)
\alpha^\dagger_k\alpha_k
\left(1+iK(k)\alpha^\dagger_{-k}\alpha^\dagger_{k}\right)|0;h\rangle\nn
\fl\qquad
&=
\frac{K^2(k)}{1+K^2(k)}\ ,\quad {\tt a}=\NS,\R.
\eea
This implies that in the case where $K(k)$ is uniformly small
in $k$ we have
\be
\langle\Psi_0|\alpha^\dagger_k\alpha_k|\Psi_0\rangle=K^2(k)+{\cal
  O}\left(K^4\right),
\label{densityofexc}
\ee
where $|\Psi_0\rangle$ is the initial state of our quantum
quench. Physically the small parameter characterizing the expansion in
powers of $K(k)$ is therefore the density of excitations of the
post-quench Hamiltonian $H(h)$ induced by the quantum quench.

\section{Proof of the product formula}
\label{App:prod}

We need to evaluate the product of the $2\times2$ matrices
\be
\Pi_{2n}(\{a_i\})\equiv\prod_{i=1}^{2n}{\mathbb M}(a_i), \qquad {\rm with}\; {\mathbb M}(a)= A \s_x +B \s_ye^{i a \s_x}\,.
\ee
Compared to Eq. (\ref{fdef}) in the main text we have $A=n_x(k_0)$, $B=|\vec n_\perp(k_0)|$, $a_i=2\eps_{i-1} t$
 and we choose (without loss of generality) $\hat n_\perp=\hat y$.
Using the algebra of the Pauli matrices, it is straightforward to calculate
\be\fl
\Pi_2(a_1,a_2)={\mathbb M(a_1)}\cdot{\mathbb M(a_2)}=
A^2 {\mathbb I}+B^2 e^{i(a_2-a_1) \s_x} +i A B \s_z (e^{i a_2 \s_x}-e^{ia_1 \s_x})\,,
\ee 
where ${\mathbb I}$ is the 2 by 2 identity matrix.
A slightly longer exercize is required to calculate $\Pi_4$ and to obtain
\bea\fl
\Pi_4&=& A^4 {\mathbb I}+i A^3 B\s_z\sum_{j=1}^4 (-1)^j e^{i a_i \s_x}+
A^2B^2\sum_{1\leq j_1<j_2 \leq 4} (-1)^{j_1+j_2+1} e^{i(a_{j_2}-a_{j_1})\s_x}
\\\fl &&
+iAB^3\s_z\sum_{1\leq j_1<j_2<j_3\leq4}(-1)^{j_1+j_2+j_3+1}e^{i(a_{j_3}-a_{j_2}+a_{j_{1}})\s_x}+
B^4 e^{i(a_4-a_3+a_2-a_1)\s_x}.\nonumber
\eea
From these two first examples, it should be clear that the general structure of $\Pi_{2n}$ is
\be\fl
\Pi_{2n}=\sum_{p=0}^{2n} A^{2n-p} B^p (i\s_z)^p\sum_{1\leq j_1<j_2<\dots j_p\leq 2n} 
(-1)^{\sum_{k=1}^p j_k} \exp\Big(i{\sum_{k=1}^p (-1)^{p-k} a_{j_k} \s_x}\Big).
\ee
Having this conjecture, it is straightforward (but require some algebra) to prove it by induction showing that it
is compatible with the recurrence relation
\be
\Pi_{2n+2}(a_1,\dots a_{2n+2})=\Pi_{2n}(a_1,\dots a_{2n}) \cdot \Pi_2(a_{2n+1},a_{2n+2})\,.
\ee

\section{Useful relations}
\label{a:useful}
%%%%%%%%%%%%%%%%%%%%%%%%%%%%%%%%%%%%%%%%%%%%%%%%%%%%%%%%%%%
\underline{\emph{Lemma 1:}}
\label{lemma1}
For any function $f(z)$ that is $2\pi$ periodic and analytic in a
strip around the real axis we have for $0<k\in{\rm R}$
\bea
\fl
\frac{1}{L}\sum_{q_n\in{\rm NS}>0}
\frac{f(q_n)e^{2it\veps(q_n)}}{(\cos k-\cos q_n)^2}
&=&
\left[\frac{L}{4}-t\veps'(k)\right]
\frac{e^{2it\veps_k}f(k)}{\sin^2k}
+i\frac{e^{2i\veps_kt}f'(k)}{2\sin^2(k)}\nn
&&-\oint \frac{dz}{2\pi}\frac{f(z)e^{2i\veps(z)t}
}{(\cos z-\cos k)^2(1+e^{iLz})},
\label{id1}
\eea
where the integration is along a closed contour encircling the
interval $[0,\pi]$.

\vskip .25cm
\noindent\underline{\emph{Lemma 2a:}}
\label{lemma2a}
For any function $f(z)$ that is $2\pi$ periodic and analytic in a
strip around the real axis we have for $0<k\in{\rm R}$
\bea
\fl
\frac{1}{L}\sum_{q_n\in{\rm R}>0\atop q_n\neq k}
\frac{f(q_n)e^{2it\veps(q_n)}}{(\cos k-\cos q_n)^2}
&=&
\left[\frac{L}{12}\left(1-\frac{6i\cot k}{L}\right)-t\veps'(k)\right]
\frac{e^{2it\veps_k}f(k)}{\sin^2k}\nn
&+&i\frac{e^{2i\veps_kt}f'(k)}{2\sin^2(k)}
-\oint \frac{dz}{2\pi}\frac{f(z)e^{2i\veps(z)t}
}{(\cos z-\cos k)^2(1-e^{iLz})}+{\cal O}\big(L^{-1}\big).
\label{id1b}
\eea
\vskip .25cm
The analogous equation for momenta in the NS sector is 

\noindent\underline{\emph{Lemma 2b:}}
\label{lemma2b}
For any function $f(z)$ that is $2\pi$ periodic and analytic in a
strip around the real axis we have for $0<k\in{\rm NS}$
\bea
\fl
\frac{1}{L}\sum_{q_n\in{\rm NS}>0\atop q_n\neq k}
\frac{f(q_n)e^{2it\veps(q_n)}}{(\cos k-\cos q_n)^2}
&=&
\left[\frac{L}{12}\left(1-\frac{6i\cot k}{L}\right)-t\veps'(k)\right]
\frac{e^{2it\veps_k}f(k)}{\sin^2k}\nn
&+&i\frac{e^{2i\veps_kt}f'(k)}{2\sin^2(k)}
-\oint \frac{dz}{2\pi}\frac{f(z)e^{2i\veps(z)t}
}{(\cos z-\cos k)^2(1+e^{iLz})}+{\cal O}\big(L^{-1}\big).
\label{id1c}
\eea

\noindent\underline{\emph{Lemma 3:}}
\label{lemma3}
For large $j\gg 1$ and any function $f(z)=f(z+2\pi)$
that is analytic in a strip around the real axis we have ($k\in\R$)
\bea
\lim_{L\to\infty}\frac{1}{L}\sum_{q_n\in{\rm NS}} \frac{f(q_n)e^{ijq_n}}
{\cos k-\cos q_n}= \frac{e^{-ikj}f(-k)-e^{ikj}f(k)}{2i\sin k}+{\cal
  O}\Big(e^{-\gamma j}\Big),
\label{id3}
\eea
where $\gamma$ is a positive constant. 

\begin{figure}[ht]
\begin{center}
\epsfxsize=0.8\textwidth
\epsfbox{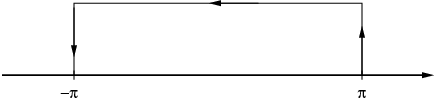}
\end{center}
\caption{Deformed integration contour.}
\label{fig:contour3}
\end{figure}

\noindent\emph{Proof:} Using contour integration we have
\bea
\frac{1}{L}\sum_{q_n} \frac{f(q_n)e^{ijq_n}}{\cos k-\cos q_n}&=&\oint 
\frac{dz}{2\pi}\frac{f(z)e^{ijz}}{(\cos z-\cos k)(1+e^{iLz})}\nn
&+&\frac{e^{-ikj}f(-k)-e^{ikj}f(k)}{2i\sin k},
\label{id3a}
\eea
where we have used the quantization conditions \fr{NS}, \fr{R} and
where the integration is along a counterclockwise countour encircling
the interval $[-\pi,\pi]$ in a manner such that no singularities of
$f(z)$ lie within it. The contribution of the part of the
contour below the real axis tends to zero in the limit $L\to\infty$
because $|e^{iz(L\pm j)}|\gg 1$. As $j>0$ the part of the contour
above the real axis can be deformed as shown in Fig.~\ref{fig:contour3}.
The contributions of the pieces parallel to the imaginary axis cancel
due to the periodicity of $f(z)$, while the remaining part has the
exponentially small bound given in \fr{id3}.
%If $j<0$ we have to move the contour into the lower half plane and in
%doing so we pick up a contribution from the simple poles at $z=\pm
%k$. The remaining integral is again exponentially small in $|j|$,
%while the pole contribution combines with the second contribution on
%the right hand side of \fr{id3a} to give the result quoted in \fr{id3}.

\noindent\underline{\emph{Lemma 4:}}
\label{lemma4}
For any function $f(z)$ that is $2\pi$ periodic and analytic in a
strip around the real axis we have 
\bea
\fl\qquad
\frac{1}{L}\sum_{q_n\in{\rm NS}}
\frac{f(q_n)e^{ijq_n}}{(\cos k-\cos q_n)^2}
=-\frac{e^{-ikj}f'(-k)+e^{ikj}f'(k)}{2i\sin^2k}\nn
\qquad\qquad\qquad+
\left[\frac{L}{4}-\frac{j}{2}\right]
\frac{e^{-ikj}f(-k)+e^{ikj}f(k)}{\sin^2k}
+{\cal O}\Big(e^{-\gamma j}\Big).
\label{id4a}
\eea
\noindent\emph{Proof:} Using contour integration we have
\bea
\fl
\frac{1}{L}\sum_{q_n\in{\rm NS}} \frac{f(q_n)e^{ijq_n}}{(\cos k-\cos
  q_n)^2}
&=&\frac{e^{-ikj}f(-k)+e^{ikj}f(k)}{\sin^2(k)}\left[\frac{L}{4}-\frac{j}{2}
\right]\nn
&+&i\frac{e^{-ikj}f'(-k)+e^{ikj}f'(k)}{2\sin^2(k)}
-\oint 
\frac{dz}{2\pi}\frac{f(z)e^{ijz}
}{(\cos z-\cos k)^2(1+e^{iLz})},
\label{id4}
\eea
where the integration is along a closed contour encircling the
interval $[-\pi,\pi]$. As for ${\rm Im}(z)<0$ we have
\be
\lim_{L\to\infty}e^{i(j-L)z}=0,
\ee
only the upper part of the contour contributes in the $L\to\infty$
limit. Using that $f(z)$ is analytic in some strip around the real axis
the part of the contour above the real axis can be deformed as shown
in Fig.~\ref{fig:contour3}. The contributions of the pieces parallel to
the imaginary axis cancel due to the periodicity of $f(z)$, while the
remaining part is exponentially small in $j$.
In the limit of large $L$ and $j$ the first term in \fr{id4a} can be
neglected, so that
\bea
\frac{1}{L}\sum_{q_n} \frac{f(q_n)e^{ijq_n}}{(\cos k-\cos q_n)^2}
\approx
\frac{e^{-ikj}f(-k)+e^{ikj}f(k)}{\sin^2(k)}\left[\frac{L}{4}-\frac{j}{2}
\right].
\label{id4b}
\eea
%\begin{figure}[ht]
%\begin{center}
%\vepsfxsize=0.4\textwidth
%\vepsfbox{contour3.eps}
%\end{center}
%\caption{Deformed integration contour.}
%\label{fig:contour3}
%\end{figure}
\noindent\underline{\emph{Lemma 5:}}
\label{lemma5}
For any function $f(z)$ that is $2\pi$ periodic and analytic in a
strip around the real axis we have 
\bea
\lim_{L\to\infty}\frac{1}{L}\sum_{0<q\in{\rm NS}}
\frac{f(q)\ e^{2i\sigma\veps(q)t}}{\cos k-\cos q}&=&
\frac{i\sigma f(k)\ e^{2i\sigma\veps_kt}}{2\sin k}
+\int_{\sigma i0}^{\pi+\sigma i0}\frac{dz}{2\pi}
\frac{f(z)\ e^{2i\sigma\veps(z)t}}{\cos k-\cos z},
\label{id5}
\eea
where $\sigma=\pm$ and $0<k\in{\rm R}$.

%%%%%%%%%%%%%%%%%%%%%%%%%%%%%%%%%%%%%%
\section{``Pair Ensemble'' Averages}
\label{app:PE}
%%%%%%%%%%%%%%%%%%%%%%%%%%%%%%%%%%%%%%
The ``diagonal'' terms in the Lehmann representation based on
the eigenstates of the post quench Hamiltonian $H(h)$ give rise to a
\emph{time-independent} contribution to the expectation value $\langle
\Psi_0(0)|{\cal O}|\Psi_0(0)\rangle$, which we call
``pair ensemble''. It is defined as the average
\bea
\quad\fl
\langle {\cal O}\rangle_{\rm PE}\equiv&
\frac{1}{{}_{\NS}\langle B|B\rangle_{\NS}}\
\sum_{n=0}^\infty\frac{1}{n!}
\ \sumprime_{0<k_1,\dots,k_n\in\NS}\
\left[\prod_{j=1}^nK^2(k_j)\right]\nn
\fl
&\qquad\times\ {}_{\rm NS}\langle -k_n,k_n,\ldots,-k_1,k_1;h|{\cal O}
|k_1,-k_1\ldots,k_n,-k_n;h\rangle_{\rm NS}\ .
\label{F1_again}
\eea
Averages of the form \fr{F1_again} can be represented using a density
matrix as $\langle{\cal O}\rangle_{\rm PE}={\rm tr}\big(\rho_{\rm
  PE}{\cal O}\big)$, where
\be
\rho_{\rm PE}=(1-n_0)(1-n_\pi)\prod_{0<k\in \NS}\Bigl(\frac{\mathbb{I}_k \mathbb{I}_{-k}-n_k  \mathbb{I}_{-k}-\mathbb{I}_{k}  n_{-k}}{1+ K^2(k)}+n_{k}n_{-k}\Bigr)\, .
\ee
Here $n_q=\alpha^\dagger_q\alpha_q$ are the Bogoliubov fermion number
operators in momentum space and $\mathbb{I}_k$ is the identity in the
subspace with momentum $k$. The only non-vanishing expectation
values of the pair ensemble are 
\be
\langle n_{q_1}n_{q_2}\dots n_{q_m}\rangle_{\rm PE}=\prod_{|q_j|\in S}
\frac{K^2(|q_j|)}{1+K^2(|q_j|)}\ ,
\label{pair1}
\ee
where $S$ is the set of all momenta $q_j$ with \emph{mutually
  distinct} magnitudes, i.e.
\be
0<|q_r|\neq|q_s|<\pi\ \forall q_r,q_s\in S.
\ee
We note that $\braket{n_k n_{-k}{\cal O}}=\braket{n_k{\cal O}}$ since
the ensemble describes pairs of particles with opposite momenta.

As shown in section \ref{s:Determinant approach} lattice spin
operators can be expressed as products of the Majorana fermions
$a_j^x$, $a_j^y$, which are related to the Bogoliubov fermions
diagonalizing the post-quench Hamiltonian $H(h)$ by
\bea
a_j^x&=\frac{1}{\sqrt{L}}\sum_k e^{-ikj}
e^{i\theta_k/2}\left[\alpha_k+\alpha^\dagger_{-k}\right]\ ,\nn
a_j^y&=\frac{1}{\sqrt{L}}\sum_k e^{-ikj}
e^{-i\theta_k/2}i\left[\alpha_{-k}^\dagger-\alpha_{k}\right].
\eea

As the only non-zero expectation values in the pair ensemble are of
the form \fr{pair1} we can express a general average in the form
\bea
\Big\langle\prod_{r=1}^{2n}a^{b_r}_{j_r}\Big\rangle_{\rm PE}=
\frac{1}{L^n}\sum_{k_1,\ldots, k_n}\sum_{s=0}^{n}
\chi^{(s)}_{\vec{b}}(k_1,\ldots,k_n|\vec{j})\ \langle\prod_{u=1}^sn_{k_u}
\rangle_{\rm PE},
\label{pair2}
\eea
where $\chi^{(s)}_{\vec{b}}(k_1,\ldots,k_n|\vec{j})$ are well-behaved
functions of the momenta. 

The generalized Gibbs ensemble for the Ising model is defined as
\bea
\langle {\cal O}\rangle_{\rm GGE}\equiv&\frac{1}{Z_{\rm GGE}}
{\rm tr}\left[e^{-\sum_q \lambda_qn_q}{\cal O}\right],
\label{GGE}
\eea
where $Z_{\rm GGE}={\rm tr}\left[e^{-\sum_q \lambda_qn_q}\right]$ and
\be
\lambda_q=-\log\big(K^2(q)\big)\, .
\ee
By taking the trace over a basis of eigenstates of $H(h)$ we conclude
that the only non-zero averages are
\be
\langle n_{q_1}n_{q_2}\dots n_{q_m}\rangle_{\rm GGE}
=\prod_{j=1}^n
\frac{K^2(q_j)}{1+K^2(q_j)}\ .
\ee
The corresponding density matrix is
\be
\rho_{GGE}=\prod_{k\in \NS}\frac{\mathbb{I}_k
-\big(1-K^2(k)\big)n_k}{1+K^2(k)}\ .
\ee 
We observe  that $\braket{n_{k_1} n_{k_2}\cdots n_{k_m}}_{\rm
  GGE}=\braket{n_{k_1}}_{\rm GGE}\braket{n_{k_2}\cdots n_{k_m}}_{\rm GGE}$ for
any set of distinct momenta $k_1,k_2,\ldots, k_m$, and unlike in the
pair ensemble Wick's theorem applies in the generalized Gibbs
ensemble. General averages can be expressed as 
\bea
\Big\langle\prod_{r=1}^{2n}a^{b_r}_{j_r}\Big\rangle_{\rm GGE}=
\frac{1}{L^n}\sum_{k_1,\ldots, k_n}\sum_{s=0}^{n}
\chi^{(s)}_{\vec{b}}(k_1,\ldots,k_n|\vec{j})\ \langle\prod_{u=1}^sn_{k_u}
\rangle_{\rm GGE},
\eea
where the functions $\chi^{(s)}_{\vec{b}}(k_1,\ldots,k_n|\vec{j})$ are
the same as in \fr{pair2}. As the averages
$\langle\prod_{u=1}^sn_{k_u} \rangle_{\rm GGE}$ and
$\langle\prod_{u=1}^sn_{k_u} \rangle_{\rm PE}$ are the same unless 
at least two of the $|k_u|$'s coincide, and such
contributions are suppressed by factors of $1/L$, we conclude that
\be
\lim_{L\to\infty}\Big\langle\prod_{r=1}^{2n}a^{b_r}_{j_r}\Big\rangle_{\rm
PE}=
\lim_{L\to\infty}\Big\langle\prod_{r=1}^{2n}a^{b_r}_{j_r}\Big\rangle_{\rm GGE}.
\ee
The above arguments show that averages of operators that are local in
space are the same in both ensembles. However, non-local operators
have in general different averages, e.g.
$\braket{n_k n_{-k}}_{\rm GGE}=\braket{n_k n_{-k}}_{\rm PE}^2$.

\section*{References}

%%%%%%%%%%%%%%%%%%%%%%%%%%%%%%

\end{document}